\documentclass[preprint,3p,12pt]{elsarticle}
\usepackage{mathrsfs}
\usepackage{amsmath}
\usepackage{stmaryrd}
\usepackage{bbding}
\usepackage{dcolumn}
\usepackage{graphicx}
\usepackage{amsfonts}
\usepackage{amssymb}
\usepackage{psfrag}
\usepackage{wrapfig}
\usepackage{subfigure}
\usepackage{makeidx}
\usepackage{bm}
\usepackage{epsf}
\usepackage{epsfig}
\usepackage{setspace}
\usepackage{graphicx}
\usepackage{epstopdf}
\usepackage{psfrag}
\usepackage{subfigure}
\usepackage{color}

\newtheorem{rmk}{Remark}

\begin{document}
\title{Compact Higher-order Gas-kinetic Schemes with Spectral-like Resolution for Compressible Flow Simulations}

\author[HKUST1]{Fengxiang Zhao}
\ead{fzhaoac@connect.ust.hk}

\author[HKUST2]{Xing Ji}
\ead{xjiad@connect.ust.hk}

\author[HKUST1]{Wei Shyy}
\ead{weishyy@ust.hk}

\author[HKUST1,HKUST2]{Kun Xu\corref{cor}}
\ead{makxu@ust.hk}

\address[HKUST1]{Department of Mechanical and Aerospace Engineering, Hong Kong University of Science and Technology, Clear Water Bay, Kowloon, HongKong}
\address[HKUST2]{Department of Mathematics, Hong Kong University of Science and Technology, Clear Water Bay, Kowloon, HongKong}
\cortext[cor]{Corresponding author}

\begin{abstract}

In this paper, a class of compact higher-order gas-kinetic schemes (GKS) with spectral resolution will be presented.
Based on the high-order gas evolution model in GKS,
both the interface flux function and conservative flow variables can be evaluated explicitly from the time-accurate gas distribution function.
 As a result, inside each control volume both the cell-averaged flow variables and their cell-averaged gradients
 can be updated within each time step. The flow variable update and slope update are coming from the same physical solution at the cell
 interface. Different from many other approaches, such as HWENO, there are no additional governing equations in GKS
 for the slopes or equivalent degrees of freedom independently inside each cell.
 Therefore, based on both cell averaged values and their slopes, compact 6th-order and 8th-order linear and nonlinear reconstructions
 can be developed.
 As analyzed in this paper, the local linear compact reconstruction can achieve a spectral-like resolution at large wavenumber than the
 well-established compact scheme of Lele with globally coupled flow variables and their derivatives.
 For nonlinear gas dynamic evolution, in order to avoid spurious oscillation in discontinuous region,
 the above compact linear reconstruction from the symmetric stencil can be divided into sub-stencils
 and apply a biased nonlinear WENO-Z reconstruction.
 Consequently discontinuous solutions can be captured through the 6th-order and 8th-order compact WENO-type nonlinear reconstruction.
 In GKS, the time evolution solution of the gas distribution function at a cell interface is based on an integral solution of the kinetic
 model equation, which covers a physical process from an initial non-equilibrium state to a final equilibrium one.
 Since the initial non-equilibrium state is obtained based on the nonlinear WENO-Z reconstruction,
  and the equilibrium state is basically constructed from the linear symmetric reconstruction,
 the GKS evolution models unifies the nonlinear and linear reconstructions in gas evolution process for the determination of a
   time-dependent gas distribution function, which gives great advantages in capturing both discontinuous shock wave
   and the linear aero-acoustic wave in a single computation due to dynamical adaptation of non-equilibrium and equilibrium state from GKS
   evolution model in different regions.
 This dynamically adaptive model helps to solve a long lasting problem in the development of high-order schemes
 about the choices of the  linear and nonlinear reconstructions.
 Compared with discontinuous Galerkin (DG) scheme, the current compact GKS uses the same local and compact stencil,
 achieves the 6th-order and 8th-order accuracy, uses a much larger time step with CFL number $\geq 0.3$,
 has the robustness as a 2nd-order scheme, and gets accurate solutions in both shock and smooth regions
 without introducing trouble cell and additional limiting process.
 The nonlinear reconstruction in the compact scheme is solely based on the WENO-Z reconstruction.
 At the same time, the current scheme solves the Navier-Stokes equations.
 Due to the use of multi-stage multi-derivative (MSMD) time-stepping technique,
 for achieving a 4th-order time accuracy, the current scheme uses only two stages instead of four in the traditional Runge-Kutta method.
 As a result, the current GKS becomes
 much more efficient than the corresponding same order DG method.
 A variety of numerical tests are presented to validate the compact 6th and 8th-order GKS.
 The current scheme presents a state-of-art numerical solutions under a wide range of flow conditions,
 i.e., strong shock discontinuity, shear instability, aero-acoustic wave propagation, and NS solutions.
 It promotes the development of high-order scheme to a new level of maturity.
  The success of the current scheme crucially depends on the high-order gas evolution model,
  which cannot be achieved by any other approach once the 1st-order Riemann flux function is still used in the high-order algorithm development.
\end{abstract}


\maketitle

\section{Introduction}

High-order methods with spectral resolutions and shock capturing capability
are needed in many engineering applications, such as turbulent flows, aeroacoustics,
and various complex flows with shock and boundary layer interactions.
With the great potentials of the high-order methods in solution accuracy and computational efficiency,
extensive effort has been paid to the development of high-order schemes in the past decades.
However, the spectral accuracy and shock capturing seem to be contradictory in the current CFD algorithms, where
the linear schemes for the smooth flow and nonlinear schemes
for discontinuous flow play different roles under different flow conditions \cite{bai}.
It seems hard to possess both properties in a single high-order method.

For smooth flows, the compact schemes \cite{lele,mahesh} and discontinuous Galerkin methods (DG) \cite{reed,cockburn1,cockburn2} are very attractive.
The compact finite difference scheme constructs implicit relation between derivatives and flow variables on compact stencils.
Dispersion and dissipation properties of the scheme have been fully analyzed \cite{lele}, and the scheme influences greatly on the development of linear schemes.
The reason for designing compact scheme is its flexibility in complex geometries and effective parallelization.
The DG has a second-order scheme stencil, but achieves high accuracy by using
high-order piecewise polynomials within elements and evolving the multiple degrees of freedom (DOFs).
Each element in DG only interacts
with its neighboring elements and the scheme becomes very efficient for parallel computation \cite{shu}.
But, both compact and DG methods seem very successful for the smooth flow, and problems are still remained for flow simulation with
shock discontinuity.
Theoretically, compact scheme is physically sounded because under the CFL condition, the domain of dependence is indeed only related to the
neighboring cells. All non-compact schemes must be associated with dynamical weakness because they seek for help from the
cells far away which may not have any physical connection in terms of the wave propagating speed.

Nonlinear schemes have been designed for the flows with discontinuous shocks.
The successful nonlinear schemes include total variation diminishing (TVD) \cite{harten1},
essentially non-oscillation (ENO) \cite{harten2}, and weighted ENO (WENO) \cite{liu,jiang}.
 In past twenty years, the WENO-based methods have received the most attention among nonlinear schemes.
 The central ingredients in WENO scheme is to construct several low order
polynomials and to design smoothness indicators to adaptively  assemble them to get a higher one.
Most current effort is about the selection of optimal stencil and the design of weighting functions.
WENO scheme can achieve very high-order accuracy in the smooth region and maintain non-oscillatory property across shock waves
\cite{shu1}. But, their numerical dissipation is still very  higher than the linear schemes \cite{hill,taylor}, and the modified
WENO schemes, such as WENO-M and WENO-Z \cite{WENO-M,WENO-Z}, have been proposed recently.
The hybrid linear and nonlinear schemes have been investigated as well \cite{ren,zhang,choi}.

In the existing DG and WENO methods, the first-order Riemann flux plays a key role for the flow evolution of compressible flow.
Recently, beyond the 1st-order Riemann solver, many schemes based on the time-dependent flux functions have been developed,
such as the Lax-Wendroff method \cite{lax}, the generalized Riemann problem (GRP) \cite{grp1,grp2,grp3}, and ADER framework \cite{toro,dumbser}.
An outstanding method is the two-stage fourth-order scheme for the Euler equations \cite{li},
where both the flux and its time derivative are used in the construction of higher-order scheme.
The two-stage fourth-order algorithms have been developed under the multi-stage multi-derivative (MSMD) framework \cite{hairer,seal,christlieb}.
Similar to the GRP method, the gas-kinetic scheme (GKS) is also based on a time accurate flux function at a cell interface \cite{xu1,xu2,xu3}.
The flux evaluation in GKS is based on a time evolution solution of kinetic model equation, which provides a physical process for the gas evolution from the initial non-equilibrium state to an equilibrium one.
The comparison between GRP and GKS has been presented in \cite{li2}.
Under the same MSMD framework, based on the WENO reconstruction a two-stage fourth-order (S2O4) GKS
has been successfully developed for the Euler and Navier-Stokes equations \cite{pan1,pan2}.
The robustness of the S2O4 GKS is as good as the second-order shock capturing scheme \cite{pan0}.
Due to the use of only two reconstructions in the 4th-order scheme, the GKS becomes very efficient
in comparison with schemes based on the Runge-Kutta time stepping technique.
By combining the second-order or third-order GKS fluxes and MSMD technique again, a family of higher-order gas-kinetic methods has been constructed \cite{ji1,ji2}.

The time dependent gas-distribution function in GKS at the cell interface provides not only the flux and its time derivative,
but also time accurate flow variables at the cell interface.
The design of compact GKS based on the cell averaged and cell interface values has been conducted  \cite{xu,pan3,pan4,ji2}.
In  this paper, we are going to develop 6th-order and 8th-order compact GKS in rectangular mesh for
compressible flow simulations.
The construction of the compact high-order schemes is based on the following ingredients.
Firstly, the high-order gas evolution model at cell interface is used for the flux and flow variable evaluation. Consequently, the cell
averaged flow variables and cell averaged slopes can be updated. Secondly, based on the time-dependent flux and its time derivative,
the two-stage fourth-order MSMD technique is used for the 4th-order temporal accuracy. Thirdly, based on the flow variables
and their slopes within each cell, compact nonlinear and linear 6th-order and 8th-order spatial
reconstructions will be designed for the determination of the initial non-equilibrium state and the evolved equilibrium one.
Based on the GKS evolution model, in the smooth flow region the linear reconstruction will be achieved
quickly in the determination of interface gas distribution function, and the spectral-like resolution can be obtained.
At the same time, in the discontinuous region the nonlinear reconstruction will persist in the determination of
gas evolution due to long existence of the non-equilibrium state, and the scheme becomes robust
in capturing shock and discontinuous solutions.
Since the gas evolution model in GKS is based on the physical process from the initial non-equilibrium state to the final equilibrium one,
the nonlinear  and linear reconstructions are naturally unified in a
time evolution process with automatically identified non-equilibrium and equilibrium regions.
As a result, both discontinuous shocks and smooth aeroacoustic waves can be accurately captured by the current compact 6th-order and 8th-order GKS.
The compact GKS provides the Navier-Stokes solutions directly.
 Even for the 6th-order and 8th-order accuracy, the current compact scheme can use a time step determined by the CFL condition
 with a large CFL number $(\geq 0.5)$ in all test cases in this paper.
The current 6th-order and 8th-order compact GKS take advantages of both
nonlinear WENO-Z reconstruction for the introduction of numerical dissipation in discontinuous region
and linear reconstruction with much reduced dissipation in resolving small amplitude waves.
The current compact GKS are ideal schemes for the study of multiscale and complex flow interactions,
such as the flow transition and turbulence.

The idea of using flow variables and their gradients has been investigated in finite volume or finite difference methods.
The finite volume Hermite WENO (HWENO) schemes  update both flow variables and their first derivative \cite{qiu1,qiu2,qiu3}.
The HWENO approach is also used in the hybrid schemes \cite{balsara}, and a monotonicity preserving strategy for detecting
troubled zones is proposed.
The HWENO schemes on unstructured grids have been extensively investigated \cite{luo1,luo2,luo3}.
The HWENO approach needs additional limiters as both DG and
reconstructed DG methods.
The major advantage of these HWENO approaches is the compactness of stencils in the reconstructions.
The current compact GKS is different from the HWENO approach is mainly on the update of gradients.
In GKS, there is no explicit evolution equation for the gradients and the gradients are obtained rigorously from the evolution solutions at cell interfaces.
Therefore, the gradients in GKS are not introduced as additional degrees of freedom with their own evolution equations.
 The cell-averaged gradients in GKS like cell averaged conservative variables, which are obtained from the same
gas distribution functions at cell interfaces.
 The robustness and improved resolution in GKS is due to the physically reliable gradients
 from the evolution solution of the original governing equation.
In other words, the updated gradients in GKS are not coming from the "equations" by taking additional derivatives to the original ones,
such as the HWENO and multi-layer compact (MLC) schemes \cite{bai}.
In the HWENO and MCL, the gradients have their own evolution equations which are only reliable for the
smooth flow, where new governing equations can be continuously created by taking spatial derivatives to the original equation.

This paper is organized as follows. The GKS and MSMD method will be introduced in Section 2.
In Section 3, the compact linear and nonlinear 6th-order and 8th-order reconstructions will be presented.
The dissipation and dispersion will be analyzed as well. In Section 4, the compact GKS will be tested in
wide range of flow problems from the strong shock interaction to the linear acoustic wave propagation.
The last section is the conclusion.

\section{Gas-kinetic scheme and multi-stage multi-derivative method}

In the past years, the gas-kinetic scheme (GKS) has been developed systematically \cite{xu1,xu2}.
The gas evolution model used in this paper for the flux and interface flow variable evaluation is almost identical to that of the
fourth-order compact scheme \cite{ji2}. Therefore, only a brief introduction about GKS will be presented in this section.

\subsection{Gas-kinetic scheme}

The gas kinetic evolution model in GKS is based on the BGK equation \cite{BGK-1},
\begin{equation}\label{bgk}
f_t+\textbf{u}\cdot\nabla f=\frac{g-f}{\tau},
\end{equation}
where $f$ is the gas distribution function, $g$ is the corresponding equilibrium state that $f$ approaches, and $\tau$ is defined as the collision time.
In two-dimensional case, the equilibrium state is the Maxwellian distribution
\begin{equation*}
\begin{split}
g=\rho(\frac{\lambda}{\pi})^{\frac{K+2}{2}}e^{\lambda((u-U)^2+(v-V)^2+\xi^2)},
\end{split}
\end{equation*}
where $\lambda =m/2kT $, and $m, k, T$ represent the molecular mass, the Boltzmann constant, and temperature,
$K$ is the number of internal degrees of freedom, i.e. $K=(4-2\gamma)/(\gamma-1)$ for two-dimensional flow,
and $\gamma$ is the specific heat ratio. $\xi$ is the internal variable with $\xi^2=\xi^2_1+\xi^2_2+...+\xi^2_K$.
The collision term satisfies the following compatibility condition
\begin{equation}\label{compatibility}
\int \frac{g-f}{\tau}\psi \text{d}\Xi=0,
\end{equation}
where $\psi=(\psi_1,\psi_2,\psi_3,\psi_4)^T=(1,u,v,\displaystyle \frac{1}{2}(u^2+v^2+\xi^2))$, $\text{d}\Xi=\text{d}u\text{d}v\text{d}\xi_1...\text{d}\xi_{K}$.

The macroscopic mass $\rho$, momentum ($\rho U, \rho V$), and energy $\rho E$ at a cell interface can be evaluated from the
gas distribution function,
\begin{equation}\label{g-to-convar}
{{W}} =
\left(
\begin{array}{c}
\rho\\
\rho U\\
\rho V\\
\rho E\\
\end{array}
\right)
=\int \psi f d\Xi,
\end{equation}
and the flux in the $x$ direction can be evaluated as well,
\begin{equation}\label{g-to-flux}
{{F}} =\int uf\psi d\Xi.
\end{equation}
Therefore, the central point in GKS is to evaluate the time-dependent gas distribution function $f$ at a cell interface.

By direct modeling on the mesh size scale \cite{xu1}, the conservations of mass, momentum and energy in a control volume get to
written as
\begin{align*}
\frac{\text{d}W_{ij}}{\text{d}t}=-\frac{1}{\Delta x}
(F_{i+1/2,j}(t)-F_{i-1/2,j}(t))-\frac{1}{\Delta y}
(G_{i,j+1/2}(t)-G_{i,j-1/2}(t)),
\end{align*}
where $W_{ij}$ is the cell averaged conservative variables, $F_{i\pm 1/2,j}(t)$ and $G_{i,j\pm 1/2}(t)$
are the time dependent fluxes at cell interfaces in $x$ and $y$ directions.
Like the traditional high-order finite volume scheme, the Gaussian quadrature points are used in order to get a high accuracy.
In this paper, three-points Gaussian quadrature will be used on each cell interface in the 2D case.

To model the interface gas distribution function,  the integral solution of BGK equation \eqref{bgk} is used
\begin{equation}\label{integral1}
f(x_{i+1/2},y_{j_\ell},t,u,v,\xi)=\frac{1}{\tau}\int_0^t g(x',y',t',u,v,\xi)e^{-(t-t')/\tau}dt'\\
+e^{-t/\tau}f_0(-ut,-vt,u,v,\xi),
\end{equation}
where $(x_{i+1/2}, y_{j_\ell})=(0,0)$ is the location of cell interface, and $x_{i+1/2}=x'+u(t-t')$ and $y_{j_\ell}=y'+v(t-t')$ are the trajectory of
particles. Here $f_0$ is the initial gas distribution function and $g$ is the equilibrium state in space and time.
The integral solution basically states a physical process from the particle free transport
in $f_0$ in the kinetic scale to the hydrodynamic flow evolution in the integral term of $g$.
The contributions from $f_0$ and $g$ in the determination of $f$ at the cell interface
depend on the ratio of time step to the local particle collision time,
i.e., $\exp(-\Delta t/\tau)$, and the modeling of $f_0$ and $g$ in space and time is needed to evaluate the solution $f$ from  Eq.\eqref{integral1}.
For the continuum flow simulation, such as the NS solutions,
the determination of $f_0$ and $g$ depend only on the macroscopic flow variables and their initial reconstructions.
In this paper, the WENO-Z method will be used as a nonlinear reconstruction in the determination of $f_0$,
and the linear reconstruction is adopted in the determination of $g$. Therefore,
the above integral solution not only incorporates a physical evolution process from non-equilibrium to an equilibrium state, but alo
from a nonlinear reconstruction  to a linear one. This fact is critically important for the current scheme to capture both nonlinear shock and
linear acoustic wave accurately in a single computation.

Based on the integral solution, a simplified third-order gas distribution function can be obtained \cite{zhou},
\begin{align}\label{3rd-simplify-flux}
f(x_{i+1/2},y_{j_\ell},t,u,v,\xi)&=g_0+{\overline{A}} g_0t+\frac{1}{2}\overline{a}_{tt}g_0t^2\nonumber\\
&-\tau[(\overline{a}_1u+\overline{a}_2v+\overline{A})g_0+(\overline{a}_{xt}u+\overline{a}_{yt}v+\overline{a}_{tt})g_0t]\nonumber\\
&-e^{-t/\tau}g_0[1-(\overline{a}_{1}u+\overline{a}_{2}v)t]\nonumber\\
&+e^{-t/\tau}g_l[1-(a_{1l}u+a_{2l}v))t]H(u)\nonumber\\
&+e^{-t/\tau}g_r[1-(a_{1r}u+a_{2r}v))t](1-H(u)),
\end{align}
where the terms related to $g_0$ are from the integral of the equilibrium state and
the terms related to $g_l$ and $g_r$ are from the initial term $f_0$ in the Eq.(\ref{integral1}).
All other terms, such as  $\overline{A}, \overline{a}_{tt}, \overline{a},...$, are coming from the derivatives of a Maxwellian distribution.
All these coefficient in the above equation can be determined from the initially reconstructed  macroscopic flow variables.

\subsection{Multi-stage multi-derivative time stepping method}

With the above time accurate gas distribution function at a cell interface, in order to improve the temporal accuracy of the scheme,
the multi-stages multi-derivative
(MSMD) technique will be used for the evaluation of the flux function and the interface values \cite{ji2}.
The implementation of  MSMD needs the flux function and its time derivative.
For conservation laws, the semi-discrete finite volume scheme is written as
\begin{align*}
\frac{\text{d} W_{ij}}{\text{d} t}=-\frac{1}{\Delta x}
(F_{i+1/2,j}(t)-F_{i-1/2,j}(t))-\frac{1}{\Delta y}
(G_{i,j+1/2}(t)-G_{i,j-1/2}(t)):=\mathcal{L}(W_{ij}),
\end{align*}
where $\mathcal{L}_{ij}(W)$ is the numerical operator for spatial derivative of flux, $F$ and $G$ are obtained fluxes at the Gaussian quadrature
points.

 A fourth-order temporal accurate solution for $W(t)$ at $t=t_n
+\Delta t$ can be obtained by
\begin{equation}\label{step-hyper-1}
\begin{aligned}
W^*=W^n+\frac{1}{2}\Delta t\mathcal
{L}(W^n)+\frac{1}{8}\Delta t^2\frac{\partial}{\partial
t}\mathcal{L}(W^n),
\end{aligned}
\end{equation}
\begin{equation}\label{step-hyper-2}
\begin{aligned}
W^{n+1}=W^n+\Delta t\mathcal
{L}(W^n)+\frac{1}{6}\Delta t^2\big(\frac{\partial}{\partial
	t}\mathcal{L}(W^n)+2\frac{\partial}{\partial
	t}\mathcal{L}(W^*)\big),
\end{aligned}
\end{equation}
where $\mathcal{L} $ and $\frac{\partial}{\partial	t}\mathcal{L} $ are related to the fluxes and the time derivatives of
the fluxes evaluated from the time-dependent gas distribution function $f(t)$ at cell interfaces.
And the middle state $W^*$ is obtained at time $t^* = t^n + \Delta t /2$.
Again, with the time accurate gas distribution function $f(t)$, along the same line with MSMD technique the gas distribution function $f$
at a cell interface at $t^{n+1}$ becomes
\begin{align*}
f^{n+1}=& f^n +\Delta t f_t^n + \frac{1}{6}\Delta
t^2( f_{tt}^n + 2 f_{tt}^* ) ,
\end{align*}
where $f^*$ is for the middle state at time $t^* = t^n + \Delta t /2$,
$$f^* = f^n + \frac{1}{2} \Delta t f_t^n + \frac{1}{8} (\Delta t)^2 f_{tt}^n .$$
Therefore, based on the cell interface $ f_{j+1/2}^{n+1} $, the flow variables $W_{j+1/2}^{n+1}$ can be explicitly obtained, i.e.,
$W_{j+1/2}^{n+1} = \int \psi f_{j+1/2}^{n+1} d \Xi$,
from which the slope inside each cell, such as in the 1D case, can be updated as
$$ (W_x)_j^{n+1} = (W_{j+1/2}^{n+1} - W_{j-1/2}^{n+1})/\Delta x .$$
More detailed formulation for the above GKS part can be found in \cite{ji2}.

\section{Compact WENO reconstruction}
In the classical high-order finite volume method \cite{harten2,liu}, the pointwise value is reconstructed based on cell averages,
and the Riemann solver is used for the flux evaluation at the interface, and the Runge-Kutta time stepping method is used for the
time accuracy.
However, for the gas kinetic scheme, as presented in the last section the cell-averaged conservative flow variables and
cell-averaged slopes inside each cell can be updated. Based on these updated data, a 6th-order and a 8th-order compact linear and nonlinear
reconstruction will be designed consistently in this section, which can be subsequently used in the determination of equilibrium state and initial non-equilibrium
one in the gas kinetic scheme.

\subsection{Compact linear reconstruction}
In the section 2, the value $W_{i+1/2}$ is provided by taking moments of the gas distribution function at the cell interface.
The averaged slope in the cell $I_i$ can be obtained by the Gaussian formula
\begin{align}\label{Gausian-formula}
W^{'}_{i}=\frac{1}{\Delta x}\int_{I_{i}}\frac{\partial
W}{\partial x}\text{d}x=\frac{1}{\Delta x}(W_{i+1/2}-W_{i-1/2}),
\end{align}
where $W^{'}_{i}$ is the cell averaged slope in the cell $I_i$.

\begin{figure}[!htb]
\centering
\includegraphics[width=0.65\textwidth]{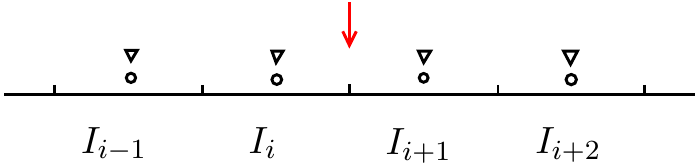}
\caption{\label{stencil-1} The large stencil for compact reconstruction of the value at the interface $x_{i+1/2}$.
The circles represent cell averages, and the gradients represent averaged slopes.}
\end{figure}

By taking advantage of the cell averaged variables and their slopes,
the compact stencil $S_{i+1/2}=\{I_{i-1},I_{i},I_{i+1},I_{i+2}\}$ shown in Fig.\ref{stencil-1} will be
used for the reconstruction of the value at the cell interface $x_{i+1/2}$.
On each cell $I_i$, the cell averages and their slopes are denoted as $W_{i}$ and $W^{'}_{i}$.
Based on the large stencil $S_{i+1/2}$, a series of high-order polynomials can be constructed,
\begin{align}\label{polynomial}
P^{n}(x)\equiv\sum_{k=0}^{n} a_{k}x^{k},
\end{align}
where $n$ is the order of the polynomial and $n=5,7$ correspond to the 6th-order and 8th-order schemes, respectively.

The polynomial $P^{7}(x)$ can be uniquely determined with the condition
\begin{equation}\label{recons-8th}
\begin{split}
&\frac{1}{\Delta x}\int_{I_{k}}P^7(x)\text{d}x=W_k, \\
&\frac{1}{\Delta x}\int_{I_{k}}(dP^7(x)/dx)\text{d}x=W^{'}_k,  I_k\in S_{i+1/2}, k=i-1,\cdots,i+2.
\end{split}
\end{equation}
In order to determine the $P^{5}(x)$, the above condition has to be modified.
For the cells $I_i$ and $I_{i+1}$, the following equations are strictly enforced
\begin{equation}\label{recons-6th-1}
\begin{split}
&\frac{1}{\Delta x}\int_{I_{k}}P^5(x)\text{d}x=W_k, \\
&\frac{1}{\Delta x}\int_{I_{k}}(dP^5(x)/dx)\text{d}x=W^{'}_k,  I_k\in S_{i+1/2}, k=i,i+1,
\end{split}
\end{equation}
and for the cells $I_{i-1}$ and $I_{i+2}$ the following conditions are satisfied in the sense of the least square
\begin{equation}\label{recons-6th-2}
\begin{split}
&\frac{1}{\Delta x}\int_{I_{k}}P^5(x)\text{d}x = W_k, \\
&\int_{I_{k}}(dP^5(x)/dx)\text{d}x =\Delta x W^{'}_k,  I_k\in S_{i+1/2}, k=i-1,i+2.
\end{split}
\end{equation}
Thus the linear reconstruction $P^n(x_{i+1/2})$ at the cell interface $x=x_{i+1/2}$ can be written as
\begin{equation}\label{recons-6th-val}
\begin{split}
P^5(x_{i+1/2})=\frac{1}{600}(&W_{i-1}+299W_{i}+299W_{i+1}+W_{i+2}- \\
  &3\Delta xW^{'}_{i-1}+111\Delta xW^{'}_{i}-111\Delta xW^{'}_{i+1}+3\Delta xW^{'}_{i+2}),
\end{split}
\end{equation}

\begin{equation}\label{recons-8th-val}
\begin{split}
P^7(x_{i+1/2})=\frac{1}{420}(&25W_{i-1}+185W_{i}+185W_{i+1}+25W_{i+2}+ \\
  &6\Delta xW^{'}_{i-1}+102\Delta xW^{'}_{i}-102\Delta xW^{'}_{i+1}-6\Delta xW^{'}_{i+2}),
\end{split}
\end{equation}
where $P^n(x_{i+1/2})$ has the truncation error of $O(\Delta x^{n+1})$.
The first-order and second-order derivatives of the linear reconstruction at the cell interface can be obtained  from $P^{n}(x)$,
\begin{equation}\label{recons-6th-deriva}
\begin{split}
P^5_x(x_{i+1/2})   =\frac{1}{1560\Delta x}(&-3W_{i-1}-3411W_{i}+3411W_{i+1}+3W_{i+2}+ \\
  &11\Delta xW^{'}_{i-1}-941\Delta xW^{'}_{i}-941\Delta xW^{'}_{i+1}+11\Delta xW^{'}_{i+2}), \\
P^5_{xx}(x_{i+1/2})=\frac{1}{40\Delta x^2}(&-W_{i-1}+W_{i}+W_{i+1}-W_{i+2}+ \\
  &3\Delta xW^{'}_{i-1}-51\Delta xW^{'}_{i}+51\Delta xW^{'}_{i+1}-3\Delta xW^{'}_{i+2}),
\end{split}
\end{equation}

\begin{equation}\label{recons-8th-deriva}
\begin{split}
P^7_x(x_{i+1/2})   =\frac{1}{108\Delta x}(&-14W_{i-1}-270W_{i}+270W_{i+1}+14W_{i+2}- \\
                      &3\Delta xW^{'}_{i-1}-99\Delta xW^{'}_{i}-99\Delta xW^{'}_{i+1}-3\Delta xW^{'}_{i+2}),\\
P^7_{xx}(x_{i+1/2})=\frac{1}{4\Delta x^2}(&-4W_{i-1}+4W_{i}+4W_{i+1}-4W_{i+2}- \\
                      &\Delta xW^{'}_{i-1}-9\Delta xW^{'}_{i}+9\Delta xW^{'}_{i+1}+\Delta xW^{'}_{i+2}).
\end{split}
\end{equation}

In order to eliminate the spurious oscillation and improve the stability, the above reconstruction is based on the characteristic variables.
The dimension-by-dimension strategy is applied for the reconstruction in the two-dimensional case.
The above compact reconstruction is mainly used to determine the equilibrium state in the normal direction in the gas kinetic scheme.
Due to multi-dimensionality in the GKS evolution model, in the tangential direction  a linear 5th-order reconstruction is used
in both 6th-order and 8th-order schemes.
The above linear reconstruction is used to determine the equilibrium state in GKS.
In the reconstruction of the initial non-equilibrium state, the WENO-based compact nonlinear reconstruction will be designed in later subsection.

\subsection{Resolution analysis}
In this section, the resolution analysis of the above compact spatial reconstruction will be presented.
The linear advection equation is used for the resolution analysis,
\begin{align}\label{advection}
W_t=-UW_x.
\end{align}
The integral form on the cell $I_i$ becomes
\begin{align}\label{advection-1}
(W_i)_t=-U(W_i)_x,
\end{align}
where $W_i$ is the cell averaged $W(x)$ in the cell $I_i$.

\begin{rmk}
In order to make Fourier expansion for the cell averages $W_i$, a continuous function $\overline{W}(x)$ is introduced \cite{harten2} as follows.
\begin{equation}\label{average-function}
\overline{W}(x)=\frac{1}{\Delta x}\int_{x-\Delta x/2}^{x+\Delta x/2}W(x+y)dy.
\end{equation}
Thus the cell average $W_i$ becomes the value of the continuous function $\overline W(x)$ at cell center $x_i$.
\end{rmk}

In order to analyse the resolution of spatial discretization in the current compact schemes,
we can compare the Fourier coefficients of the derivative given by the spatial discretization
with the Fourier coefficients of exact derivative. Using the reconstructed value at the cell interface,
the spatial discretization of $(W_i)_x$ in Eq.\eqref{advection-1} becomes
\begin{equation}\label{discrete-6th}
\begin{split}
(W_i)_x^{6th}=\frac{1}{600\Delta x}(&298(W_{i+1}-W_{i-1})+(W_{i+2}-W_{i-2})- \\
  &114\Delta x(W^{'}_{i+1}+W^{'}_{i-1})+3\Delta x(W^{'}_{i+2}+W^{'}_{i-2})+222\Delta xW^{'}_{i}),
\end{split}
\end{equation}

\begin{equation}\label{discrete-8th}
\begin{split}
(W_i)_x^{8th}=\frac{1}{420\Delta x}(&160(W_{i+1}-W_{i-1})+25(W_{i+2}-W_{i-2})- \\
  &96\Delta x(W^{'}_{i+1}+W^{'}_{i-1})-6\Delta x(W^{'}_{i+2}+W^{'}_{i-2})+204\Delta xW^{'}_{i}).
\end{split}
\end{equation}

For the purpose of Fourier analysis, the function $\overline W(x)$ is assumed to be periodic over the domain$[0,L]$,
and $\overline W(x)$ can be decomposed into its Fourier coefficients
\begin{equation*}
\overline{W}(x)=\sum_{k=0}^{k=N/2}c_k e^{(\frac{2\pi ikx}{L})}.
\end{equation*}
It is convenient to introduce new variables $\omega=2\pi k\Delta x/L=2\pi k/N$ and $s=x/\Delta x$ \cite{lele}.
Then the Fourier expansion can be written as
\begin{equation*}
\overline{W}(x)=\sum_{k=0}^{k=N/2}c_k  e^{i\omega(k)s},
\end{equation*}
and the first-order derivative of $\overline{W}(x)$ is
\begin{equation*}
\overline{W}^{'}(x)=\sum_{k=0}^{k=N/2}c_k\frac{i\omega}{h} e^{iw(k)s}=\sum_{k=0}^{k=N/2}c^{'}_k e^{i\omega(k)s},
\end{equation*}
where $c^{'}_k=c_k\frac{i\omega}{h}$.
Suppose the numerical solution given by the compact schemes can be written as
\begin{equation*}
\overline{W}_h(x)=\sum_{k=0}^{k=N/2}\hat{c}_k e^{i\omega(k)s},~~~\\
\overline{W}_h^{'}(x)=\sum_{k=0}^{k=N/2}\hat{c}^{'}_k e^{i\omega(k)s}.
\end{equation*}
Since the current spatial discretization is based on the symmetrical stencil, it can be shown that $\hat{c}^{'}_k=i\omega^{'}c_k$.
Substituting the average value in Eq.\eqref{discrete-6th} and Eq.\eqref{discrete-8th} with the form of Fourier expansion,
we get the modified wavenumber
\begin{equation}\label{wavenumber-6th}
\omega^{',6th}(\omega)=\frac{1}{300}(298 \sin(\omega) + \sin(2\omega)- 114 \omega  \cos(\omega) + 3 \omega  \cos(2\omega) + 111\omega),
\end{equation}

\begin{equation}\label{wavenumber-8th}
\omega^{',8th}(\omega)=\frac{1}{210}(160 \sin(\omega) + 25 \sin(2\omega)-96 \omega  \cos(\omega) -6 \omega  \cos(2\omega)+102\omega).
\end{equation}
In addition to the modified wavenumber, the error in the phase speed can be alternatively used
to evaluate the dispersive error. For the linear advection equation,
the phase speed for a wavenumber $\omega$ is given by the current spatial discretization (exact time advancement is assumed) as
\begin{equation}\label{wavenumber-phasespeed}
(c_p)_h=\omega^{'}(\omega)/\omega.
\end{equation}

\begin{figure}[!htb]
\centering
\includegraphics[width=0.485\textwidth]{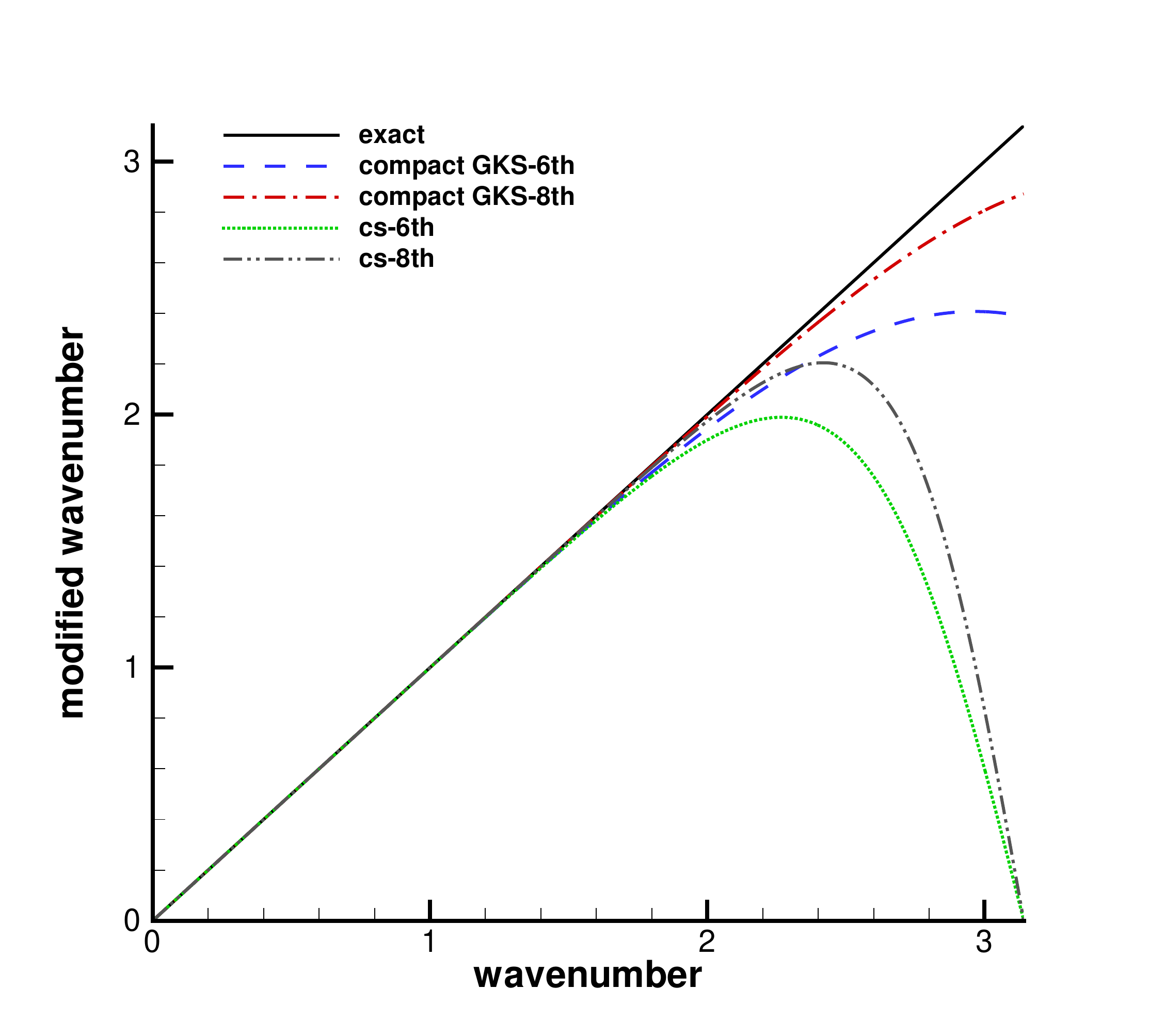}
\includegraphics[width=0.485\textwidth]{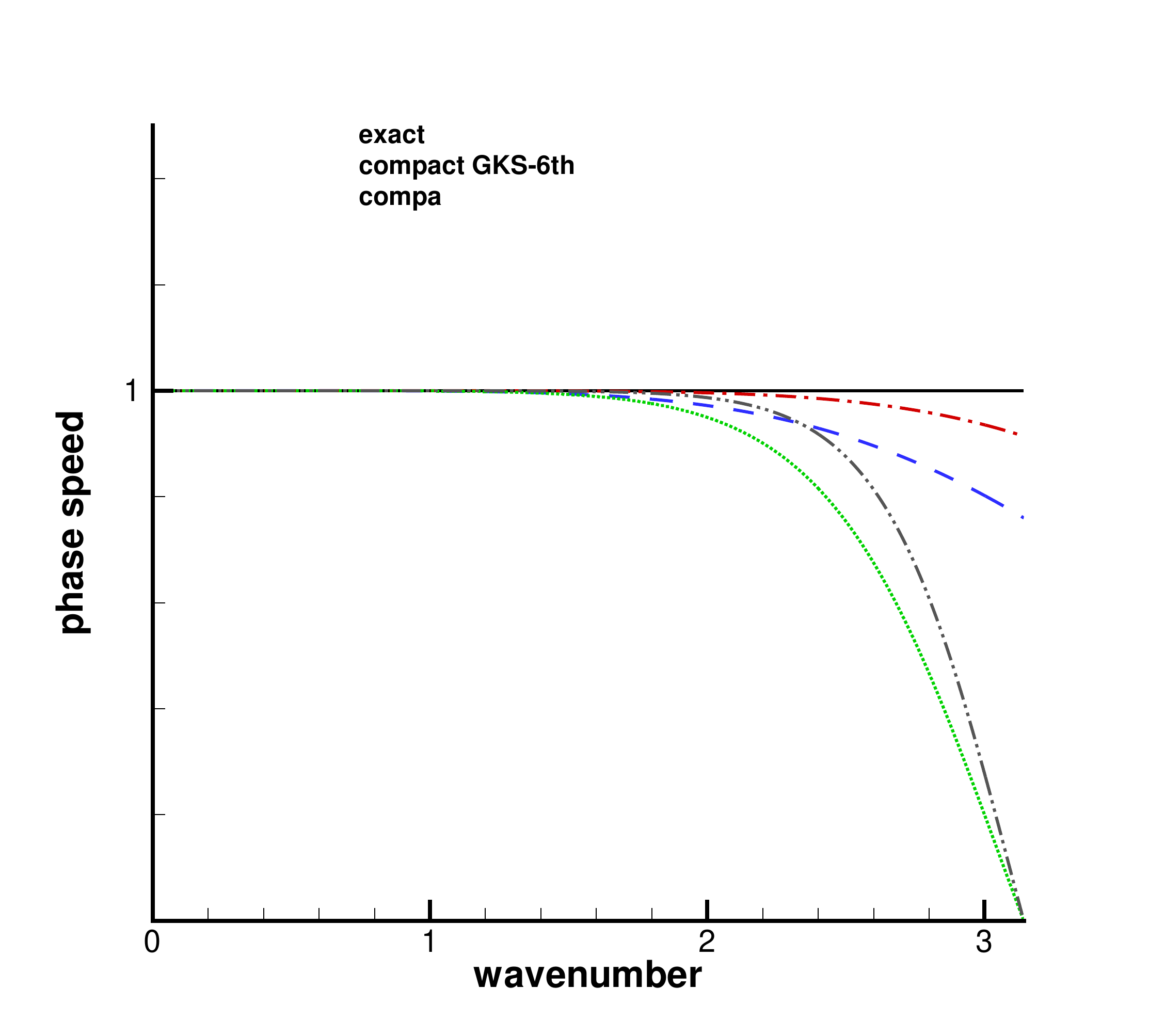}
\caption{\label{1d-resolution} Plots of modified wavenumber and phase speed vs wavenumber for different schemes.
cs-6th is the compact 6th-order tridiagonal scheme studied by Lele \cite{lele},
and the scheme has the best resolution in the series of 6th-order scheme;
cs-8th is the compact 8th-order pentadiagonal scheme \cite{lele},
and the cs-8th uses the same formula in spatial discretization as the compact 8th-order GKS.}
\end{figure}

\begin{figure}[!htb]
\centering
\includegraphics[width=0.485\textwidth]{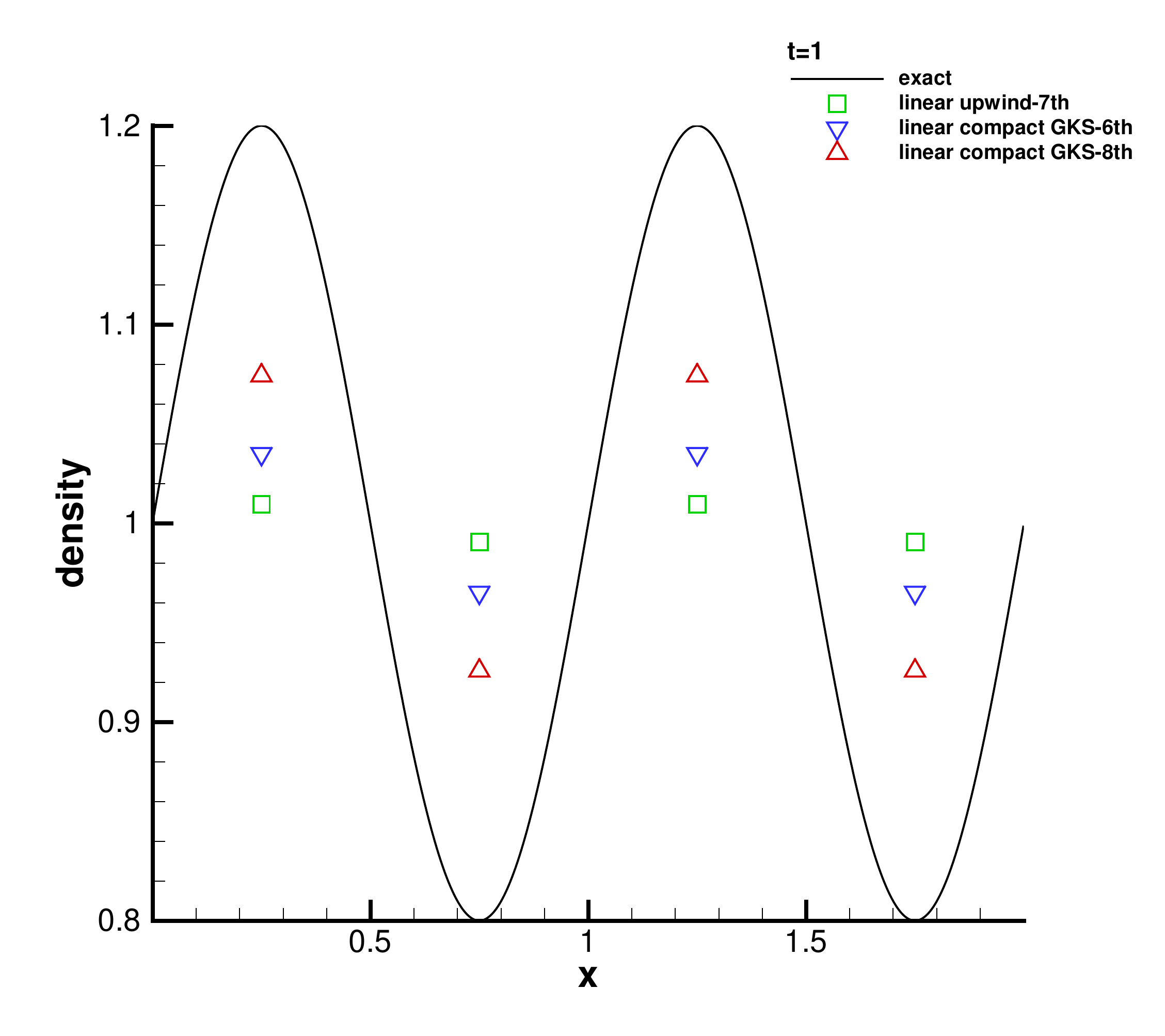}
\includegraphics[width=0.485\textwidth]{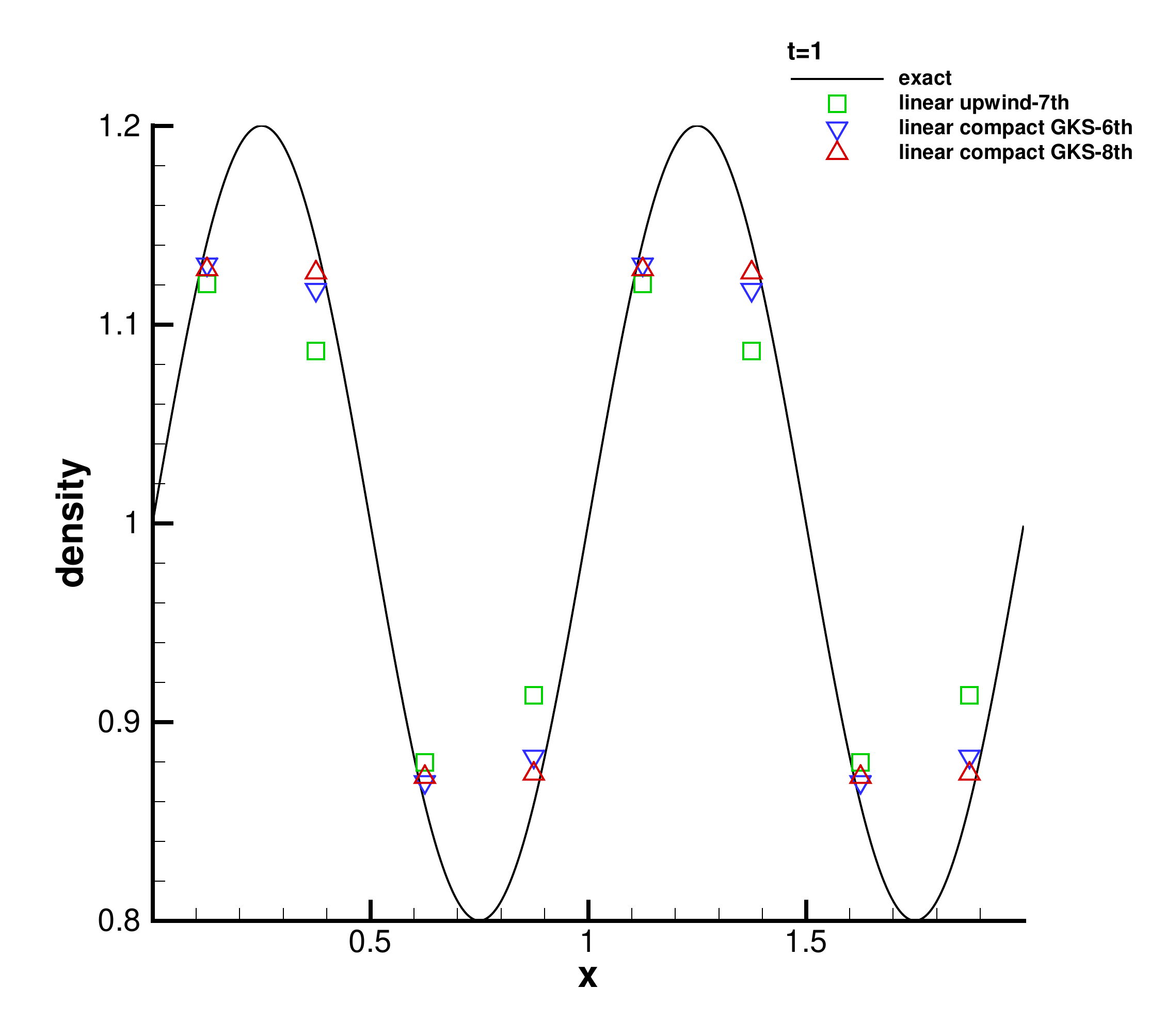}
\caption{\label{1d-resolution-test} Limit resolution for linear wave: Results of linear upwind 7th-order scheme,
linear compact 6th-order and 8th-order schemes on the meshes with $\Delta x=1/2$ and $\Delta x=1/4$ at $t=1$.
Since there are only two points for each wavelength on the left figure, the real "exact" solution should be averaged over half wave length as
well.}
\end{figure}

The comparison of the dispersion characteristics of the current compact GKS with the compact schemes by Lele \cite{lele}
is shown in Fig.\ref{1d-resolution}. The traditional compact schemes use only the node values \cite{lele}
or cell averages \cite{WCNS} as independent values in spatial discretization, and the curves of modified wavenumber
and phase speed in the range $\omega \in (\pi/2,\pi)$
deviate greatly from the curves corresponding to the analytical solution.
Therefore, these schemes have poor resolution for the wave with $\omega \in (\pi/2,\pi)$, because
the wavenumber range ($\pi >\omega >\pi /2$) corresponds to the wavelength $(2\Delta x,4 \Delta x)$,
and such wave cannot be determined by the cell averages (or point values) alone with less than four points in a wavelength.
However, the compact GKS can retain a good resolution because of the use of slopes in each cell, and these slopes are evaluated from the
gas evolution solutions at the cell interfaces which are independent from the cell averaged values. In the traditional compact schemes by Lele,
the node values and derivatives are not fully independent. The coupling increases the error since the node values cannot resolve high wavenumber
solution at all.
As an example, we test the schemes for the initial condition $(\rho,U,p)=(1+0.2\sin(2\pi x),1,1)$ in a computational domain $[0,2]$.
The results of linear density wave at $t=1$ with meshes $\Delta x=1/2$ and $\Delta x=1/4$ are shown in Fig. \ref{1d-resolution-test}.
These results demonstrate that the current compact schemes have favorable  resolution in comparison with
the non-compact 7th-order linear upwind scheme.

\subsection{WENO reconstruction}
In gas dynamics, shock and contact discontinuities can appear.
To deal with discontinuities, WENO reconstruction \cite{liu,jiang} can be used.
In section $3.1$, the linear reconstruction gives a unique reconstructed value and its slope at the cell interface.
However, in order to capture the possible discontinuity at a cell interface, the values at the left and right sides of the cell interface
have to be valuated separately. Therefore, for the nonlinear reconstruction the sub-stencils have to be defined first.
For simplicity, the WENO reconstruction procedure is given in detail for the construction of the left side value of the cell interface $x=x_{i+1/2}$.
The procedure for the right side value can be obtained similarly according to the symmetric property, which will be omitted here.
The left side value by WENO reconstruction is given by the candidate polynomials as follows
\begin{align}\label{right-left}
W^{n}_{i+1/2}=\sum_{k=0}^{l_n}\delta^n_{k}w^n_{k,i+1/2},~n=6,8,
\end{align}
where $W^n_{i+1/2}$ is the $nth-order$ WENO reconstruction for the left value of the cell interface $x_{i+1/2}$,
$l_n$ is the number of candidate polynomials, $\delta^n_{k}$ is the WENO weight, and $w^n_{k,i+1/2}$
is the point value of the candidate polynomial $w^n_{k}(x)$ at $x_{i+1/2}$.
In the later presentation, the sets of candidate polynomials
$w^n_{k}(x)$ are the same for different $nth-$order scheme, such as $n=6$ and $8$, so the superscript is omitted for simplicity.

\begin{figure}[!htb]
\centering
\includegraphics[width=0.65\textwidth]{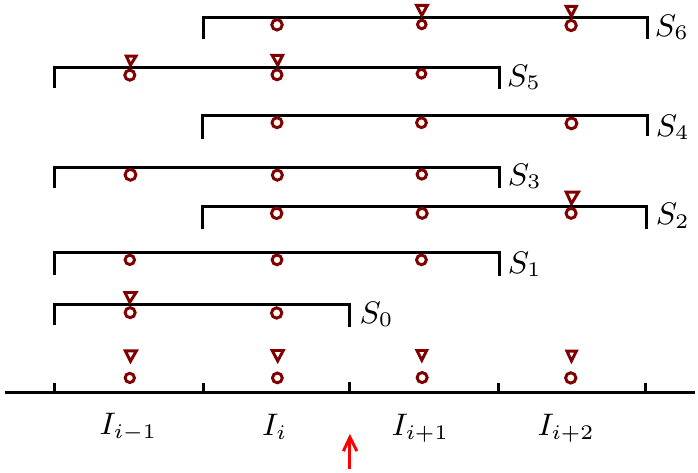}
\caption{\label{stencil-2} Sub-stencils in the reconstruction for left side value of the cell interface $x_{i+1/2}$
in the compact 6th-order and 8th-order reconstruction: the circles represent cell averages, and the gradients represent cell averaged slopes.}
\end{figure}

In the current compact WENO reconstruction, three principles are considered.
Firstly, if a discontinuity is located at one of the cell interfaces, theoretically the point-wise value at the interface cannot
be reliable to determine the cell averaged slope on the left and right side cells.
Secondly, because the smooth sub-stencils can play a dominant role in dealing with discontinuities appearing
in the large stencil, the averaged slope used in the sub-stencil should be possibly away from the interface for the data reconstruction
in order to make the sub-stencil sufficiently smooth.
Thirdly, the order of the candidate polynomial can be higher without increasing the spatial size of sub-stencils in the compact schemes,
and such higher-order candidate polynomials can achieve better resolution to solve discontinuities without spurious oscillation.
Based on the above principles, the following sub-stencils are designed,
where the 6th-order and 8th-order reconstructions are based on the same sub-stencils,
\begin{align*}
S_0&=\{W_{i-1},W_{i},W^{'}_{i-1}\} \leftrightarrow w_{0}(x), \\
S_1&=\{W_{i-1},W_{i},W_{i+1}\} \leftrightarrow w_{1}(x), \\
S_2&=\{W_{i},W_{i+1},W_{i+2},W^{'}_{i+2}\} \leftrightarrow w_{2}(x), \\
S_3&=\{W_{i-1},W_{i},W_{i+1}\} \leftrightarrow w_{3}(x), \\
S_4&=\{W_{i},W_{i+1},W_{i+2}\} \leftrightarrow w_{4}(x), \\
S_5&=\{W_{i-1},W_{i},W_{i+1},W^{'}_{i-1},W^{'}_{i}\} \leftrightarrow w_{5}(x), \\
S_6&=\{W_{i},W_{i+1},W_{i+2},W^{'}_{i+1},W^{'}_{i+2}\} \leftrightarrow w_{6}(x).
\end{align*}
$w_{2}(x)$ is a cubic polynomial, $w_{5}(x)$ and $w_{6}(x)$ are fourth-order polynomials, and others are quadratic polynomials.
According to the similar reconstruction conditions in Eq.\eqref{recons-8th}, $w_{k,i+1/2}$ can be obtained as
\begin{align}\label{sub-stencil}
\begin{cases}
w_{0,i+1/2}&=\displaystyle\frac{1}{6}(-7W_{i-1}+13W_{i}-4hW^{'}_{i-1}), \\
w_{1,i+1/2}&=\displaystyle\frac{1}{6}(-W_{i-1}+5W_{i}+2W_{i+1}), \\
w_{2,i+1/2}&=\displaystyle\frac{1}{24}(5W_{i}+32W_{i+1}-13W_{i+2}+6hW^{'}_{i+2}), \\
w_{3,i+1/2}&=\displaystyle\frac{1}{6}(-W_{i-1}+5W_{i}+2W_{i+1}),\\
w_{4,i+1/2}&=\displaystyle\frac{1}{6}(2W_{i}+5W_{i+1}-W_{i+2}),\\
w_{5,i+1/2}&=\displaystyle\frac{1}{30}(10W_{i-1}+19W_{i}+W_{i+1}+3hW^{'}_{i-1}+21hW^{'}_{i}),\\
w_{6,i+1/2}&=\displaystyle\frac{1}{30}(W_{i}+19W_{i+1}+10W_{i+2}-21hW^{'}_{i+1}-3hW^{'}_{i+2}).
\end{cases}
\end{align}
In the smooth region, the convex combination with $\delta^n_{k}=d^n_k$ recovers the reconstruction in Eq.\eqref{recons-6th-val}
and Eq.\eqref{recons-8th-val}, which can be the condition to get the linear weights \cite{jiang}.
The linear weight $d^6_{k}$ of the 6th-order reconstruction can be obtained as
\begin{align*}
d^6_{0}&=\frac{33}{700},~d^6_{1}=\frac{11}{140},~d^6_{2}=\frac{22}{175},~d^6_{3}=\frac{11}{100},
~d^6_{4}=\frac{11}{100},~d^6_{5}=\frac{37}{140},~d^6_{6}=\frac{37}{140},
\end{align*}
and the $d^8_{k}$ of 8th-order reconstruction are
\begin{align*}
d^8_{0}&=\frac{3}{98},~d^8_{1}=\frac{5}{98},~d^8_{2}=\frac{4}{49},~d^8_{3}=\frac{7}{98}~d^8_{4}=\frac{7}{98},~d^8_{5}=\frac{17}{49},~d^8_{6}=\frac{17}{49}.
\end{align*}

The WENO-Z nonlinear weights are used in the current compact schemes and they are defined as \cite{WENO-Z}
\begin{equation}\label{WENO-Z-w}
\delta^n_{k}=\frac{\alpha^n_{k}}{\sum_{m=0}^{l_K}\alpha^n_{m}},
~~\alpha^n_{k}=d^n_k\Big[1+\big(\frac{Z_{ref}^n}{\beta_k+\epsilon}\big)\Big],~~k=0,...,l_n,
\end{equation}
where $Z_{ref}^n$ is the local higher-order reference value, which is related to the accuracy of the scheme and is given in detail later.
Here $\beta_{k}$ is the smooth indicator and defined as \cite{jiang}
\begin{equation}\label{smooth-indicator}
\beta_k=\sum_{q=1}^{q_k}\Delta x^{2q-1}\int_{x_{i-1/2}}^{x_i+1/2}\big(\frac{\text{d}^q}{\text{d}x^q}w_k(x)\big)^2dx,
\end{equation}
where $q_k$ is the order of $w_k(x)$.

In smooth region, the first two candidate polynomial $w_0(x)$ and $w_1(x)$ can be combined into a cubic polynomial
which is symmetric counterpart of $w_2(x)$. Then, current sub-stencils can become basically symmetric for interface $x_{i+1/2}$.
In order to maintain the symmetry for the nonlinear schemes,
the smooth indicator $\beta_1$ of $w_1(x)$ corresponding to $S_1$ is replaced by the indicator of the cubic polynomial
$\widetilde{w}_1(x)$ on $\widetilde{S}_1=\{W_{i-1},W_{i},W_{i+1},W^{'}_{i-1}\}$.
Even without showing in this paper, some tests demonstrate that the current choice $\beta_1$ can
present a better resolution in the numerical results with excellent robustness,
The detailed formulae for all $\beta_k,~k=0,...,l_n$ are given in Appendix.

\subsection{Accuracy of the compact reconstruction}
In this subsection, the accuracy of the compact reconstruction is analysed.
The nonlinear reconstruction in Eq.\eqref{right-left} can be rewritten as
\begin{align}\label{recons-accuracy}
W^n_{i+1/2}=&\sum_{k=0}^{l_n}d^n_{k}w_{k,i+1/2}+ \sum_{k=0}^{l_n}(\delta^n_{k}-d^n_{k})w_{k,i+1/2}.
\end{align}
The reconstruction can be split into two terms \cite{WENO-M}, i.e., the linear term and nonlinear term.
For the linear part $W^{n,opt}_{i+1/2}\equiv \sum_{k=0}^{l_n}d^n_{k}w_{k,i+1/2}$, the error can be given as
\begin{align*}
W^{n,opt}_{i+1/2}-W(x_{i+1/2})=A^n(x_{i+1/2})\Delta x^{n}+O(\Delta x^{n+1}),
\end{align*}
where $W(x_{i+1/2})$ is the exact solution at $x=x_{i+1/2}$.
In the second term in Eq.\eqref{recons-accuracy}, the point value $w_{k,i+1/2}$ approximates $W(x_{i+1/2})$ as follows
\begin{align*}
w_{k,i+1/2}=W(x_{i+1/2})+B_k(x_{i+1/2})\Delta x^{q_k}+O(\Delta x^{q_k+1}).
\end{align*}
Substituting $w_{k,i+1/2}$ into Eq.\eqref{recons-accuracy} and taking $\displaystyle\sum_{k=0}^{l_n}\delta^n_{k}=\sum_{k=0}^{l_n}d^n_{k}=1$ into account, we have
\begin{align*}
W^n_{i+1/2}&=W^{n,opt}_{i+1/2} + \sum_{k=0}^{l_n}(\delta^n_{k}-d^n_{k})\big(W(x_{i+1/2})+B_{k}(x_{i+1/2})\Delta x^{q_k}+O(\Delta x^{q_k+1})\big) \\
&=W^{n,opt}_{i+1/2}+\sum_{k=0}^{l_n}B_{k}(x_{i+1/2})(\delta^n_{k}-d^n_{k})\Delta x^{q_k} +\sum_{k=0}^{l_n}(\delta^n_{k}-d^n_{k})O(\Delta x^{q_k+1}).
\end{align*}
To achieve the formal order of accuracy for the 6th-order and 8th-order reconstructions, the following sufficient condition is proposed.
\begin{align}\label{weight condition}
\delta^n_{k}-d^n_{k}=O(\Delta x^{n-3+n{'}}),~ n^{'}\geq 0.
\end{align}
For the current nonlinear weights with WENO-Z weighting functions,
with the following formulation of the local higher-order reference value $Z_{ref}^n$,
the sufficient condition Eq.\eqref{weight condition} can be satisfied when
\begin{align}\label{tau}
Z_{ref}^n=\left|3(\beta_{0}-\beta_{4}) + (\beta_{4}-\beta_{3})\right|.
\end{align}
In order to prove that the sufficient condition can be met, according to the Taylor series of $\beta_{k},~k=0,3,4$,
\begin{align*}
\beta_{0}=&\big( (W_{i}^{(1)})^{2} + \frac{13}{12}(W^{(2)}_{i}\Delta x)^{2}\big)\Delta x^{2} - \frac{1}{3}W_{i}^{(1)}W_{i}^{(3)}\Delta x^{4} + J_0(W_{i}^{(l)},\Delta x)\Delta x^{5}, \\
\beta_{3}=&\big( (W_{i}^{(1)})^{2} + \frac{13}{12}(W^{(2)}_{i}\Delta x)^{2}\big)\Delta x^{2} + \frac{1}{3}W_{i}^{(1)}W_{i}^{(3)}\Delta x^{4} + J_3(W_{i}^{(l)},\Delta x)\Delta x^{5}, \\
\beta_{4}=&\big( (W_{i}^{(1)})^{2} + \frac{13}{12}(W^{(2)}_{i}\Delta x)^{2}\big)\Delta x^{2} - \frac{2}{3}W_{i}^{(1)}W_{i}^{(3)}\Delta x^{4} + J_4(W_{i}^{(l)},\Delta x)\Delta x^{5}, l=1,2,\cdots\\
\end{align*}
we have
\begin{align*}
Z_{ref}^n=\left| 3J_0 -J_3 - 2J_4 \right|\Delta x^{5},
\end{align*}
where $J_0,J_3$ and $J_4$ are the functions of the Taylor expanded terms of $\overline{W}(x)$ as $\overline{W}(x)$ is always continuous.
Suppose $O(J_k)\sim O(1)$,
the sufficient condition Eq.\eqref{weight condition} is satisfied by
\begin{align*}
\alpha^n_{k}=&d^n_{k}\Big[1+\Big(\frac{Z_{ref}^n}{\beta_{k}+\epsilon}\Big)\Big]=d^n_{k}\big[C + O(\Delta x^{5})\big],~~k=0,...,l_n,\\
\delta^n_{k}=&\frac{\alpha^n_{k}}{\sum_{m=0}^{l_n}\alpha^r_{m}}=d^n_{k}+O(\Delta x^{5}),
\end{align*}
where $C\geq1$ is a positive constant.

\section{Numerical tests}

In this section, we are going to test the 6th-order and 8th-order compact gas kinetic schemes. The cases include linear acoustic waves, blast wave,
shock-shock interactions, shock acoustic wave interaction, as well as viscous flow computations.
The mesh used in this paper is rectangular one and the time step is determined by the CFL condition with a CFL number ($\geq 0.3$) in all test cases.
In all tests, the same linear reconstruction is used for the equilibrium state and the nonlinear reconstruction for the non-equilibrium state.
There is no any additional "trouble cell" detection or any limiter designed for specific test.
The gas kinetic evolution model basically presents a dynamical process from the nonlinear to linear one, and the convergence rate $\exp(-\Delta t/\tau)$
depends on the flow condition. The collision time $\tau$ for inviscid flow at a cell interface is defined by
\begin{align*}
\tau=\epsilon \Delta t+C\displaystyle|\frac{p_l-p_r}{p_l+p_r}|\Delta
t,
\end{align*}
where $\varepsilon=0.01$, $C=1$, and $p_l$ and $p_r$ are the pressures at the left and right sides of a cell interface.
For the viscous flow, the collision time is related to the viscosity coefficient,
\begin{align*}
\tau=\frac{\mu}{p}+C \displaystyle|\frac{p_l-p_r}{p_l+p_r}|\Delta t,
\end{align*}
where  $\mu$ is the dynamic viscosity
coefficient and $p$ is the pressure at the cell interface. In
smooth flow regions, it will reduce to $\tau=\mu/p$.
The reason for including pressure
jump term in the particle collision time is to add artificial
dissipation in the discontinuous region to keep the
non-equilibrium dynamics in the shock layer.

\subsection{Accuracy tests}
The one-dimensional advection of density perturbation is tested first, and the initial conditions are given as follows
\begin{align*}
\rho(x)=1+0.2\sin(\pi x),\ \  U(x)=1,\ \ \  p(x)=1, x\in[0,2].
\end{align*}
The periodic boundary condition is adopted, and the analytic solutions are
\begin{align*}
\rho(x,t)=1+0.2\sin(\pi(x-t)),\ \ \  U(x,t)=1,\ \ \  p(x,t)=1.
\end{align*}
With the $r$th-order spatial reconstruction and 2-stage 4th-order temporal
discretization, the leading term of the truncation error is
$O(\Delta x^r+\Delta t^4)$. To keep the $r$th-order accuracy in the test,
$\Delta t=C \Delta x^{r/4}$ needs to be used for the $r$th-order scheme.
In the computation, the uniform meshes with $N$ mesh points are used.
The $L^1$, $L^2$ and $L^\infty$ errors and convergence orders at $t=2$ for the
6th-order and 8th-order compact GKS are presented in
Table \ref{1d-accuracy-6-1} to Table \ref{1d-accuracy-8-2}, respectively.
For the 8th-order scheme, the errors reduce quickly, and the order of
accuracy does not attain $8$ with $N=80$ due to the limited round-off error.

\begin{table}[!h]
	\small
	\begin{center}
		\def\temptablewidth{0.80\textwidth}
		{\rule{\temptablewidth}{0.70pt}}
		\begin{tabular*}{\temptablewidth}{@{\extracolsep{\fill}}c|cc|cc|cc}
			
			mesh length & $L^1$ error & Order & $L^2$ error & Order& $L^{\infty}$ error & Order  \\
			\hline
            1/5  & 1.761e-003 & ~    & 1.976e-003 & ~    & 2.721e-003 & ~    \\
            1/10 & 3.171e-005 & 5.80 & 3.527e-005 & 5.81 & 4.899e-005 & 5.80 \\
            1/20 & 5.017e-007 & 5.98 & 5.554e-007 & 5.99 & 7.803e-007 & 5.97 \\
            1/40 & 7.818e-009 & 6.00 & 8.676e-009 & 6.00 & 1.224e-008 & 5.99 \\
            1/80 & 1.220e-010 & 6.00 & 1.355e-010 & 6.00 & 1.915e-010 & 6.00 \\
		\end{tabular*}
		{\rule{\temptablewidth}{0.1pt}}
	\end{center}
	\vspace{-7mm} \caption{\label{1d-accuracy-6-1} 1-D accuracy test: errors and convergence orders of 6th-order
                  compact linear scheme with $\Delta t = 0.3 \Delta x ^{3/2} $.}
\end{table}

\begin{table}[!h]
	\small
	\begin{center}
		\def\temptablewidth{0.80\textwidth}
		{\rule{\temptablewidth}{0.70pt}}
		\begin{tabular*}{\temptablewidth}{@{\extracolsep{\fill}}c|cc|cc|cc}
			
			mesh length & $L^1$ error & Order & $L^2$ error & Order& $L^{\infty}$ error & Order  \\
			\hline
            1/5  & 1.196e-004 & ~    & 1.344e-004 & ~    & 1.849e-004 & ~    \\
            1/10 & 4.715e-007 & 7.99 & 5.301e-007 & 7.99 & 7.286e-007 & 7.99 \\
            1/20 & 1.806e-009 & 8.03 & 1.998e-009 & 8.05 & 2.798e-009 & 8.02 \\
            1/40 & 6.979e-012 & 8.02 & 7.744e-012 & 8.01 & 1.092e-011 & 8.00 \\
            1/80 & 5.588e-014 & 6.96 & 6.207e-014 & 6.96 & 9.614e-014 & 6.83 \\
		\end{tabular*}
		{\rule{\temptablewidth}{0.1pt}}
	\end{center}
	\vspace{-7mm} \caption{\label{1d-accuracy-8-1} 1-D accuracy test: errors and convergence orders of 8th-order
                   compact linear scheme with $ \Delta t = 0.3 \Delta x^2 $.}
\end{table}

\begin{table}[!h]
	\small
	\begin{center}
		\def\temptablewidth{0.80\textwidth}
		{\rule{\temptablewidth}{0.70pt}}
		\begin{tabular*}{\temptablewidth}{@{\extracolsep{\fill}}c|cc|cc|cc}
			
			mesh length & $L^1$ error & Order & $L^2$ error & Order& $L^{\infty}$ error & Order  \\
			\hline
            1/5  & 7.495e-003 & ~    & 8.448e-003 & ~    & 1.248e-002 & ~    \\
            1/10 & 3.624e-004 & 4.37 & 4.307e-004 & 4.29 & 7.357e-004 & 4.08 \\
            1/20 & 7.877e-006 & 5.52 & 8.677e-006 & 5.63 & 1.461e-005 & 5.65 \\
            1/40 & 4.216e-008 & 7.55 & 5.440e-008 & 7.32 & 1.235e-007 & 6.89 \\
            1/80 & 2.119e-010 & 7.64 & 2.647e-010 & 7.68 & 7.407e-010 & 7.38 \\
            1/160 & 2.100e-012 & 6.66 & 2.279e-012 & 6.86 & 3.540e-012 & 7.71 \\
		\end{tabular*}
		{\rule{\temptablewidth}{0.1pt}}
	\end{center}
	\vspace{-7mm} \caption{\label{1d-accuracy-6-2} 1-D accuracy test: errors and convergence orders of 6th-order
                   compact nonlinear schemes with $\Delta t = 0.3 \Delta x ^{3/2} $.}
\end{table}

\begin{table}[!h]
	\small
	\begin{center}
		\def\temptablewidth{0.80\textwidth}
		{\rule{\temptablewidth}{0.70pt}}
		\begin{tabular*}{\temptablewidth}{@{\extracolsep{\fill}}c|cc|cc|cc}
			
			mesh length & $L^1$ error & Order & $L^2$ error & Order& $L^{\infty}$ error & Order  \\
			\hline
            1/5  & 5.154e-003 & ~    & 5.969e-003 & ~    & 8.959e-003 & ~    \\
            1/10 & 3.825e-004 & 3.75 & 4.392e-004 & 3.76 & 6.522e-004 & 3.78 \\
            1/20 & 7.963e-006 & 5.59 & 9.252e-006 & 5.57 & 1.588e-005 & 5.36 \\
            1/40 & 4.250e-008 & 7.55 & 6.042e-008 & 7.26 & 1.503e-007 & 6.72 \\
            1/80 & 1.339e-010 & 8.31 & 2.201e-010 & 8.10 & 7.984e-010 & 7.56 \\
            1/160 & 2.827e-013 & 8.89 & 3.801e-013 & 9.18 & 1.206e-012 & 9.37 \\ 	
		\end{tabular*}
		{\rule{\temptablewidth}{0.1pt}}
	\end{center}
	\vspace{-7mm} \caption{\label{1d-accuracy-8-2} 1-D accuracy test: errors and convergence orders of 8th-order
                  compact nonlinear scheme with $ \Delta t = 0.3 \Delta x^2 $.}
\end{table}

\begin{figure}[!htb]
\centering
\includegraphics[width=0.485\textwidth]{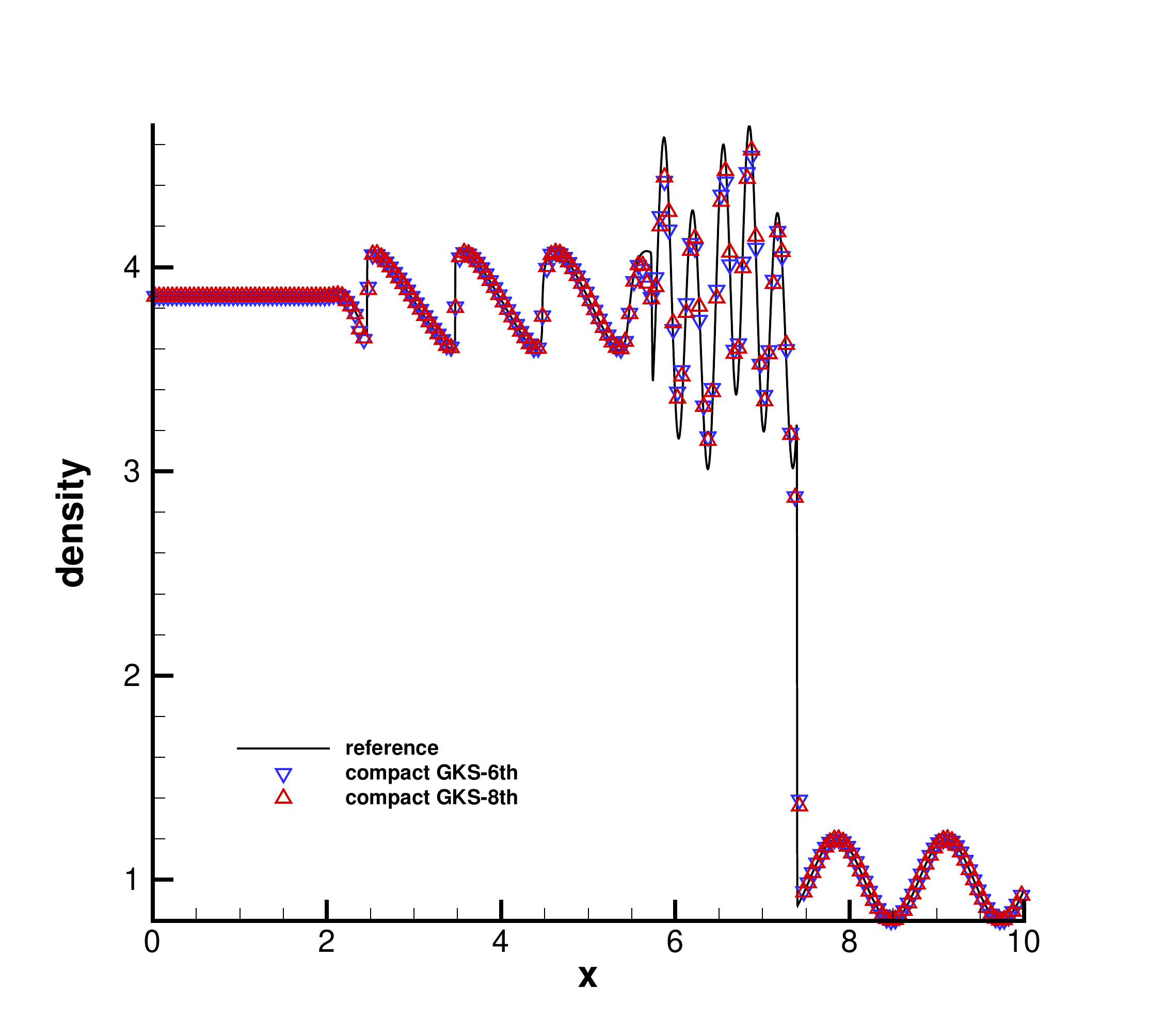}
\includegraphics[width=0.485\textwidth]{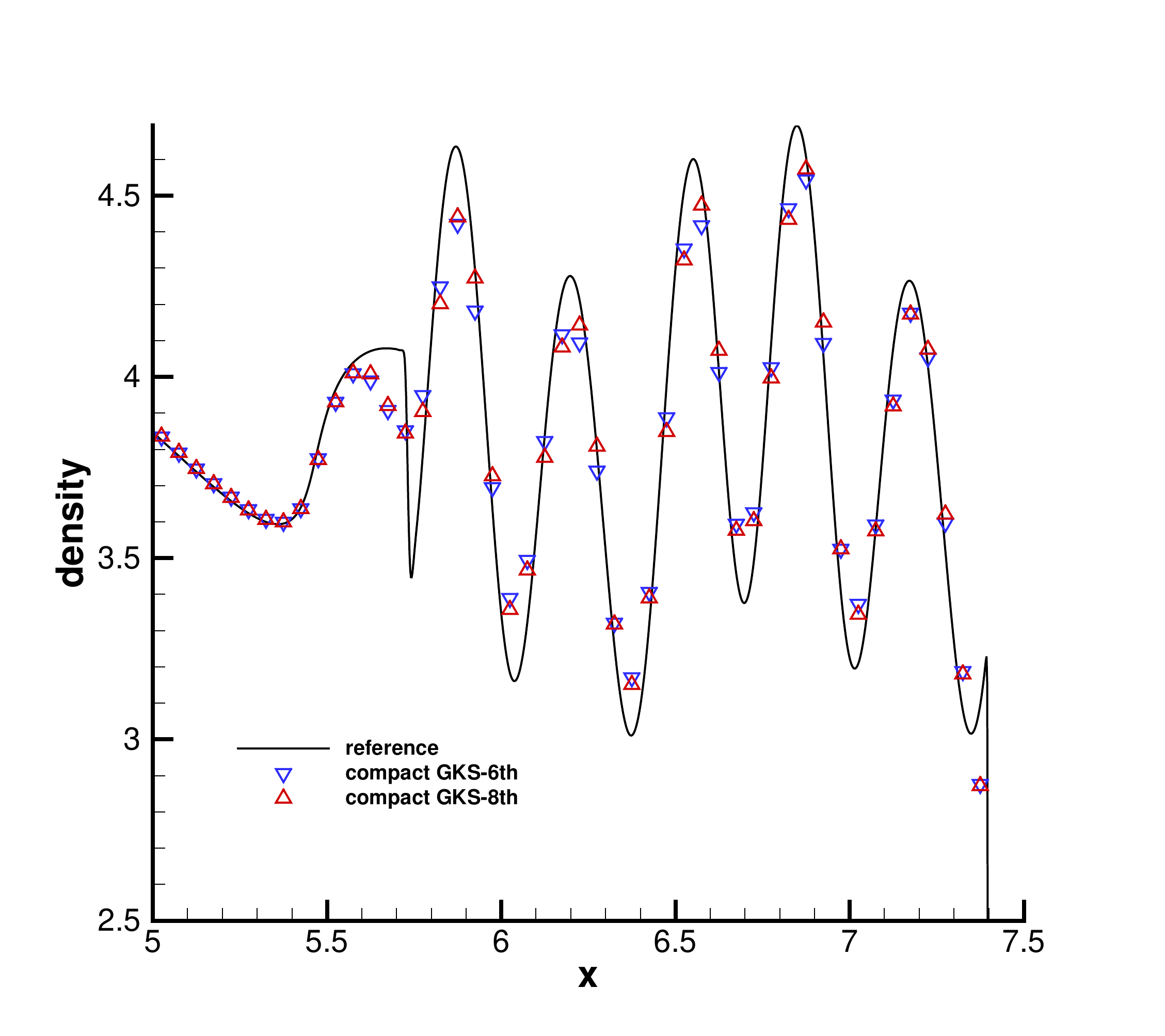}
\caption{\label{1d-riemann-Shu} Shu-Osher problem: the density distribution and local enlargement by 6th-order
and 8th-order compact GKS  at $t=1.8$ with $200$ mesh points.}
\end{figure}

\begin{figure}[!h]
\centering
\includegraphics[width=0.485\textwidth]{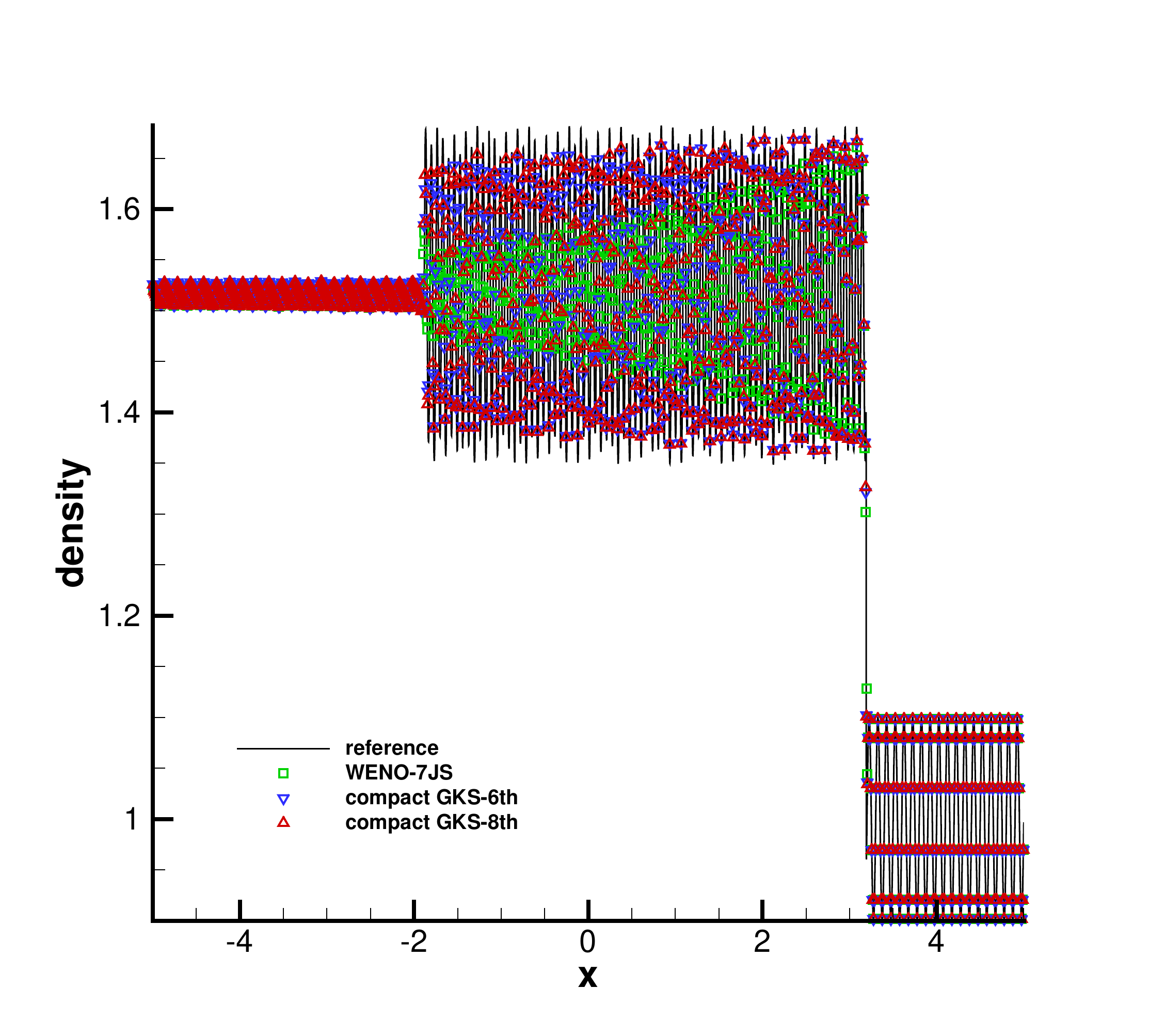}
\includegraphics[width=0.485\textwidth]{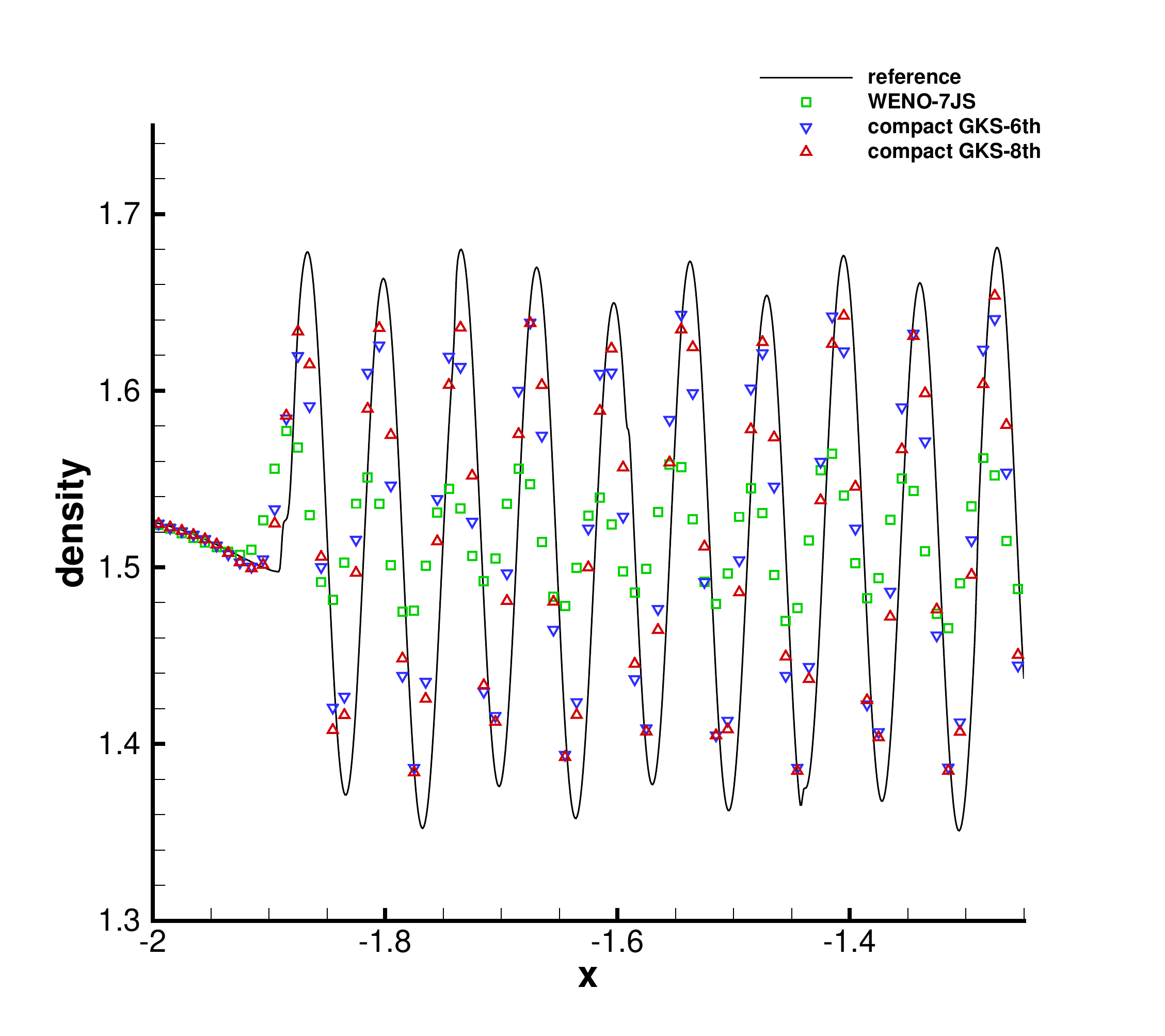}
\includegraphics[width=0.485\textwidth]{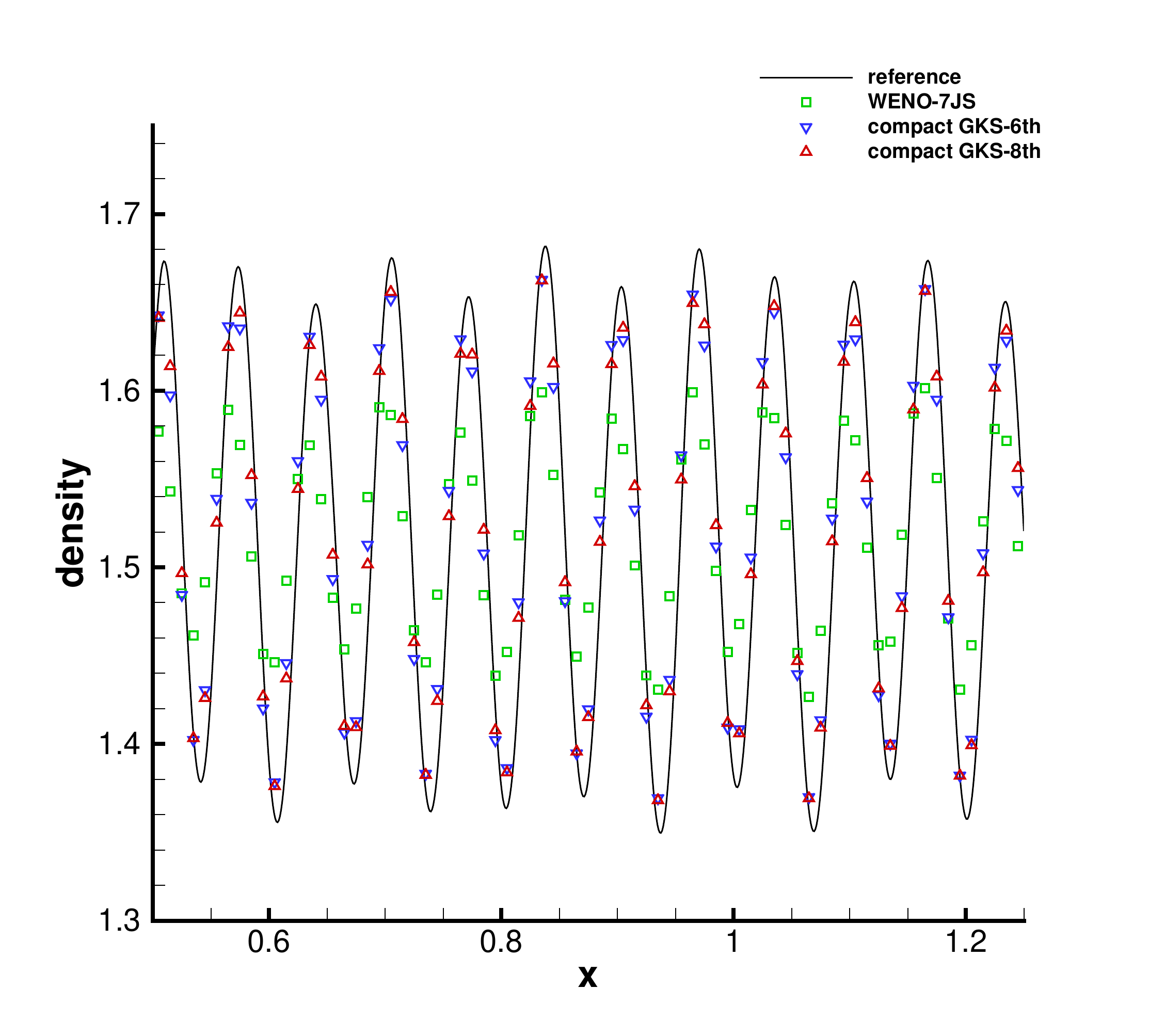}
\includegraphics[width=0.485\textwidth]{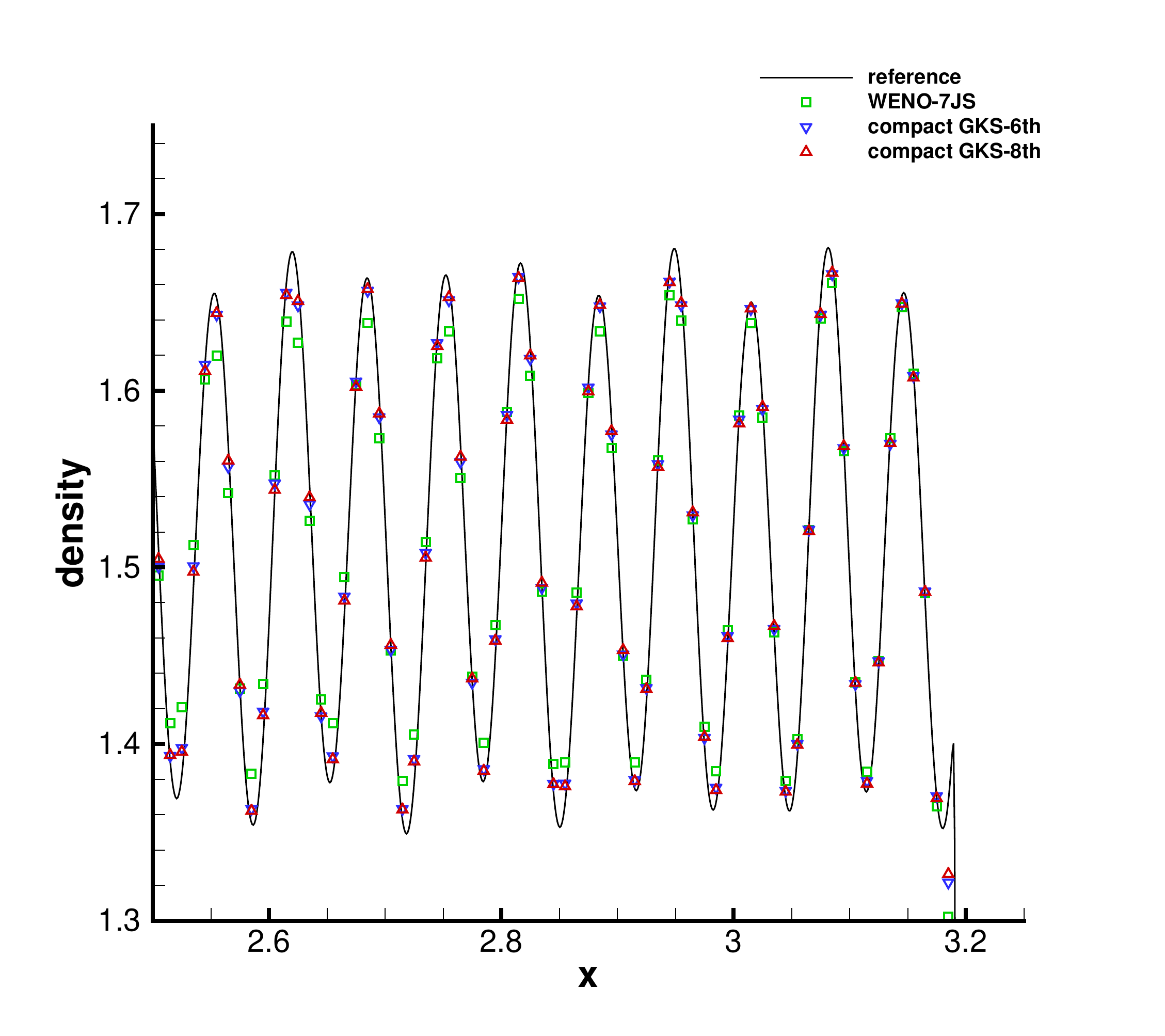}
\caption{\label{1d-riemann-ttoro} Titarev-Toro problem: the density distribution and local enlargement by 6th-order
and 8th-order compact GKS at $t=5$ with $1000$ mesh points.}
\end{figure}

\subsection{One-dimensional Riemann problems}
To test the performance of the schemes for capturing high frequency waves, the Shu-Osher
problem for density-wave shock interaction is tested \cite{Case-Shu-Osher}. The initial condition is given by
\begin{equation*}
(\rho,U,p)=\left\{\begin{array}{ll}
(3.857134, 2.629369, 10.33333),  \ \ \ \ &  x \leq 1,\\
(1 + 0.2\sin (5x), 0, 1),  &  1 <x \leq 10.
\end{array} \right.
\end{equation*}
The computational domain is $[0, 10]$ and $200$ uniform mesh points are used.
The non-reflecting boundary condition is used at both ends. The computed density
profile and local enlargement at $t=1.8$ from different schemes are shown in
Fig.\ref{1d-riemann-Shu}. Current compact 6th-order and 8th-order
schemes perform well with such a coarse mesh. Due to the improvement of order of
accuracy, the 8th-order scheme performs better then the 6th-order one.

As an extension of the Shu-Osher problem, the Titarev-Toro problem
 is tested as well \cite{Titarev-Toro}, and the initial condition in this case
is the following
\begin{align*}
(\rho,U,p) =\begin{cases}
(1.515695,0.523346,1.805),   & -5< x \leq -4.5,\\
(1 + 0.1\sin (20\pi x), 0, 1),  &  -4.5 <x <5.
\end{cases}
\end{align*}
The computational domain is $[-5, 5]$. The non-reflecting boundary condition
is imposed on left end, and the fixed wave profile is given on the right end.
Both compact 6th-order and 8th-order schemes are tested for this case. The computed
density profile with $1000$ mesh points at $t =5$, local enlargement, and the exact solution for the Titarev-Toro
problem  are shown in Fig.\ref{1d-riemann-ttoro}.
In order to show the importance and accuracy of the 6th-order and 8th-order compact reconstructions,
the results from the same GKS, but with the non-compact
standard 7th-order WENO-JS reconstruction using the cell-averaged flow variables only,
are included as well. As shown in Fig.\ref{1d-riemann-ttoro}, even with 7th-order WENO reconstruction the dissipation and dispersion errors
are much larger than those from the 6th-order compact GKS.
Based on this observation, it clearly indicates that the use of
high-order evolution model for the update of slope is favorable in the design of high-order schemes.
The compactness of the scheme is also physically sounded  because the CFL condition constrains the signal propagation
locally to the neighboring cells only within a time step. To design a reliable compact scheme depends on the high-order gas evolution model
at the cell interface. Any scheme based on the first order Riemann solver cannot achieve such a goal to have scheme  with the properties
of compactness, large CFL number, robustness, and high efficiency.

\begin{figure}[!htb]
\centering
\includegraphics[width=0.485\textwidth]{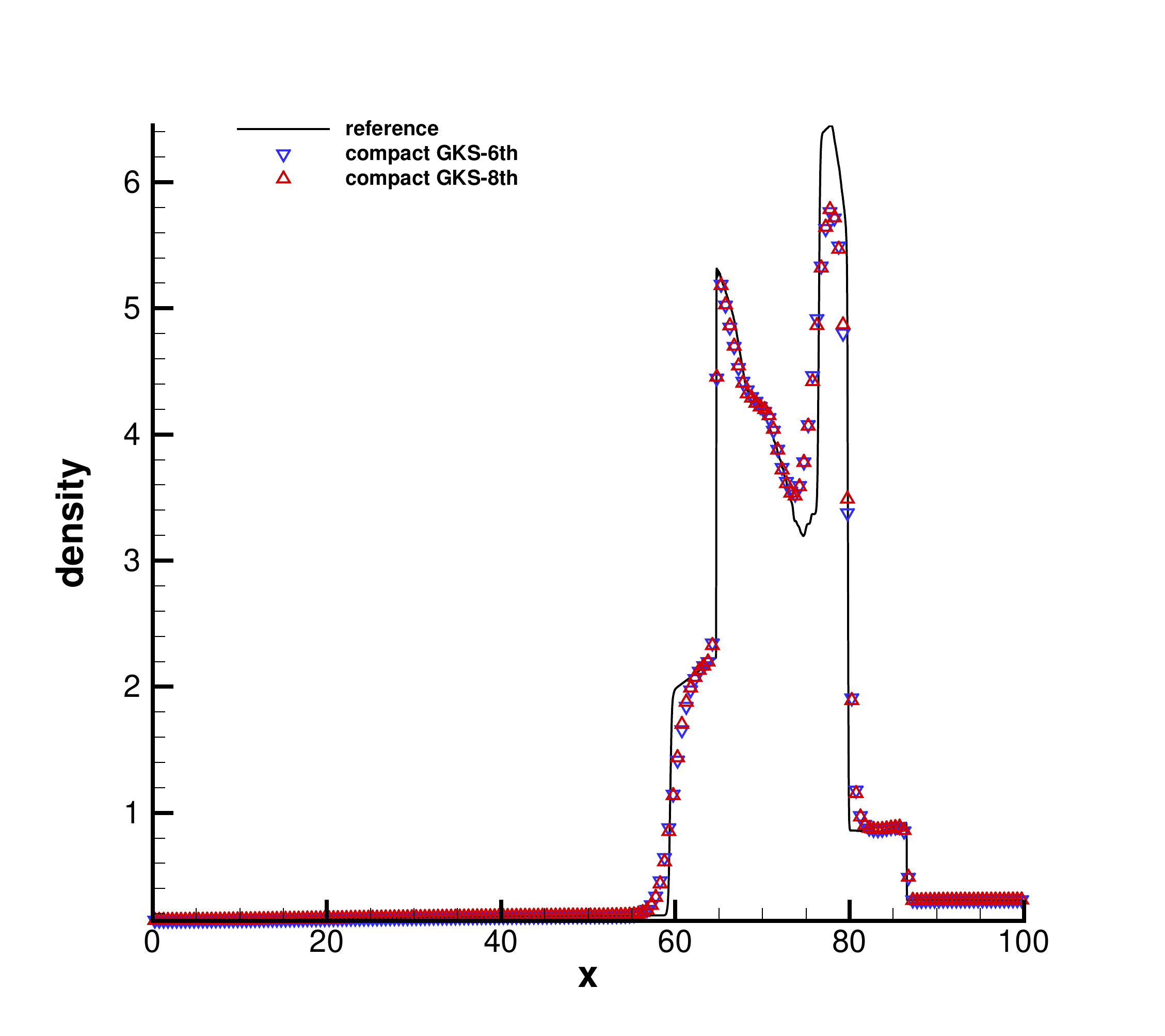}
\includegraphics[width=0.485\textwidth]{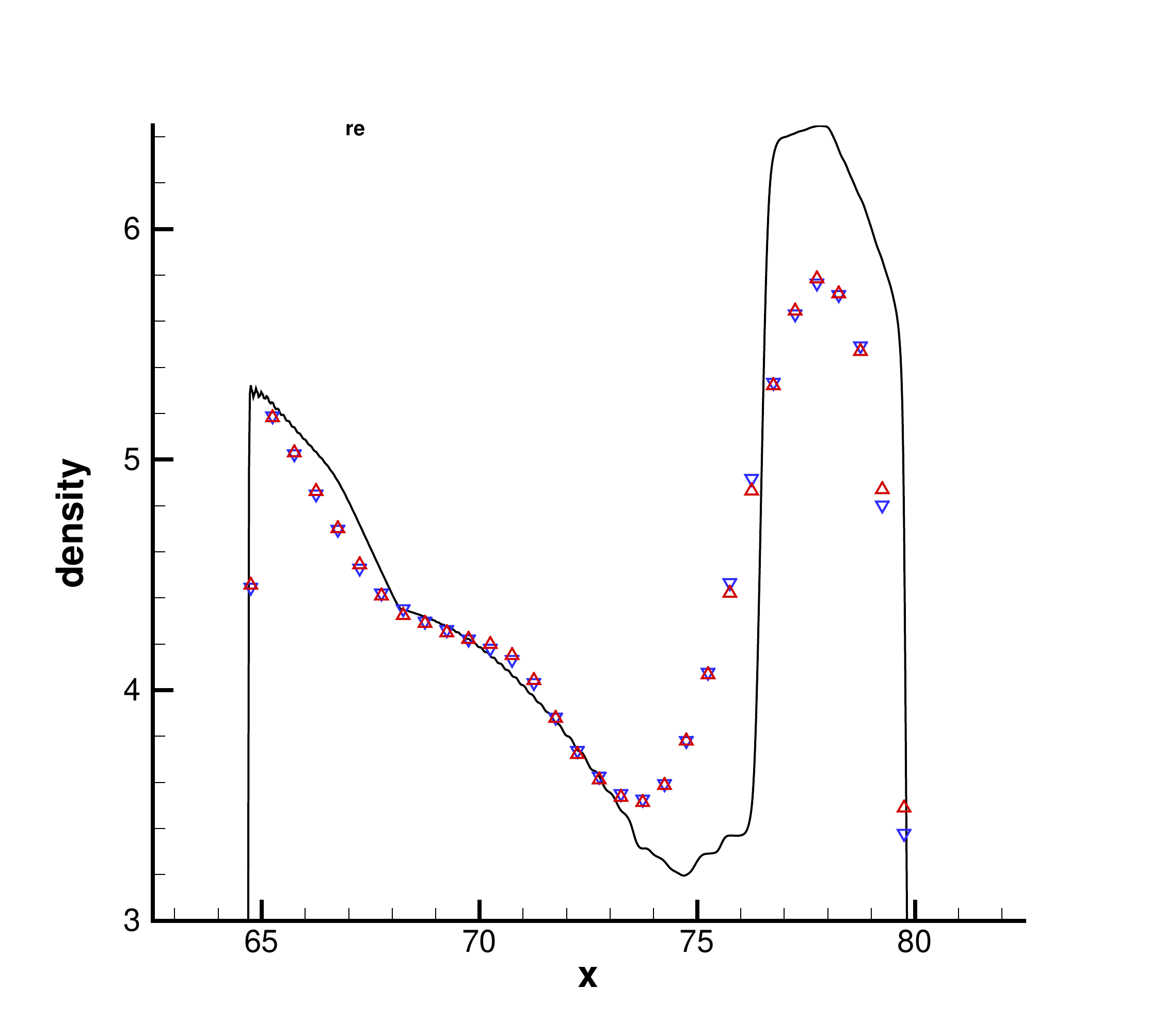}
\includegraphics[width=0.485\textwidth]{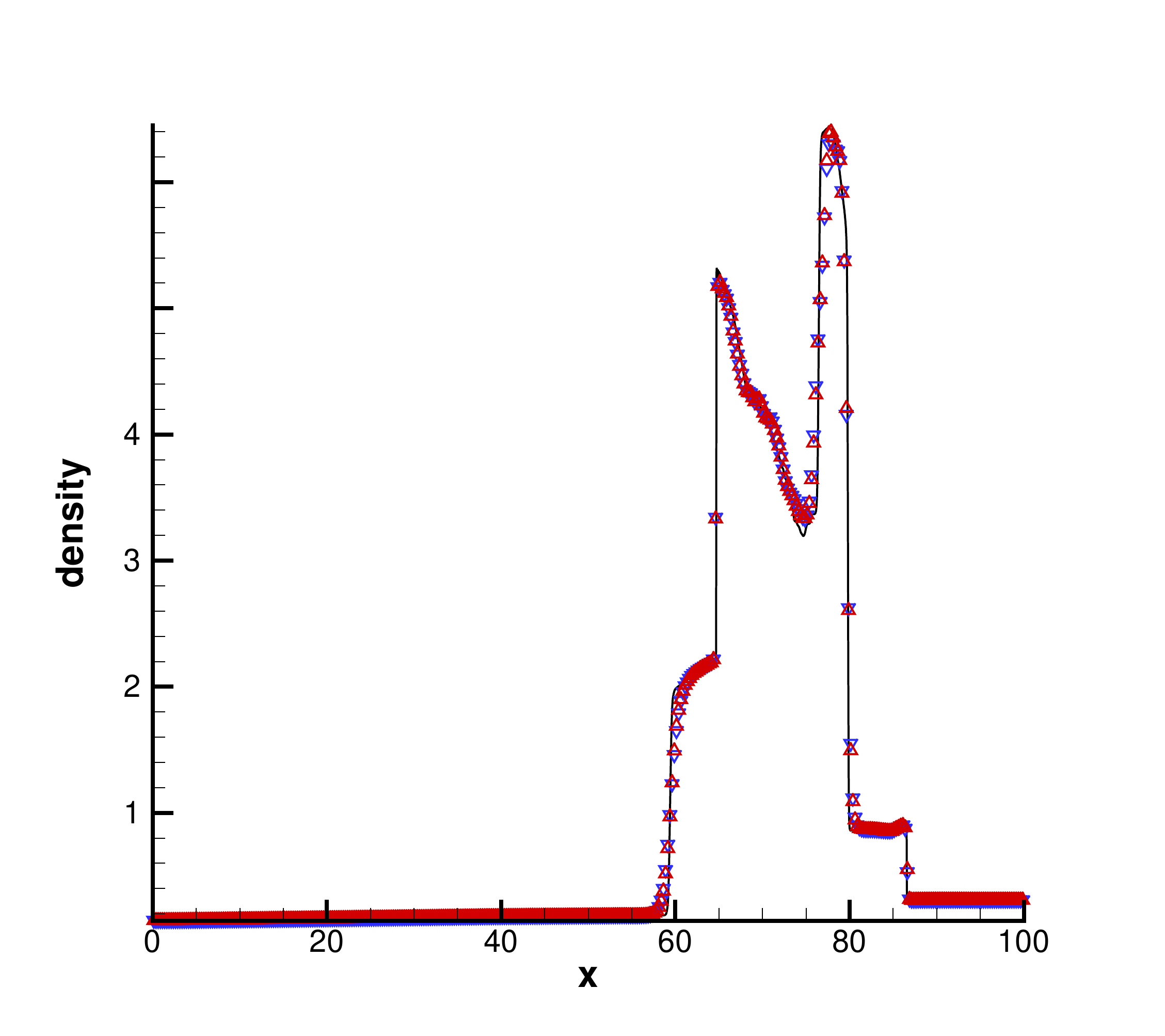}
\includegraphics[width=0.485\textwidth]{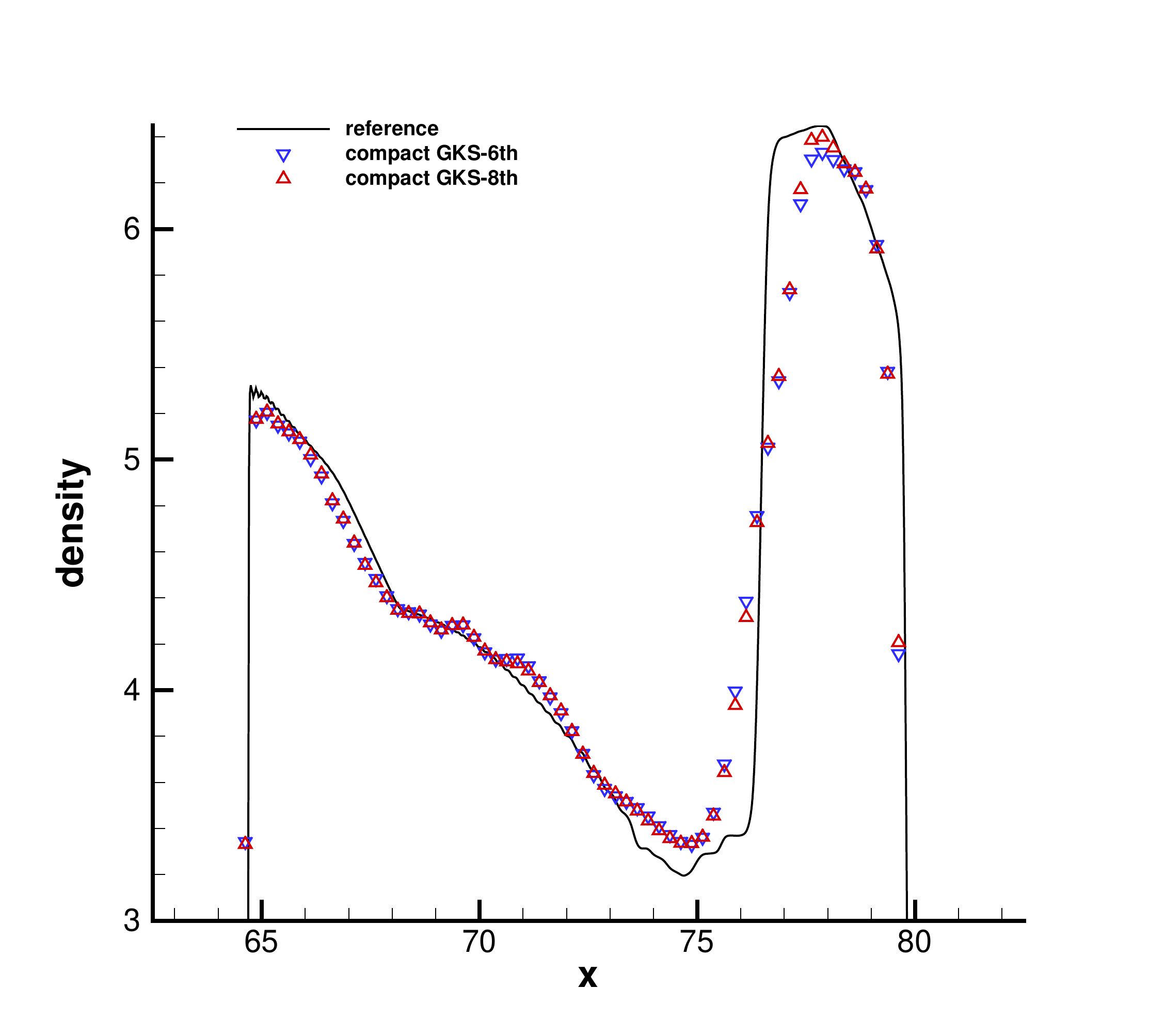}
\caption{\label{1d-riemann-Blast} Blast wave problem: the density distributions of
6th-order and 8th-order compact GKS for the blast-wave problem at $t=3.8$ with
$200$ (up) and $400$ (down) mesh points.}
\end{figure}

\begin{figure}[!htb]
\centering
\includegraphics[width=0.485\textwidth]{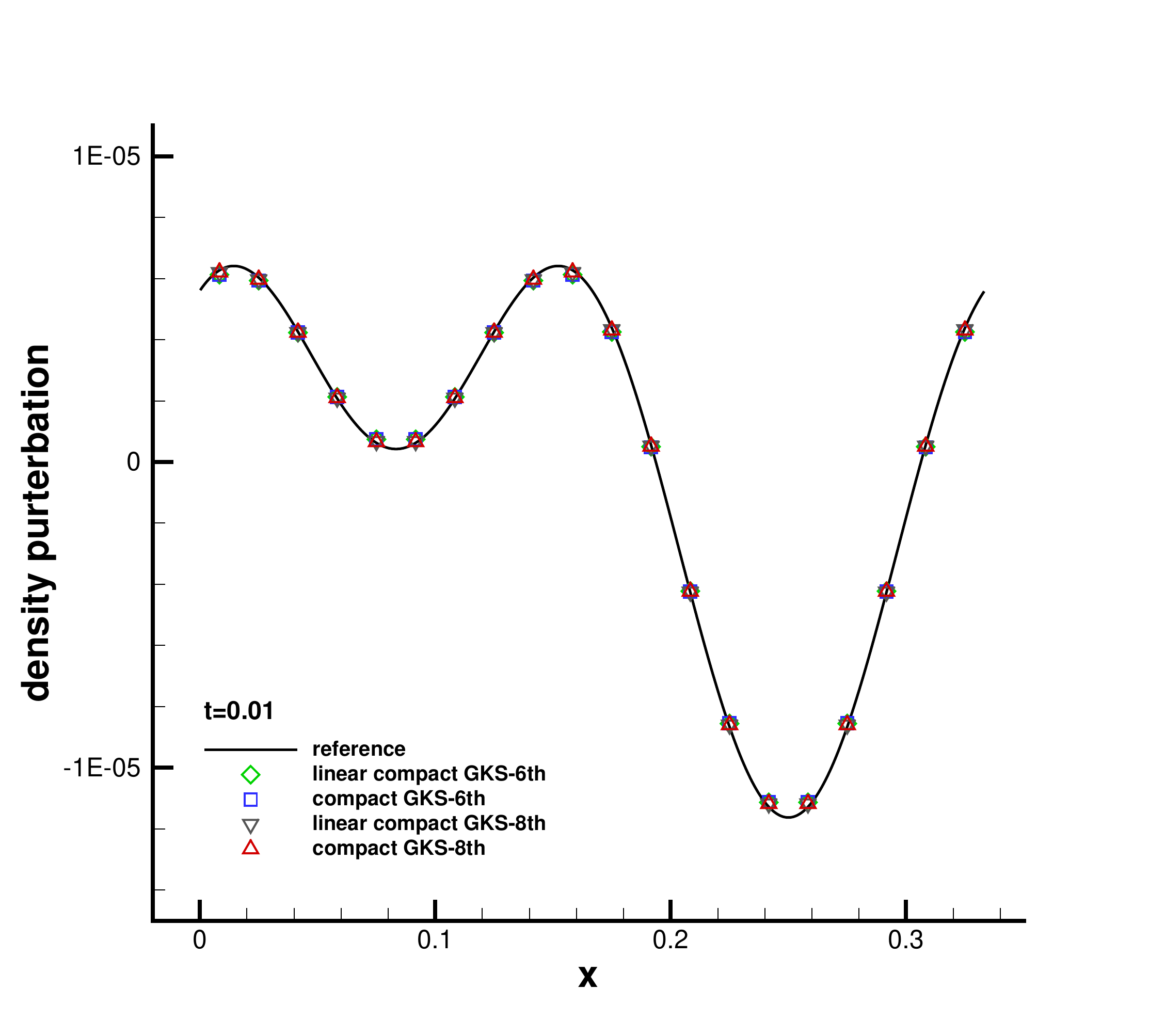}
\includegraphics[width=0.485\textwidth]{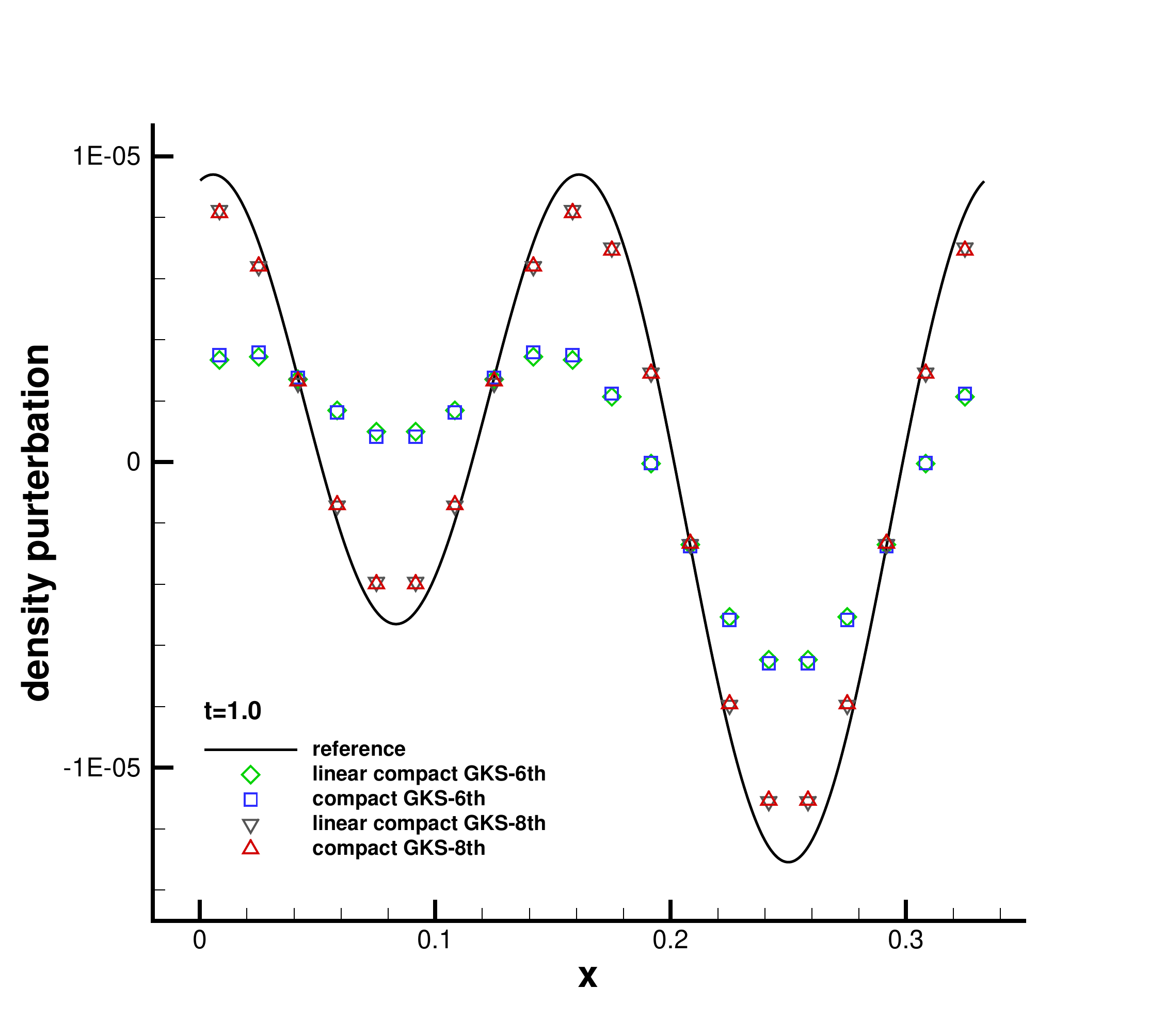}
\caption{\label{1d-acoustic1} One-dimensional acoustic problem: the distributions of density perturbation
obtained by compact GKS at $t=0.01$ (left, $1$ period) and $t=1.0$ (right, $100$ periods) with $20$ mesh points. The ratio
of non-dimensional initial density wavelength to sound speed is $\lambda_{\rho_0}/a_{\infty}=4.8\times 10^{-4}$.}
\end{figure}

\begin{figure}[!htb]
\centering
\includegraphics[width=0.485\textwidth]{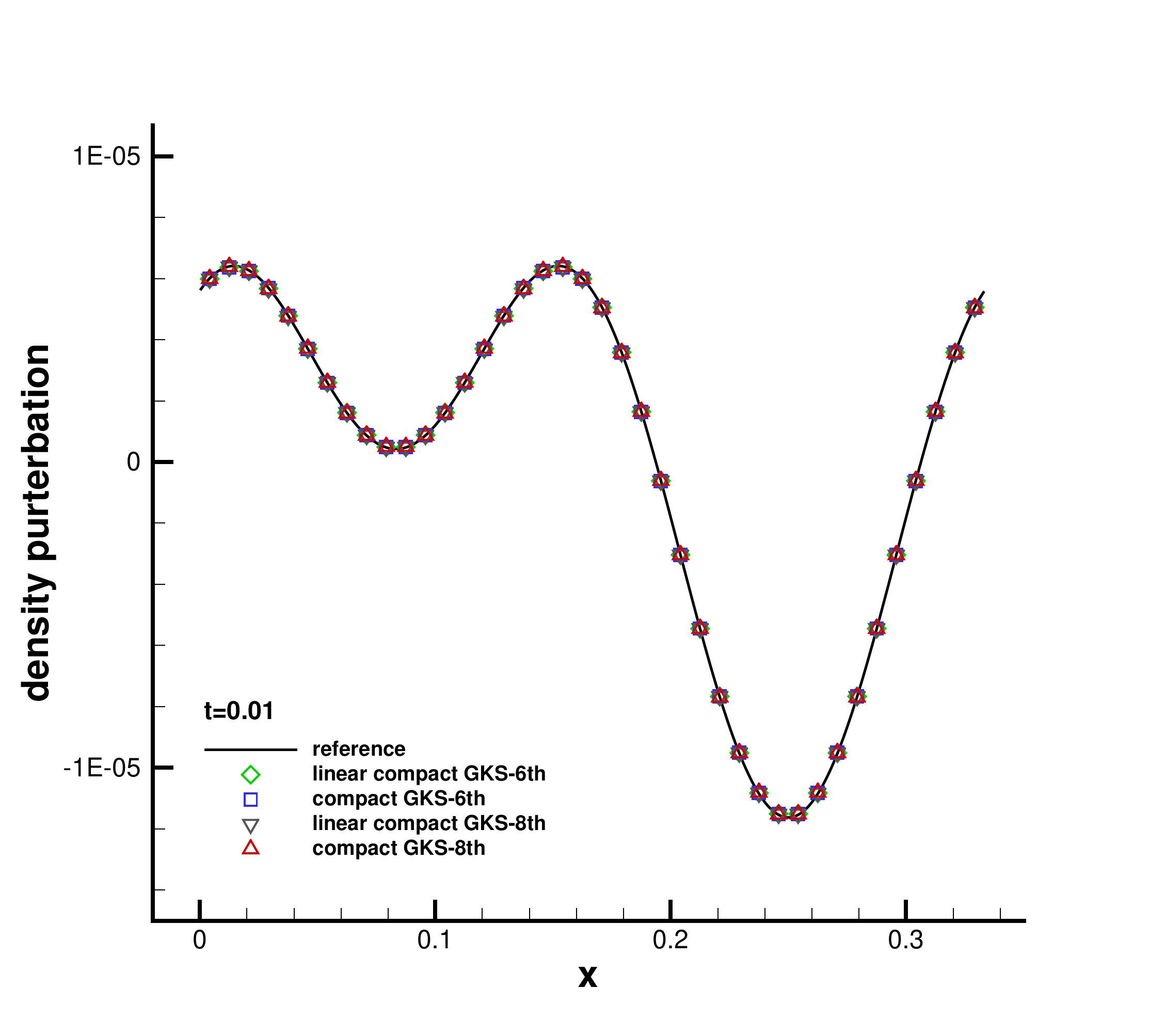}
\includegraphics[width=0.485\textwidth]{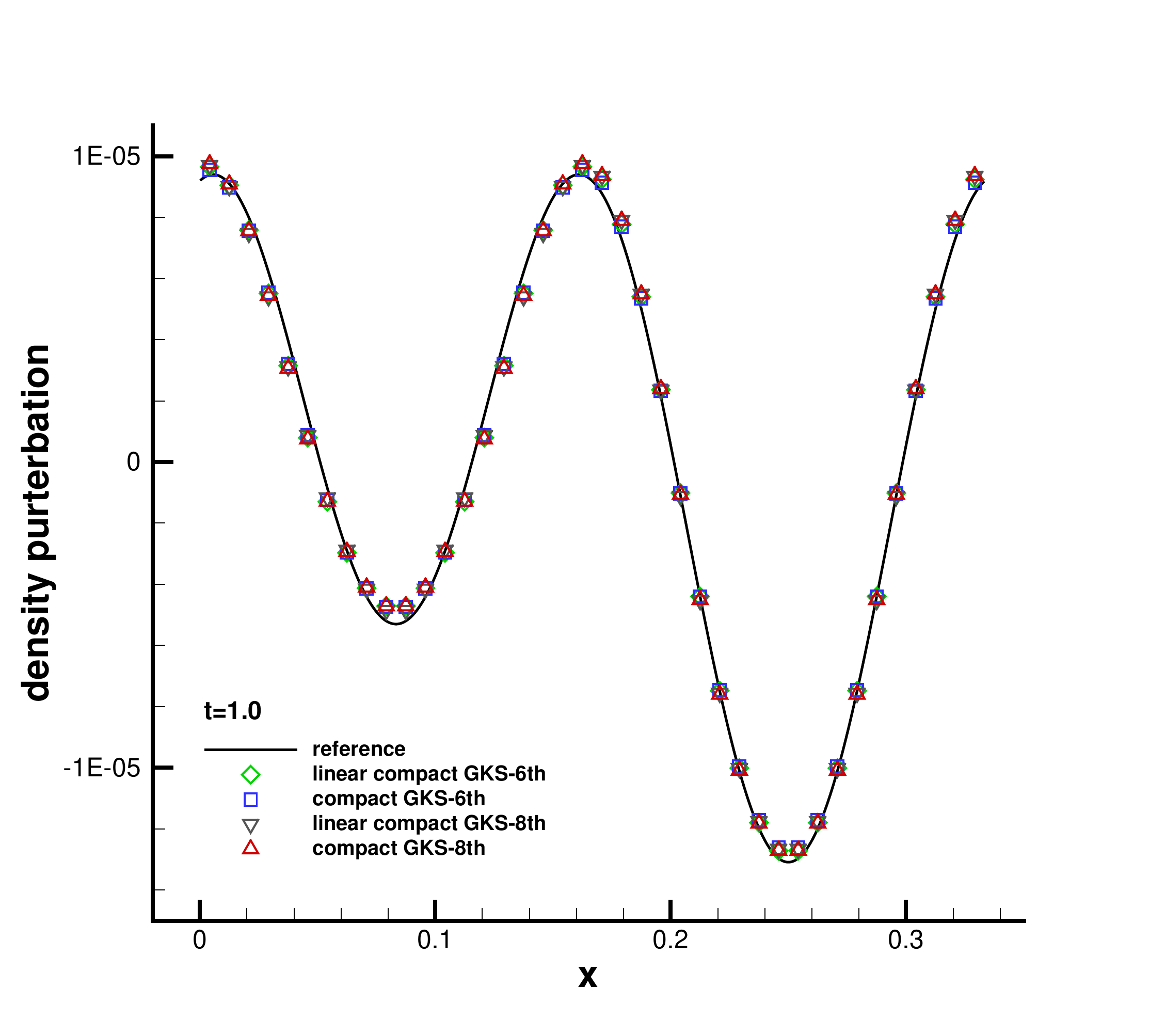}
\caption{\label{1d-acoustic2} One-dimensional acoustic problem: the distributions of density perturbation
obtained by compact GKS at $t=0.01$ (left, $1$ period) and $t=1.0$ (right, $100$ periods) with $40$ mesh
points. }
\end{figure}

The third test is the Woodward-Colella blast wave problem \cite{Case-Woodward},
and the initial conditions are given as follows
\begin{equation*}
(\rho,U,p) =\left\{\begin{array}{ll}
(1, 0, 1000), \ \ \ \ & 0\leq x<10,\\
(1, 0, 0.01), & 10\leq x<90,\\
(1, 0, 100), &  90\leq x\leq 100.
\end{array} \right.
\end{equation*}
The computational domain is $[0,100]$. The reflecting boundary conditions
are imposed on both ends. The density distributions and local enlargement from the
6th-order and 8th-order compact GKS are presented in Fig. \ref{1d-riemann-Blast} at $t=3.8$ with $200$ and $400$ mesh points.

\subsection{One-dimensional acoustic wave}
The one-dimensional acoustic wave propagation in
$x$-direction was proposed by Bai et al. in \cite{bai}. The test demonstrates
the high order and high resolution of the compact GKS to compute
acoustic wave propagating through a long-distance.
The initial conditions are given as follows
\begin{align*}
&U=U_{\infty}+\delta U, \delta U=\epsilon a_{\infty} \cos(\omega x), U_{\infty}=0\\
&\rho= \rho_{\infty}+\delta \rho,\delta \rho=\epsilon \rho_{\infty} \cos(2\omega x), \rho_{\infty}=1.1771\\
&\frac{p}{p_{\infty}}=(\frac{\rho}{\rho_{\infty}})^r, p_{\infty}=101325.0\\
&a_{\infty}=\sqrt{\gamma\frac{p_{\infty}}{\rho_{\infty}}},
\end{align*}
where $\epsilon=10^{-5}$ is the magnitude of initial perturbation, and $\omega=6\pi$
is the wavenumber of initial perturbations in velocity. The specific heat ratio is
$\gamma =1.4$. The acoustic wave given above is approximately linear because of the
very small $\epsilon$, and an analytical solution \cite{bai} is given from
the approximate acoustic wave equation,
\begin{align} \label{acoustic-1d-ref}
\begin{split}
\rho (x,t)=& \rho_{\infty}+ \frac{1}{2}\epsilon \rho_{\infty} [ \cos(2\omega(x-a_{\infty}t))+ \cos(2\omega(x+a_{\infty}t)) + \\
                                                              & \cos(\omega(x-a_{\infty}t))- \cos(\omega(x+a_{\infty}t))], \\
U (x,t)=& \frac{1}{2}\epsilon a_{\infty} [ \cos(2\omega(x-a_{\infty}t))- \cos(2\omega(x+a_{\infty}t)) + \\
                                          & \cos(\omega(x-a_{\infty}t))+ \cos(\omega(x+a_{\infty}t))], \\
p (x,t)=& p_{\infty}+ \frac{1}{2}\gamma\epsilon p_{\infty} [ \cos(2\omega(x-a_{\infty}t))+ \cos(2\omega(x+a_{\infty}t)) + \\
                                                            & \cos(\omega(x-a_{\infty}t))- \cos(\omega(x+a_{\infty}t))].
\end{split}
\end{align}
The computational domain is $[0,1/3]$. Periodic boundary conditions on both sides are adopted.

In the computation, we first obtain the cell average of the primitive variables, i.e.
$\tilde \rho _i =\int_{i-1/2}^{i+1/2} \rho(x)dx$ exactly, then convert them to the
conservative variables. Similarly, we could obtain the cell average of the derivatives,
i.e. $ \tilde \rho _{x,i} =\int_{i-1/2}^{i+1/2} \rho_x(x)dx $ exactly, from which
the derivatives of conservative variables can be obtained by the chain
rules. Theoretically, this has a 2nd-order accuracy in space. However, it is enough for
the comparison since the analytical solution in Eq. \eqref{acoustic-1d-ref} is also an
approximate solution.

Fig. \ref{1d-acoustic1} and Fig. \ref{1d-acoustic2} show the density distributions at
$t=0.01$ (left) and $t=1.0$ (right) with $20$ and $40$ uniform mesh points. The computational time
is greatly larger than $\lambda_{\rho_0}/a_{\infty}=4.8\times 10^{-4}$, where
$\lambda_{\rho_0}$ is the initial wavelength of density perturbation. Thus the
distance of acoustic wave propagation is greatly larger than the initial wavelength
of density perturbation. The results from GKS with linear and nonlinear reconstructions are in
good agreement. The results with $40$ mesh points agree well with the analytical solution, where the small deviation near the extreme
point is caused by the approximate analytical solution \cite{bai}.
This case also shows that the 6th-order GKS cannot give the accurate solution with $20$ mesh points after a long time wave propagation
($100$ periods) due to the relative large numerical error in comparison with the 8th-order scheme.

\begin{figure}[!htb]
\centering
\includegraphics[width=0.485\textwidth]{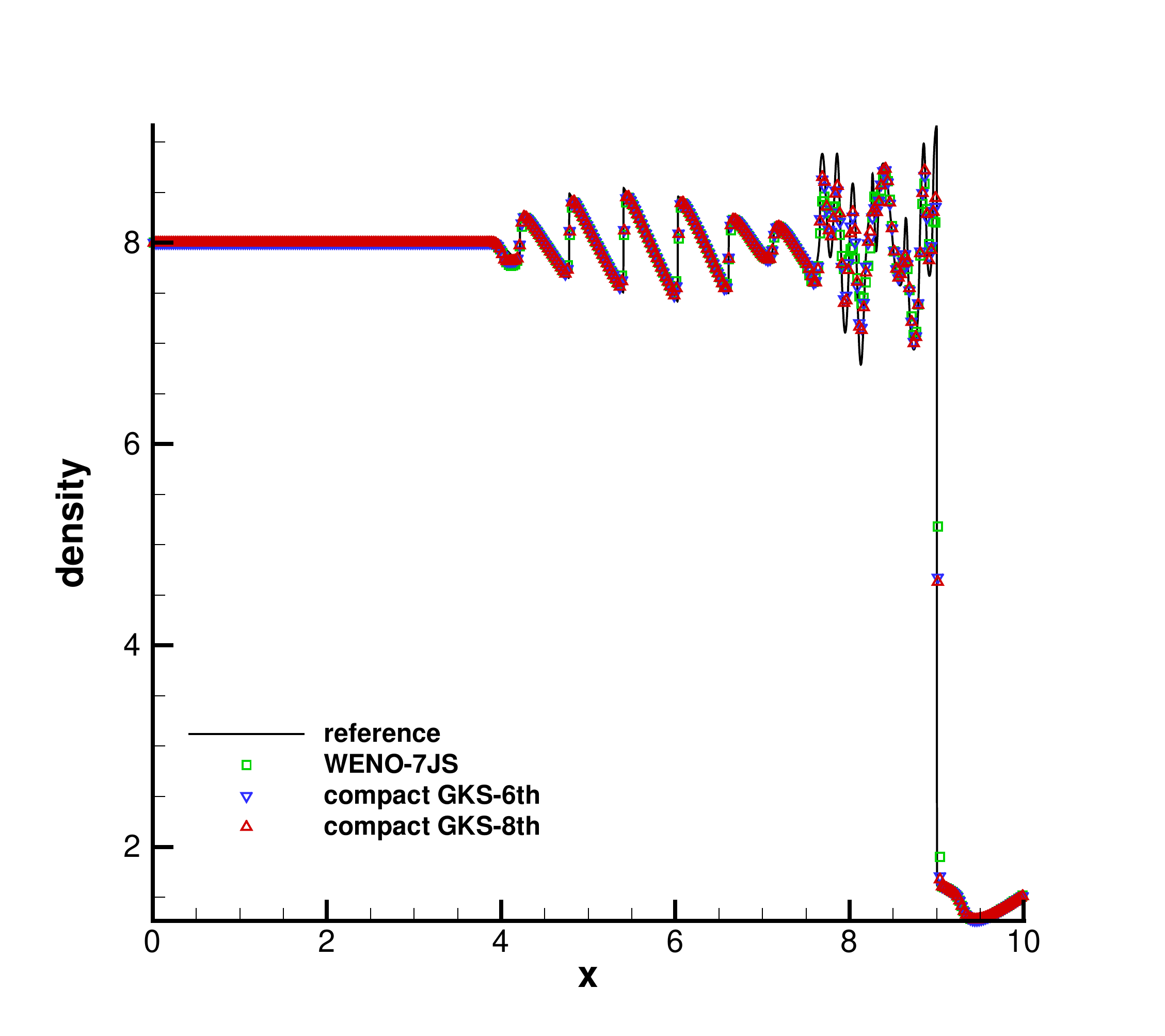}
\includegraphics[width=0.485\textwidth]{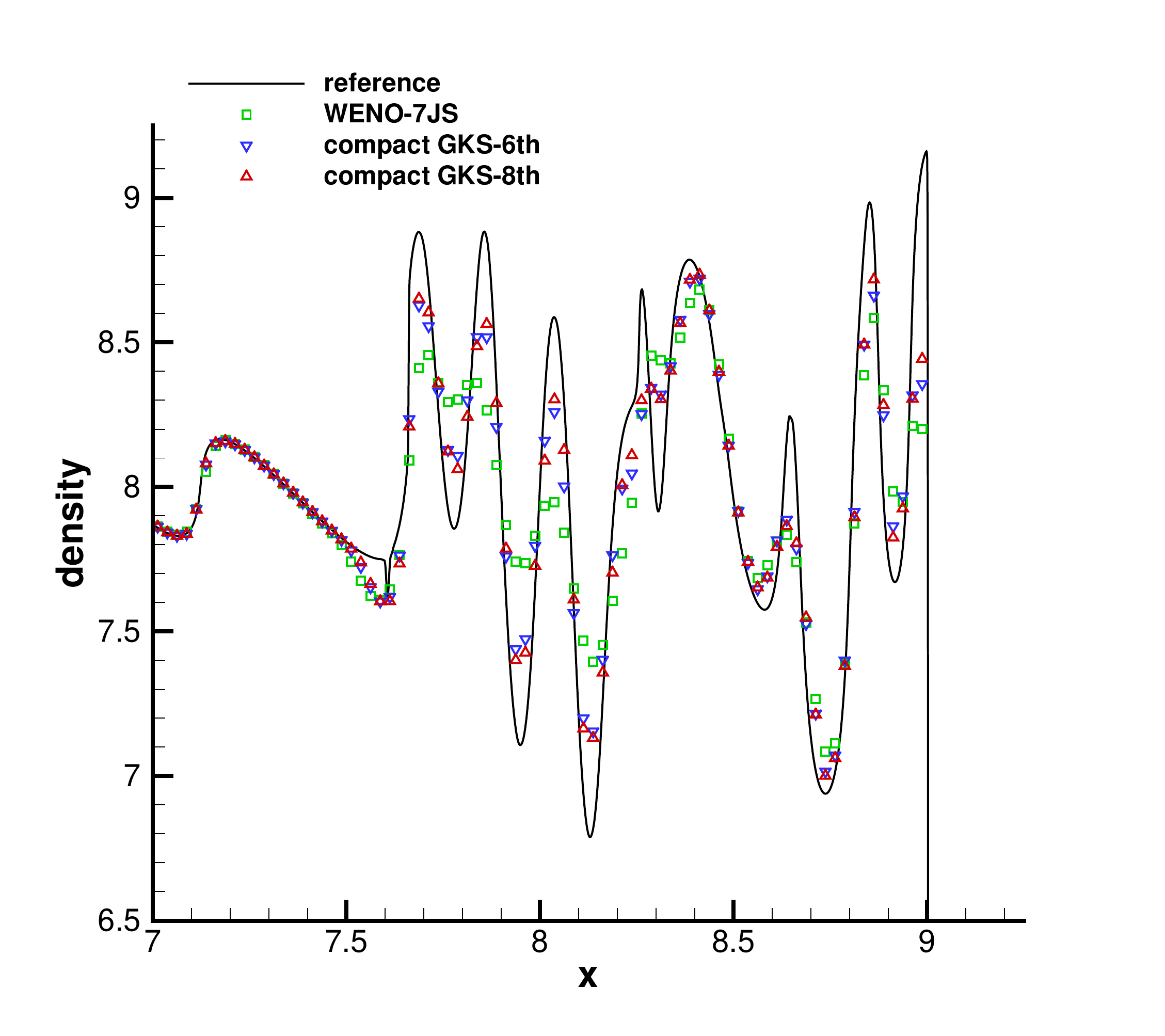}
\caption{\label{1d-shockvor-1} One-dimensional shock crossing velocity perturbation:
the density distribution and local enlargement from WENO-7JS non-compact GKS, 6th-order and 8th-order compact GKS at $t=0.8$ with $400$ mesh points.}
\end{figure}

\begin{figure}[!htb]
\centering
\includegraphics[width=0.485\textwidth]{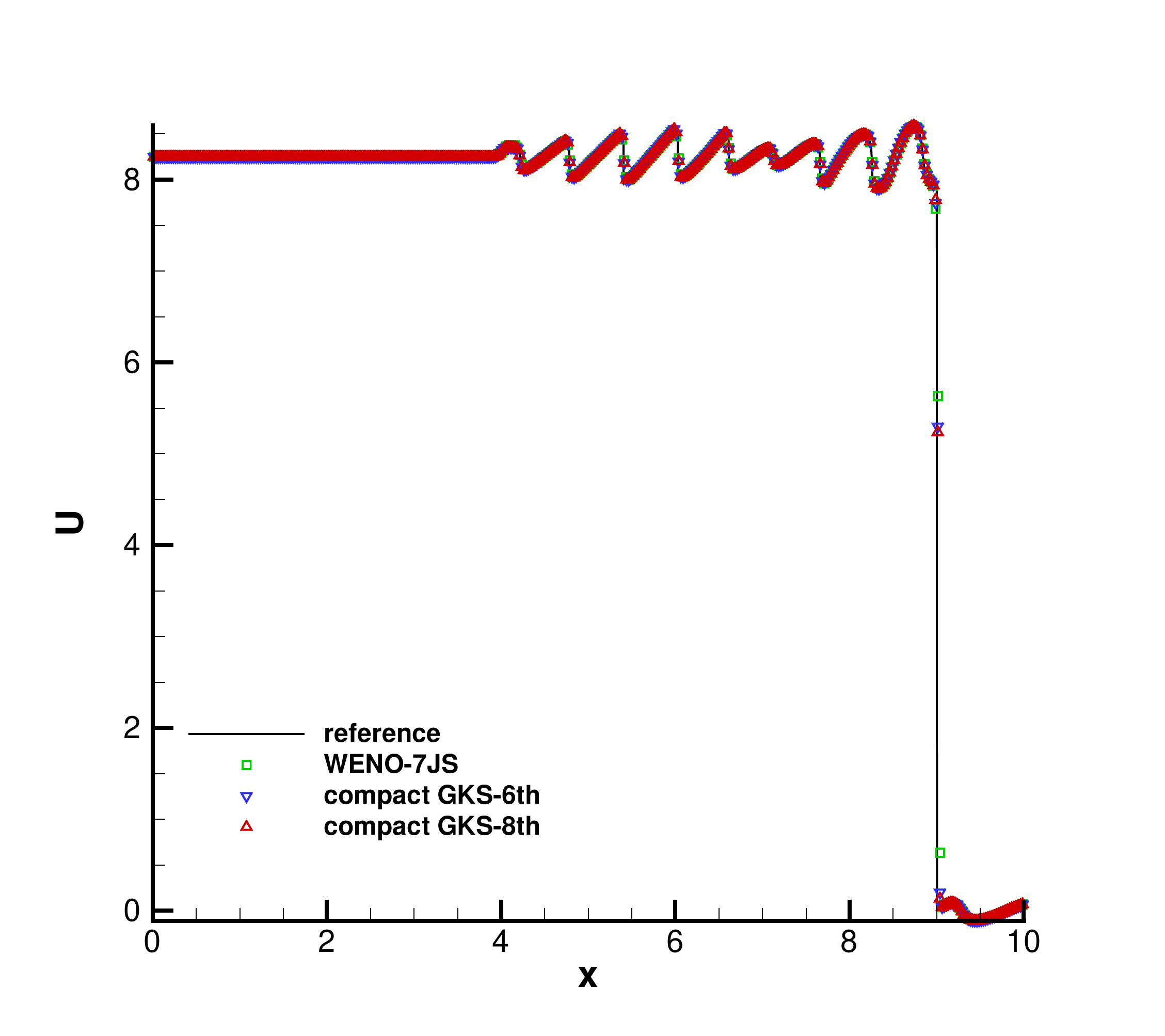}
\includegraphics[width=0.485\textwidth]{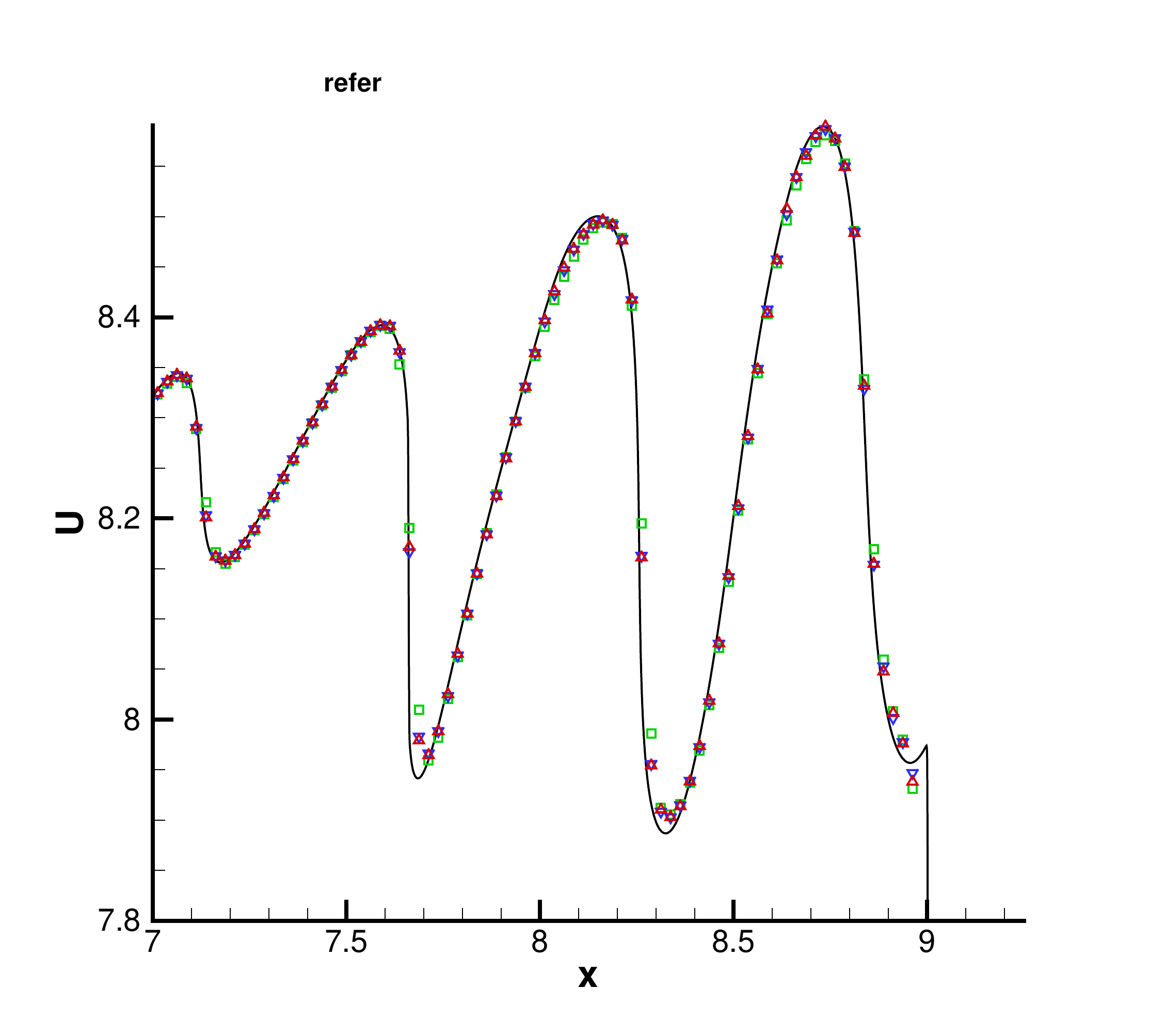}
\caption{\label{1d-shockvor-2} One-dimensional shock crossing velocity perturbation:
the velocity distribution and local enlargement from WENO-7JS non-compact GKS, 6th-order and 8th-order compact GKS at $t=0.8$ with $400$ mesh points.}
\end{figure}

\subsection{One-dimensional shock crossing velocity perturbation}
The interaction between strong shock wave and turbulence is an important flow
problem in gas dynamics, where the shock can cross the local wave packets
 which propagate faster or slower than the mean velocity.
To validate the current schemes for this kind of problems, we propose a test of shock with a Mach number $Ma=10$ crossing
a velocity perturbation.
The initial condition is given as follows
\begin{equation*}
(\rho,U,p)=\left\{\begin{array}{ll}
(8, 8.25, 116.5),  \ \ \ \ &  x \leq 1,\\
(1.4, 0.2\sin (2\pi(x-1)), 1),  &  1 <x \leq 10.
\end{array} \right.
\end{equation*}
The computational domain is $[0, 10]$.
In order to compare the resolution for capturing the multi-scale structure from
different schemes, a relative coarse mesh with $400$ points is used in the computation.
The non-reflecting boundary condition is used at both ends. The computed density
profile and the local enlargement from different schemes at $t=0.8$ are shown in
Fig.\ref{1d-shockvor-1}, and the distribution of velocity and its local enlargement are
shown in Fig.\ref{1d-shockvor-2}. The current compact 6th-order and 8th-order
schemes perform well for resolving the multi-scale flow structures, where
the non-compact WENO-7JS GKS has a relative large error. The compactness seems a preferred choice for a scheme with a better wave resolving power.

\begin{figure}[!htb]
\centering
\includegraphics[width=0.725\textwidth]{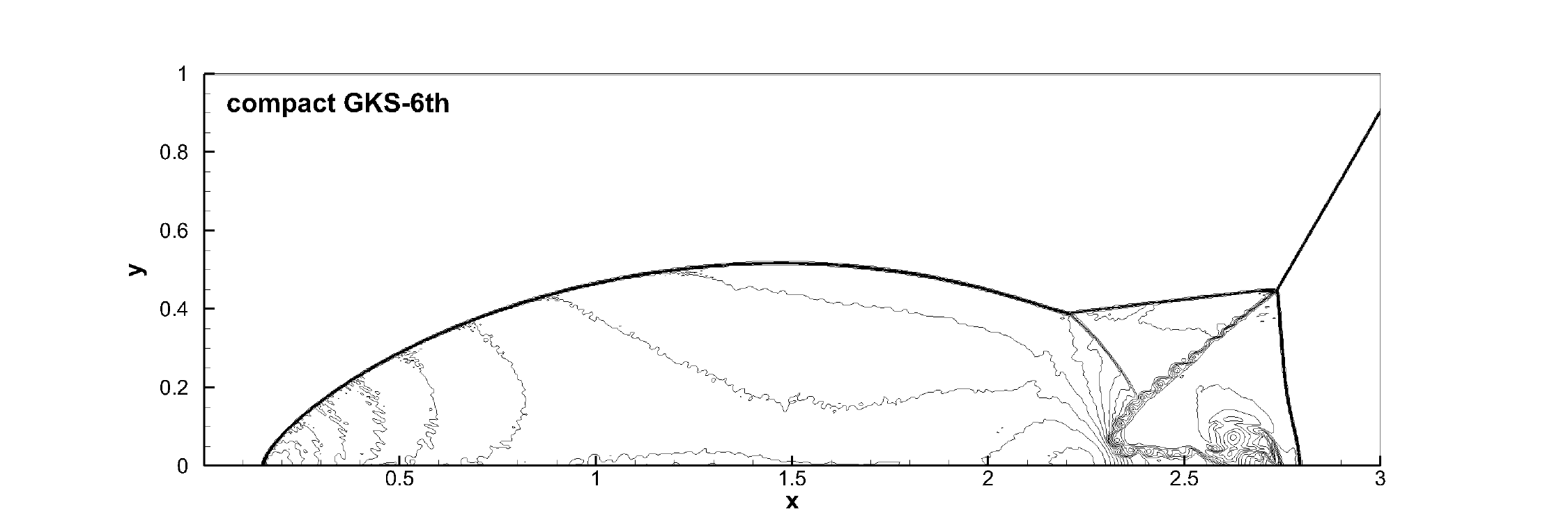}
\includegraphics[width=0.725\textwidth]{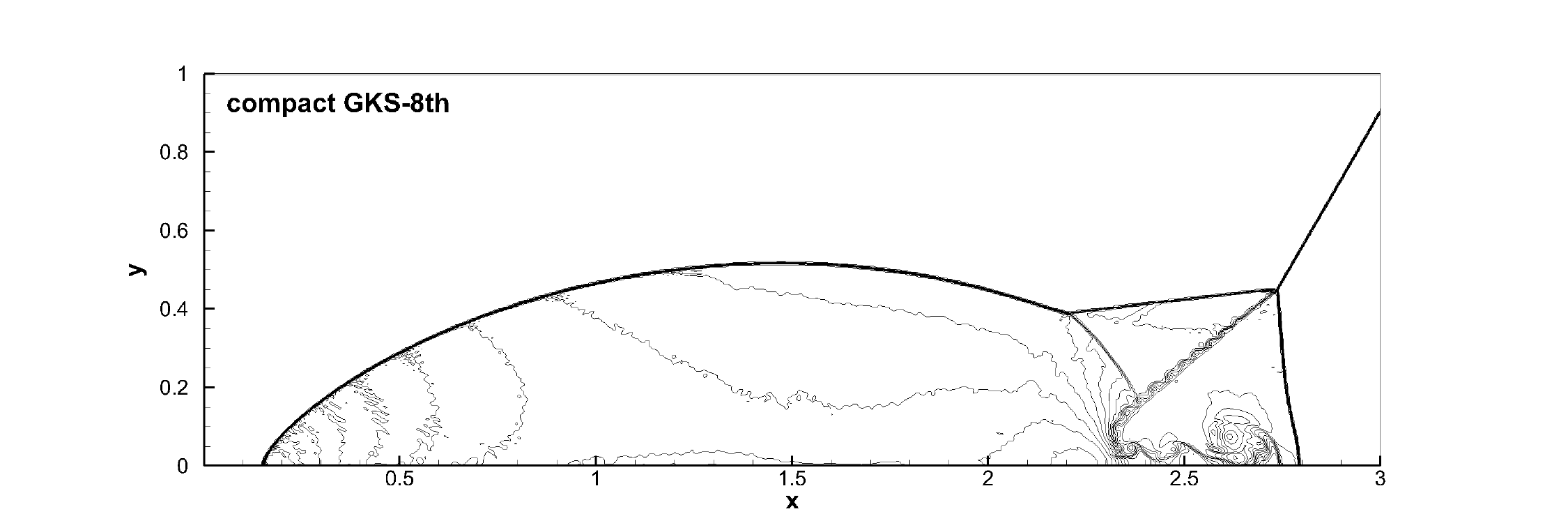}
\caption{\label{double-mach-1} Double Mach reflection: the density contours of 6th-order
and 8th-order compact GKS with  $960\times240$ mesh points.}
\end{figure}

\begin{figure}[!htb]
\centering
\includegraphics[width=0.725\textwidth]{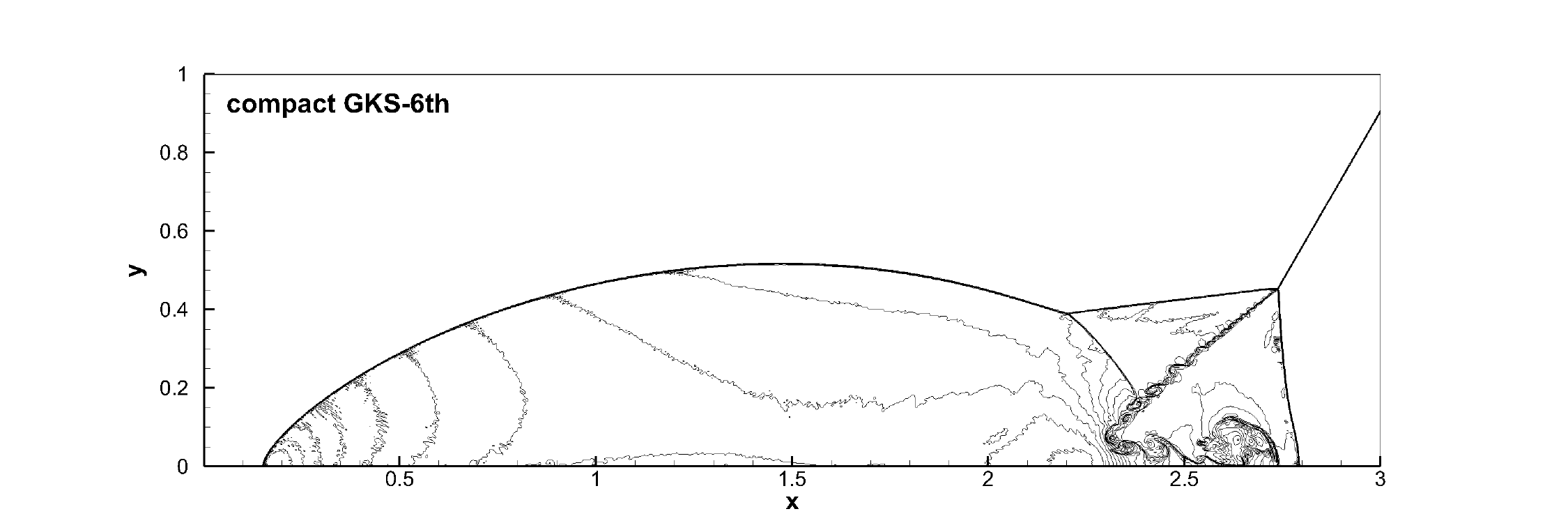}
\includegraphics[width=0.725\textwidth]{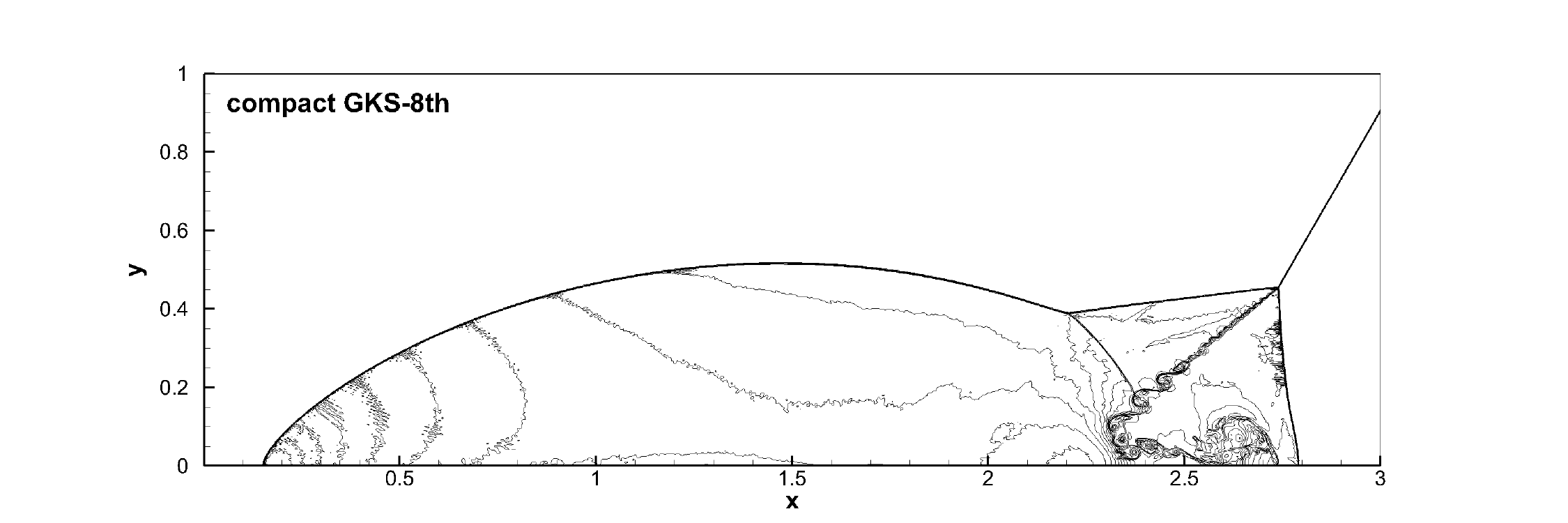}
\caption{\label{double-mach-2} Double Mach reflection: the density contours of 6th-order
and 8th-order compact GKS with  $1920\times480$ mesh points.}
\end{figure}

\subsection{Double Mach reflection problem}
\begin{figure}[!htb]
\centering
\includegraphics[width=0.485\textwidth]{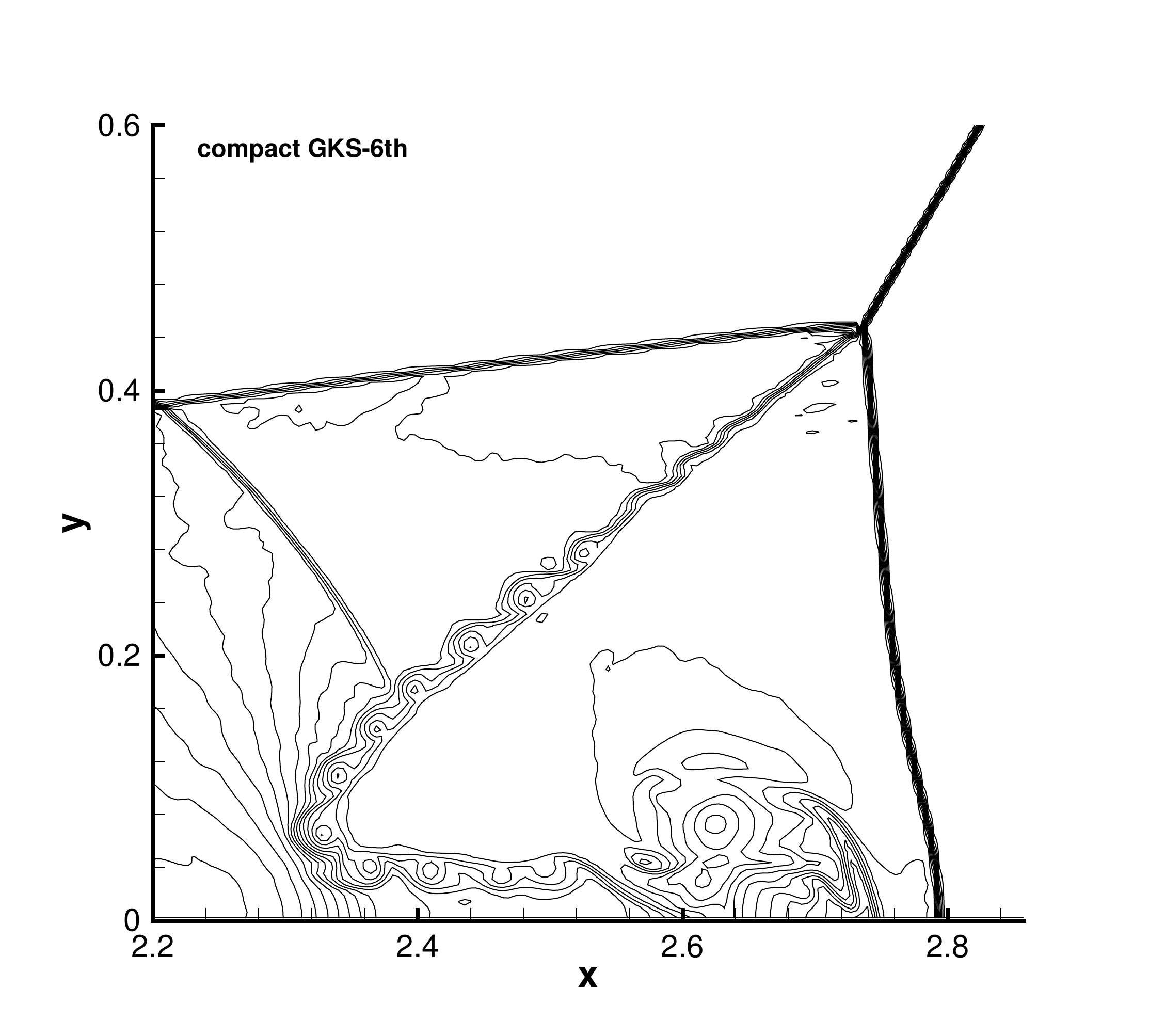}
\includegraphics[width=0.485\textwidth]{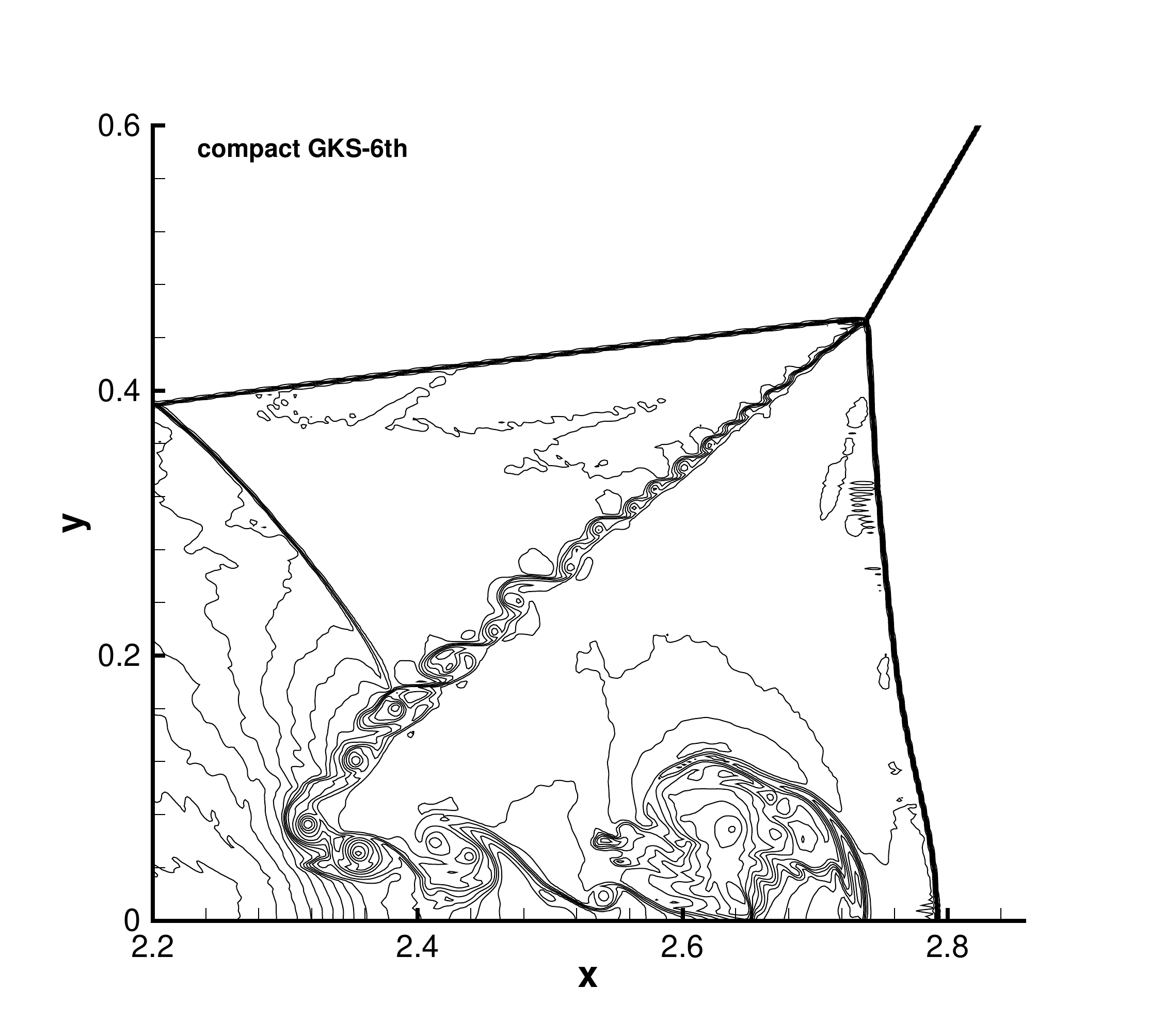}\\
\includegraphics[width=0.485\textwidth]{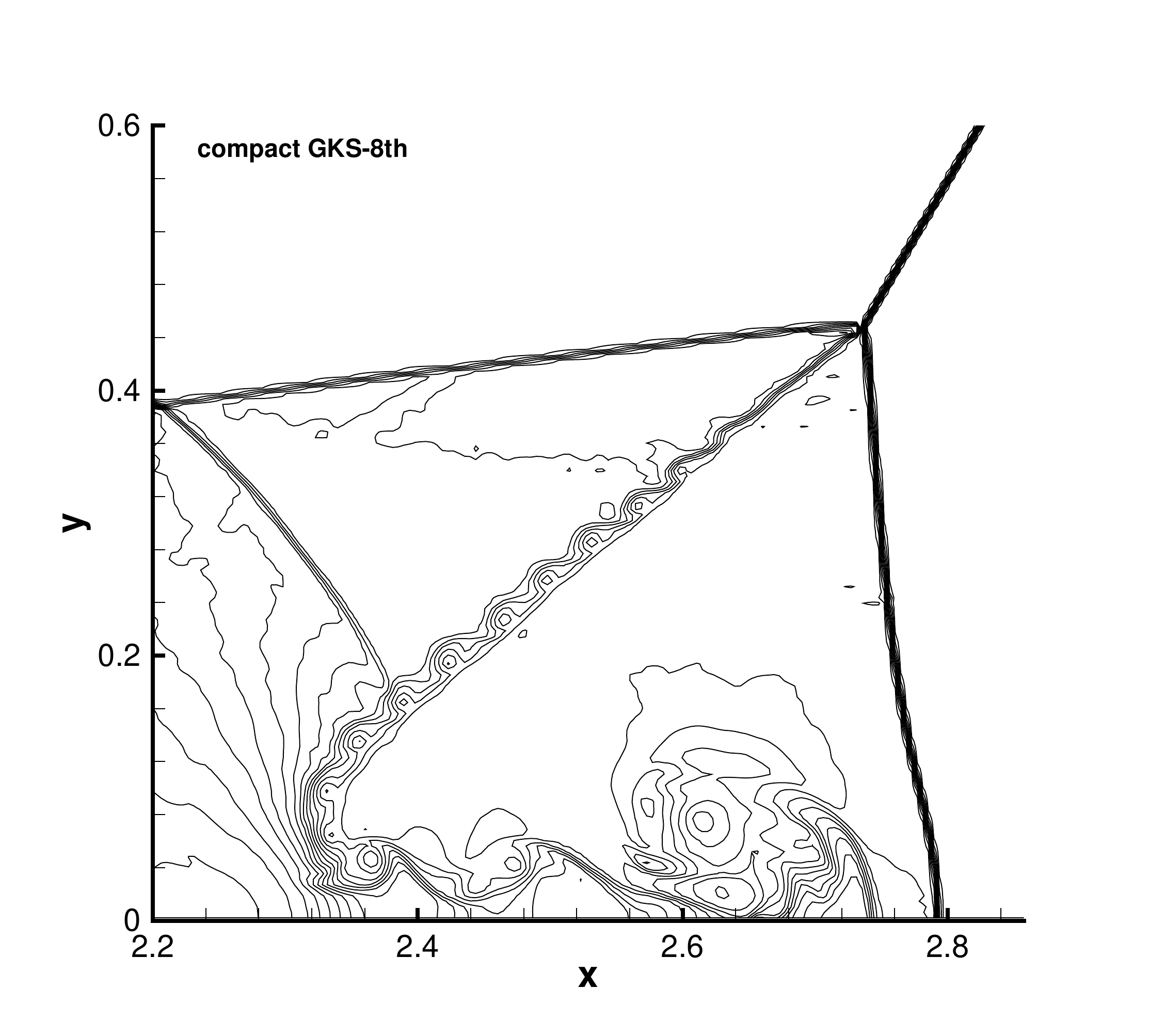}
\includegraphics[width=0.485\textwidth]{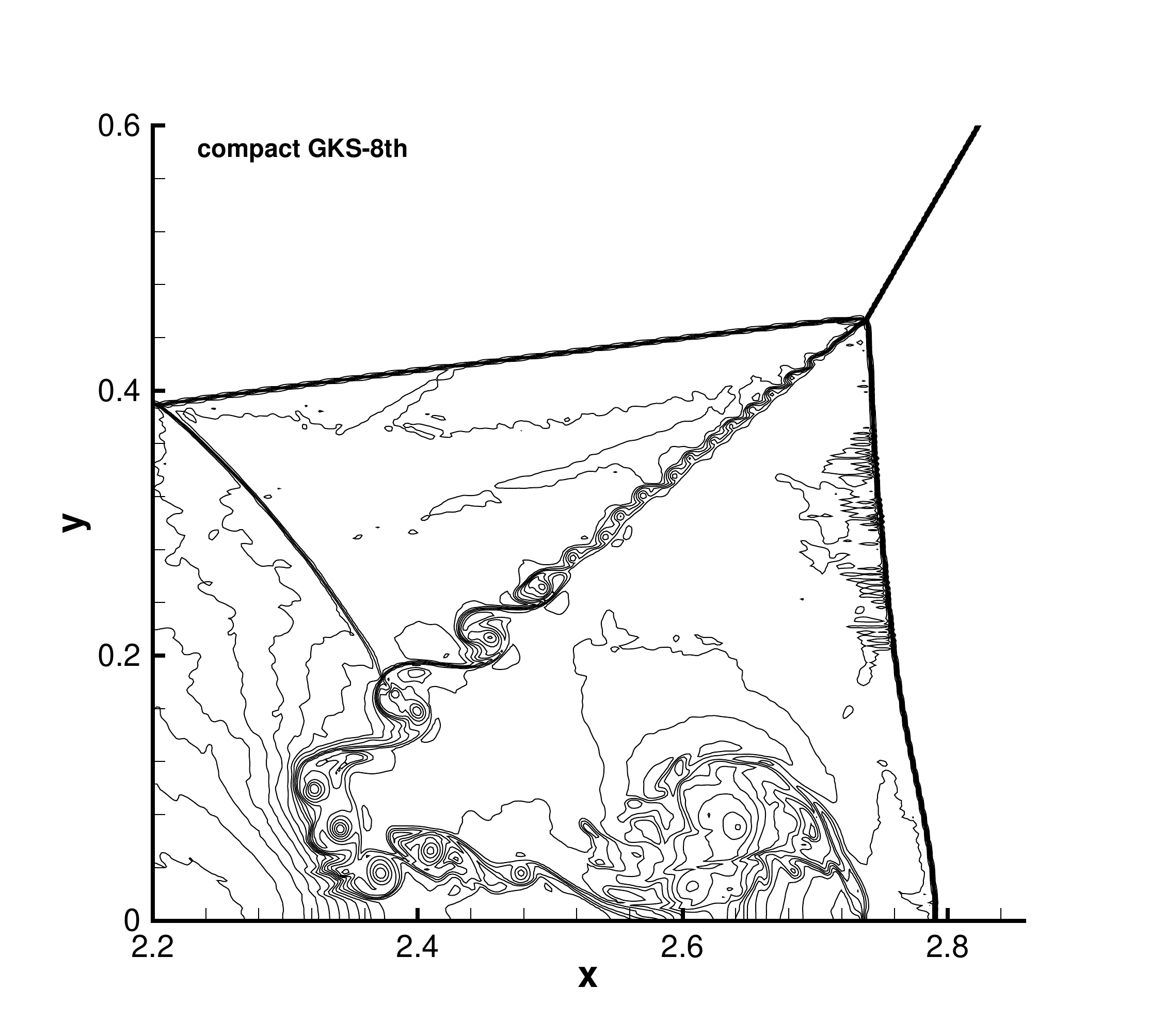}
\caption{\label{double-mach-3} Double Mach reflection: the local enlarged density distributions
around the triple point of the 6th-order (up)  and 8th-order (down) compact GKS with $960\times240$ (left)
and $1920\times480$ (right) mesh points.}
\end{figure}

In this subsection, the double Mach reflection problem is tested.
The test was extensively studied by Woodward and Colella
\cite{Case-Woodward} for the inviscid flow. The computational domain
is $[0,4]\times[0,1]$, and a solid wall lies at the bottom of the
computational domain starting from $x =1/6$. Initially a right-moving
Mach 10 shock is positioned at $(x,y)=(1/6, 0)$, which has a $60^\circ$
angle with the x-axis. The initial pre-shock and post-shock conditions are
\begin{align*}
(\rho, U, V, p)&=(8, 4.125\sqrt{3}, -4.125,
116.5),\\
(\rho, U, V, p)&=(1.4, 0, 0, 1).
\end{align*}
The reflecting boundary condition is used at the wall, and the exact
post-shock condition is imposed for the rest of bottom boundary. At
the top boundary, the flow variables are set to describe the exact
motion of the Mach $10$ shock. In this case, the compact 6th-order
and 8th-order compact GKS are tested. The density distributions with
$960\times240$ and $1920\times480$ uniform mesh points at $t=0.2$ are
shown in Fig. \ref{double-mach-1} and Fig. \ref{double-mach-2},
and the corresponding local density enlargements are shown in
Fig. \ref{double-mach-3}. The 6th-order and 8th-order schemes
resolve the flow structure under the triple Mach stem very well.

\subsection{Two-dimensional Riemann problems}
\begin{figure}[!htb]
\centering
\includegraphics[width=0.485\textwidth]{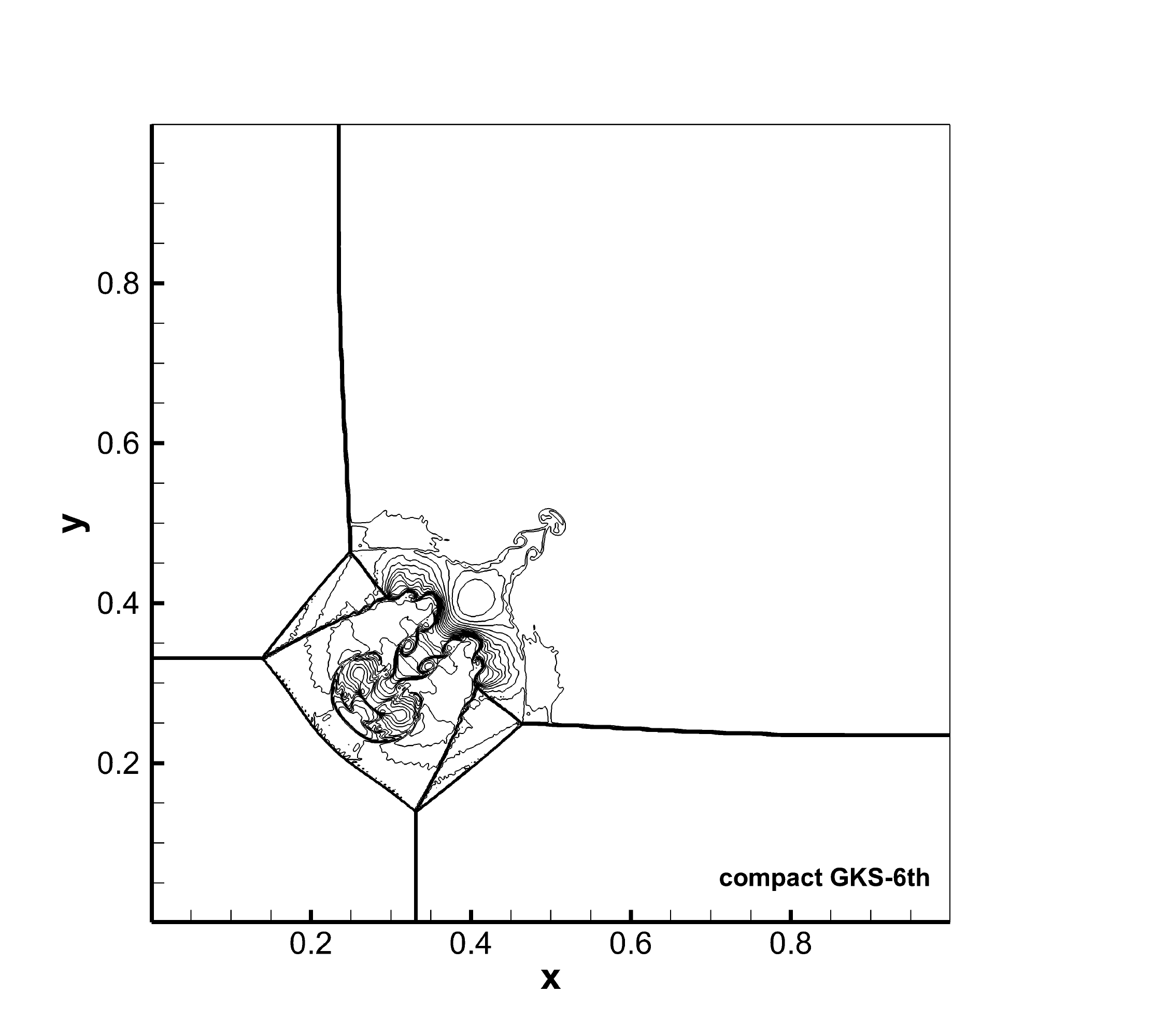}
\includegraphics[width=0.485\textwidth]{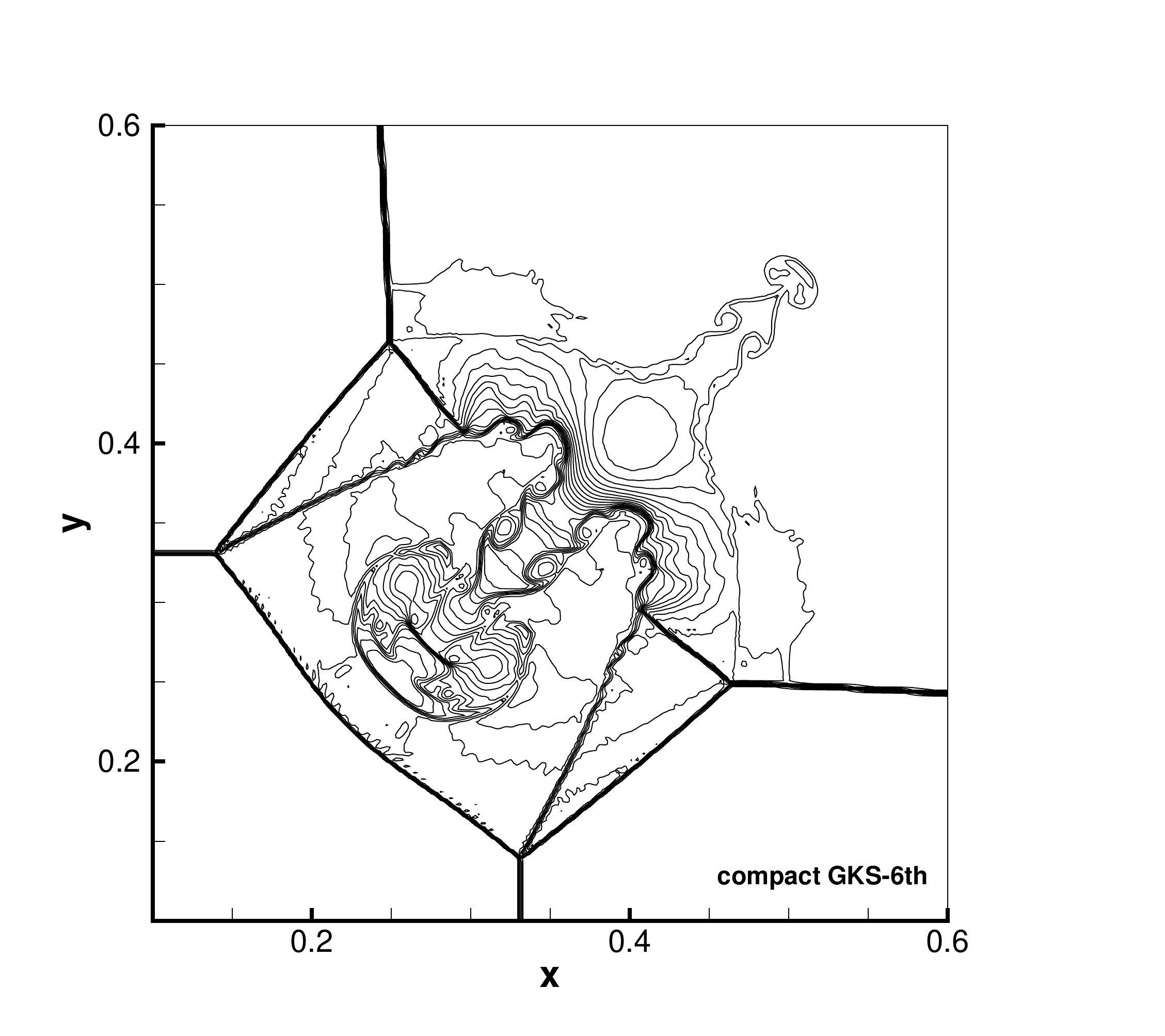}
\includegraphics[width=0.485\textwidth]{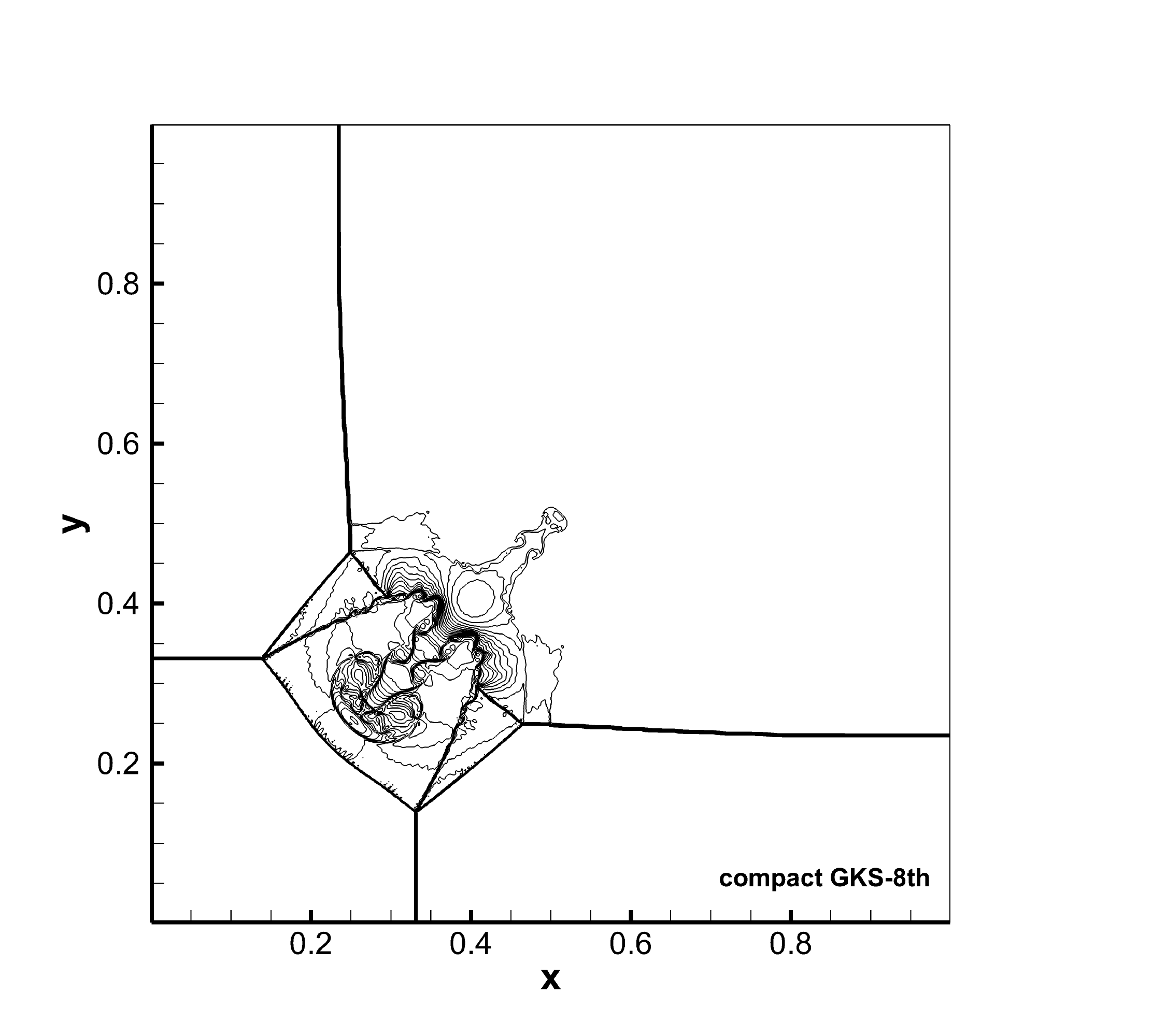}
\includegraphics[width=0.485\textwidth]{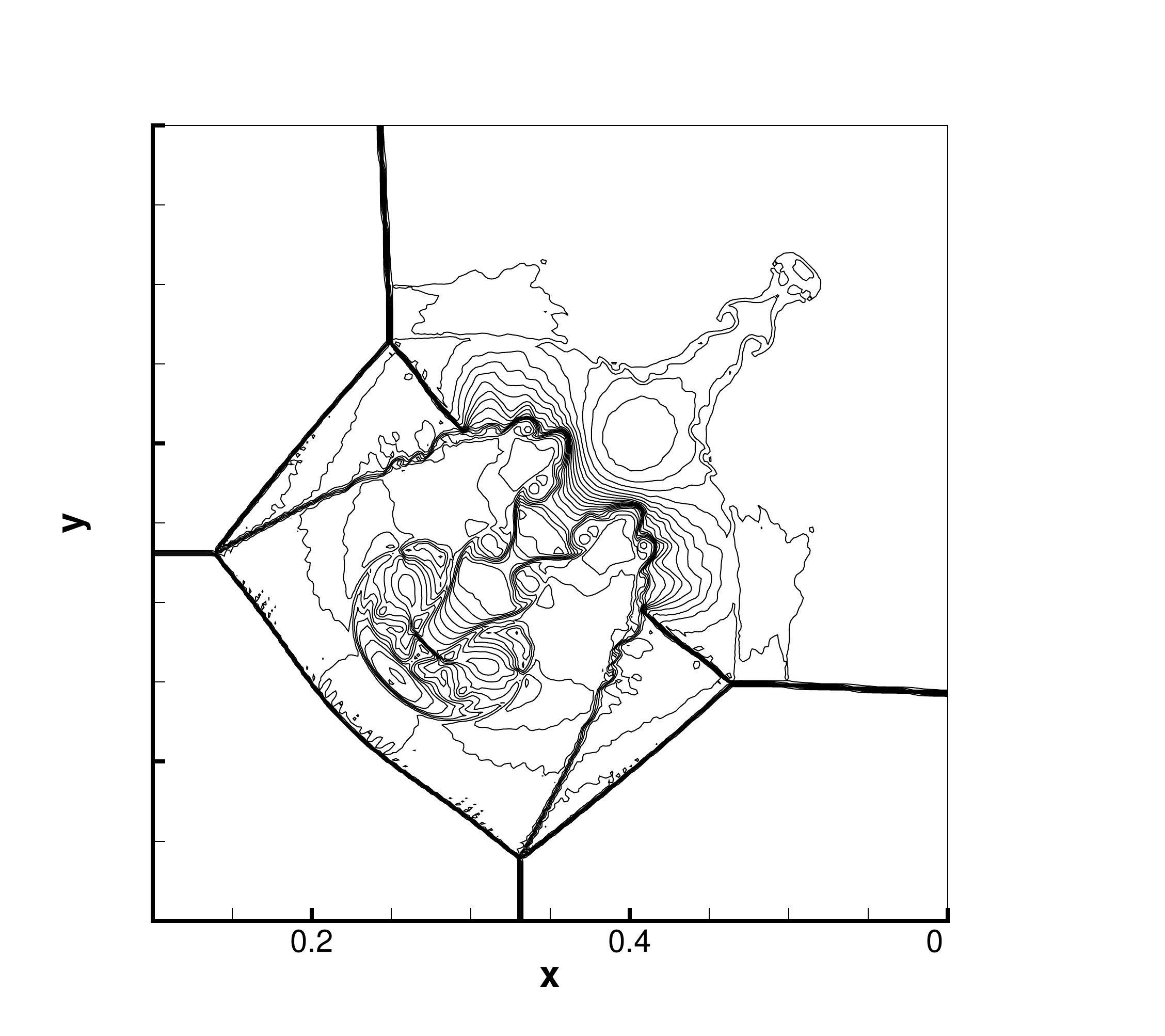}
\caption{\label{2d-riemann-1} 2D Riemann problem: the density distributions by 6th-order and 8th-order compact
GKS for the interaction of four shock waves at $t=0.4$ with $500\times500$ mesh points.}
\end{figure}

In the following, the two-dimensional Riemann problems are considered.
The first one is the interaction of four shocks
$\overleftarrow{S_{21}}~ \overleftarrow{S_{32}}~
\overleftarrow{S_{41}}~ \overleftarrow{S_{34}}$
\cite{Case-Riemann-1,Case-Riemann-2}, where the backward shock
waves connecting the areas $\Omega_l$ and $\Omega_r$ are denoted by
$\overleftarrow{S_{lr}}$. The initial conditions are given as follows
\begin{equation*}
(\rho,U,V,p)=
\left\{\begin{aligned}
&(1.5,0,0,1.5), &\Omega_1: x>0.7,y>0.7,\\
&(0.5323,1.206,0,0.3), &\Omega_2: x<0.7,y>0.7,\\
&(0.138,1.206,1.206,0.029), &\Omega_3: x<0.7,y<0.7,\\
&(0.5323,0,1.206,0.3), &\Omega_4: x>0.7,y<0.7.
\end{aligned} \right.
\end{equation*}
This case is just the mathematical formation of the double Mach
problem \cite{Case-Woodward} and the symmetric line $x=y$ can be
regarded as the rigid wall. The computational domain is
$[0,1]\times[0,1]$, and the non-reflecting boundary conditions are
used in all boundaries. The 6th-order and 8th-order compact GKS
are used, and the numerical results are given in
Fig. \ref{2d-riemann-1} at $t=0.4$ with $\Delta x=\Delta y=1/500$.

\begin{figure}[!htb]
\centering
\includegraphics[width=0.485\textwidth]{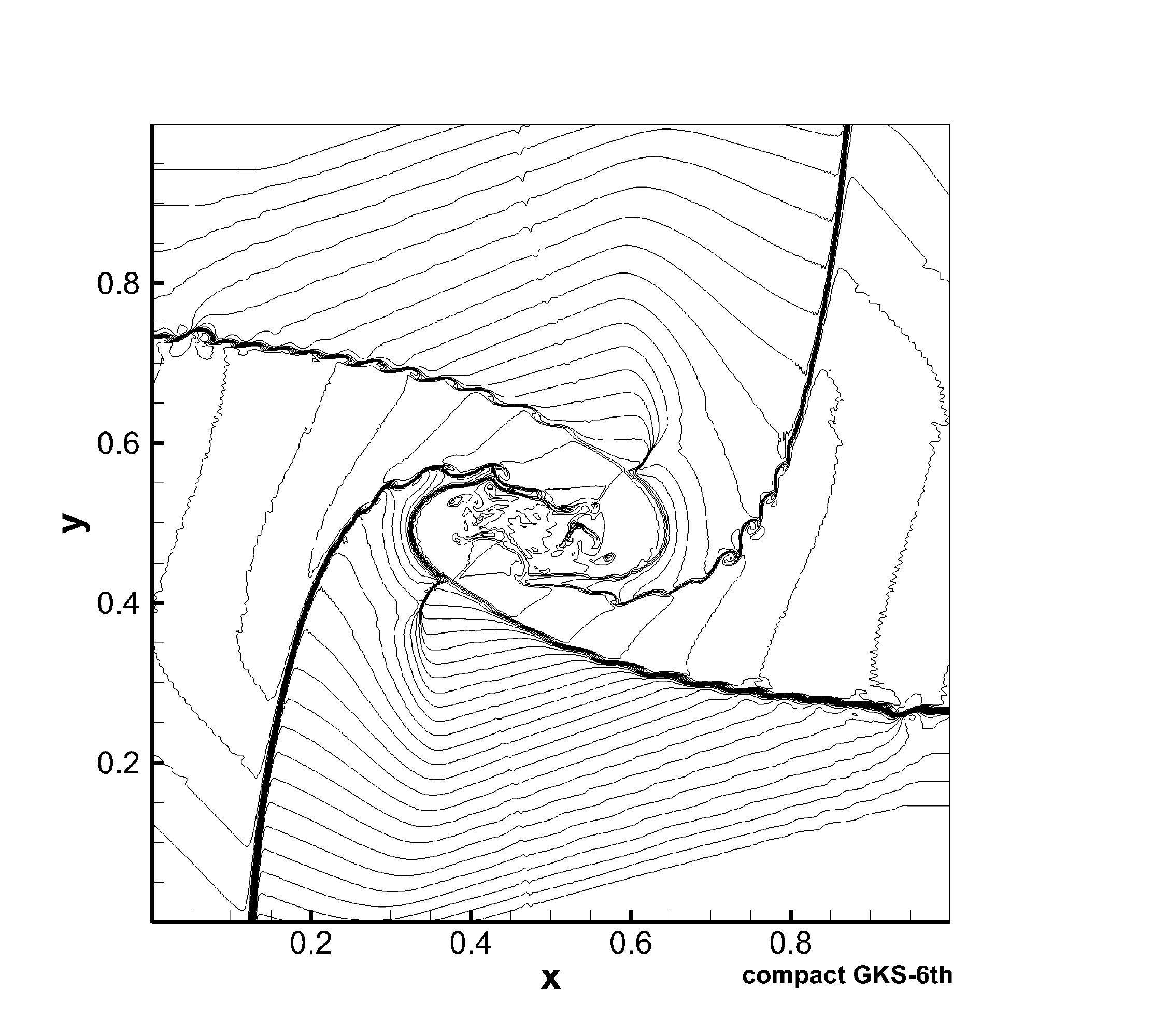}
\includegraphics[width=0.485\textwidth]{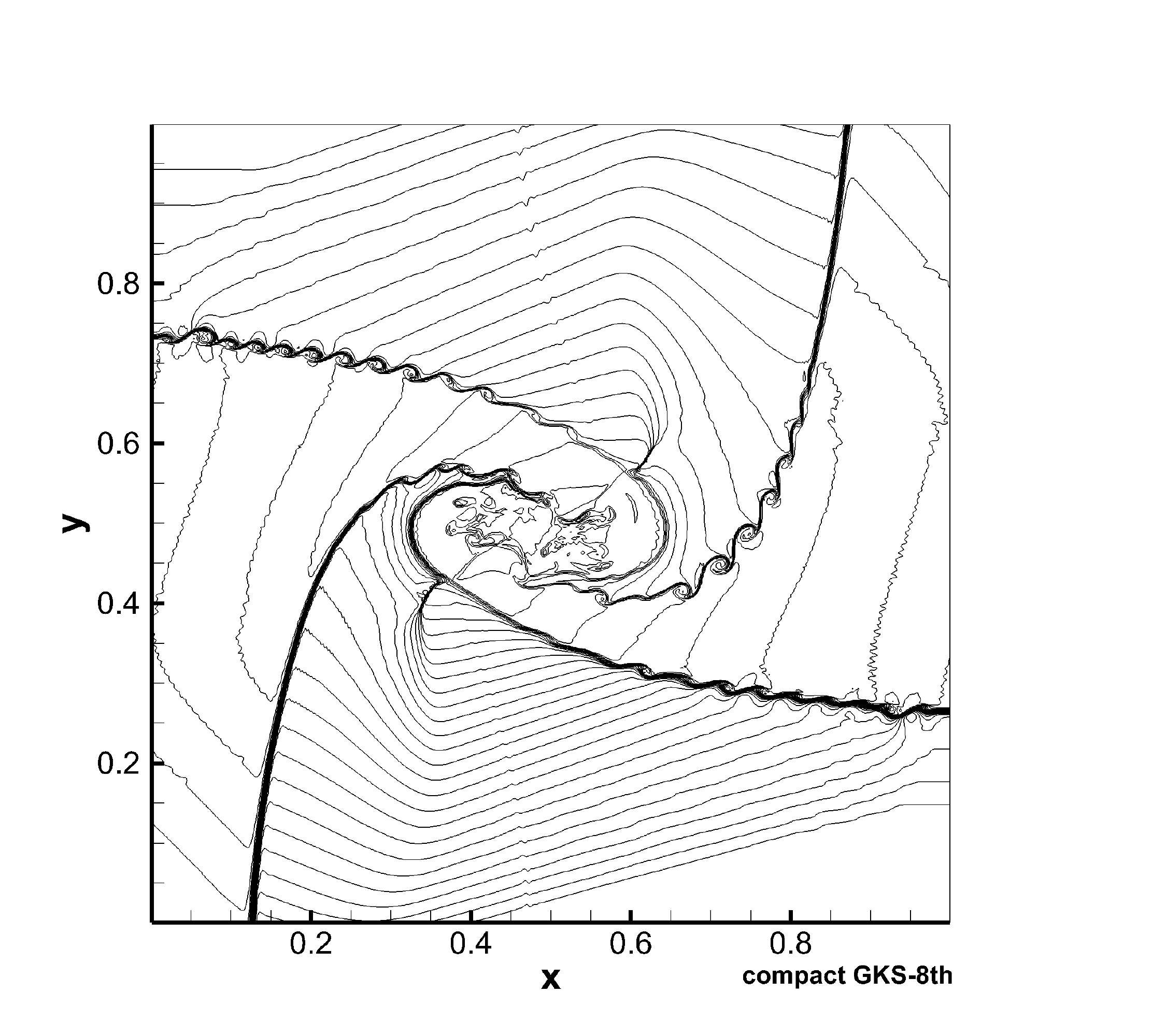}
\caption{\label{2d-riemann-2} 2D Riemann problem: the density distributions from 6th-order and 8th-order compact
GKS for the interactions
of four contact discontinuities at $t=0.5$ with $500\times500$ mesh points.}
\end{figure}

In the second case, the two-dimensional Riemann problem with four
contact discontinuities $J_{21}^-J_{32}^- J_{41}^- J_{34}^-$ is
tested \cite{Case-Riemann-1,Case-Riemann-2}, where the backward
contact discontinuities connecting the areas $\Omega_l$ and
$\Omega_r$ are denoted as $J_{lr}^-$. The initial conditions are
given as follows
\begin{equation*}
(\rho,U,V,p)=\left\{\begin{aligned}
         &(1 ,0.75,-0.5,1), &\Omega_1: x>0.5,y>0.5,\\
         &(2,0.75,0.5,1), &\Omega_2: x<0.5,y>0.5,\\
         &(1,-0.75,0.5,1), &\Omega_3: x<0.5,y<0.5,\\
         &(3,-0.75,-0.5,1), & \Omega_4: x>0.5,y<0.5.
                          \end{aligned} \right.
                          \end{equation*}
The computational domain is $[0,1]\times[0,1]$, and the
non-reflecting boundary conditions are also used in all boundaries.
The instantaneous interaction of the contact discontinuities results
in entropy wave and vortex sheets. To obtain the detailed flow
structure, the uniform mesh with $\Delta x=\Delta y=1/500$ are
used. The numerical results from compact GKS are given in Fig. \ref{2d-riemann-2} at $t=0.5$.

\begin{figure}[!h]
\centering
\includegraphics[width=0.485\textwidth]{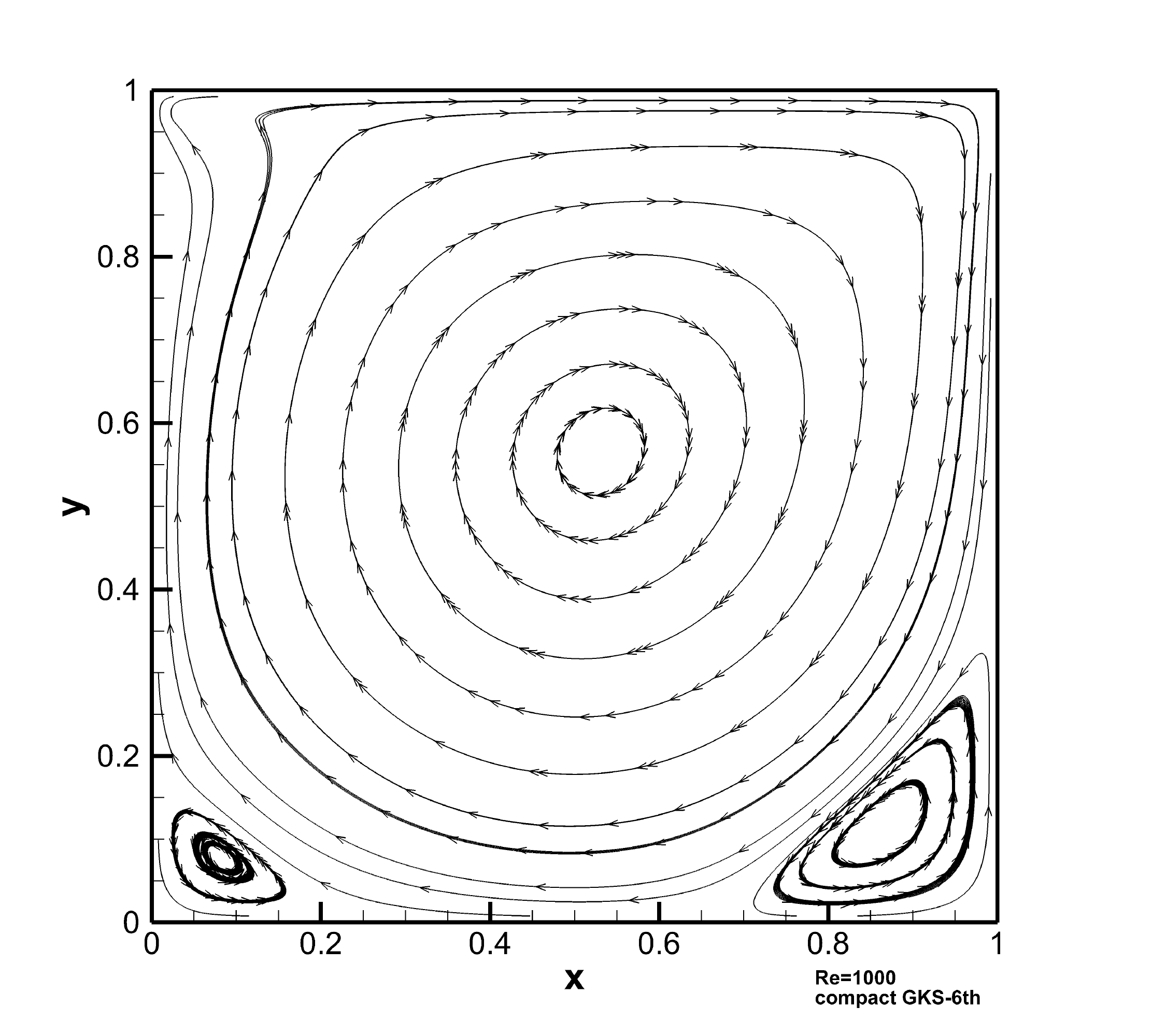}
\includegraphics[width=0.485\textwidth]{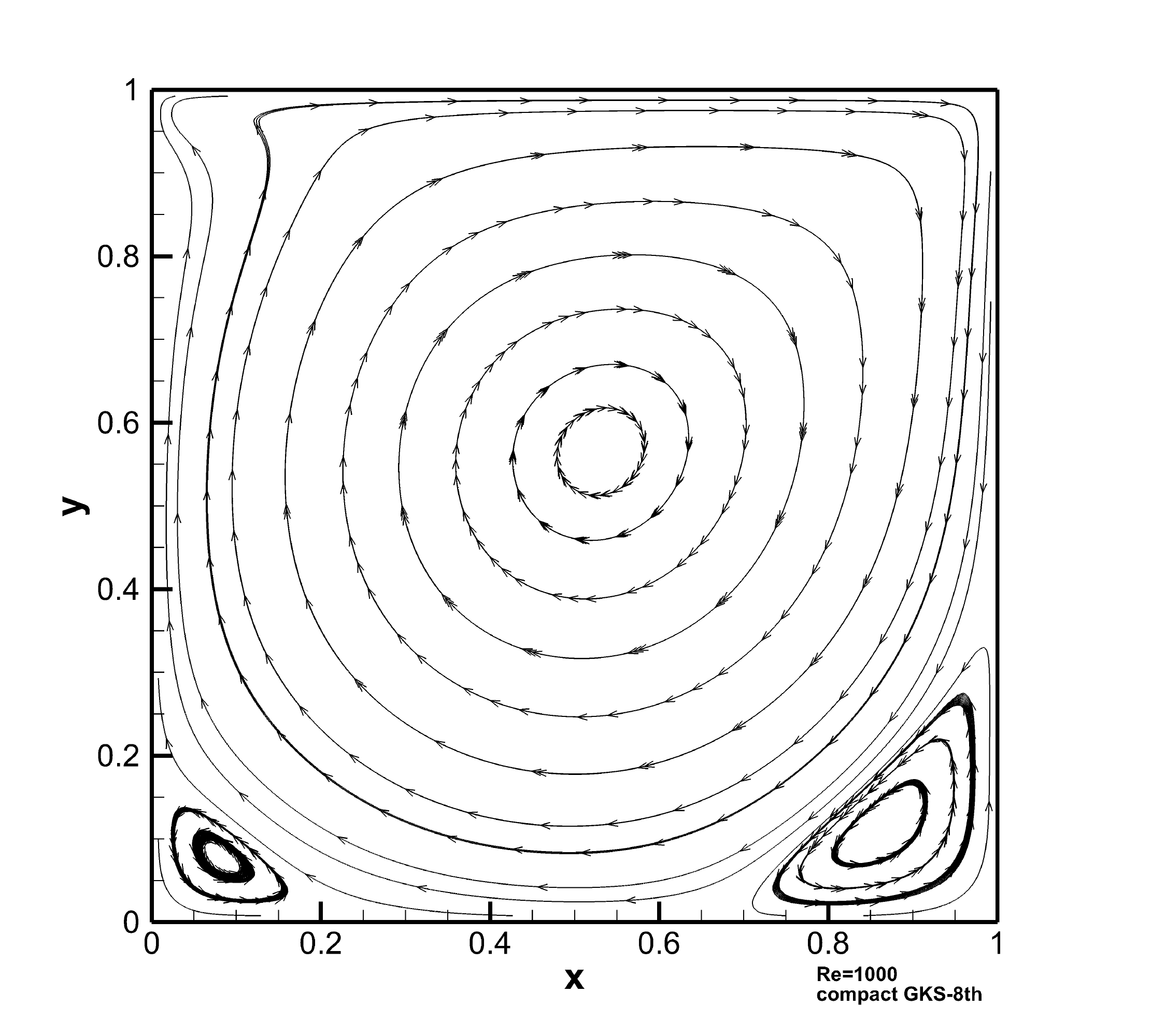}
\caption{\label{cavity-1} Lid-driven cavity flow: streamlines with
$65\times65$ uniform mesh points  for $Re=1000$ by 6th-order and 8th-order
compact GKS.}
\end{figure}

\begin{figure}[!htb]
\centering
\includegraphics[width=0.485\textwidth]{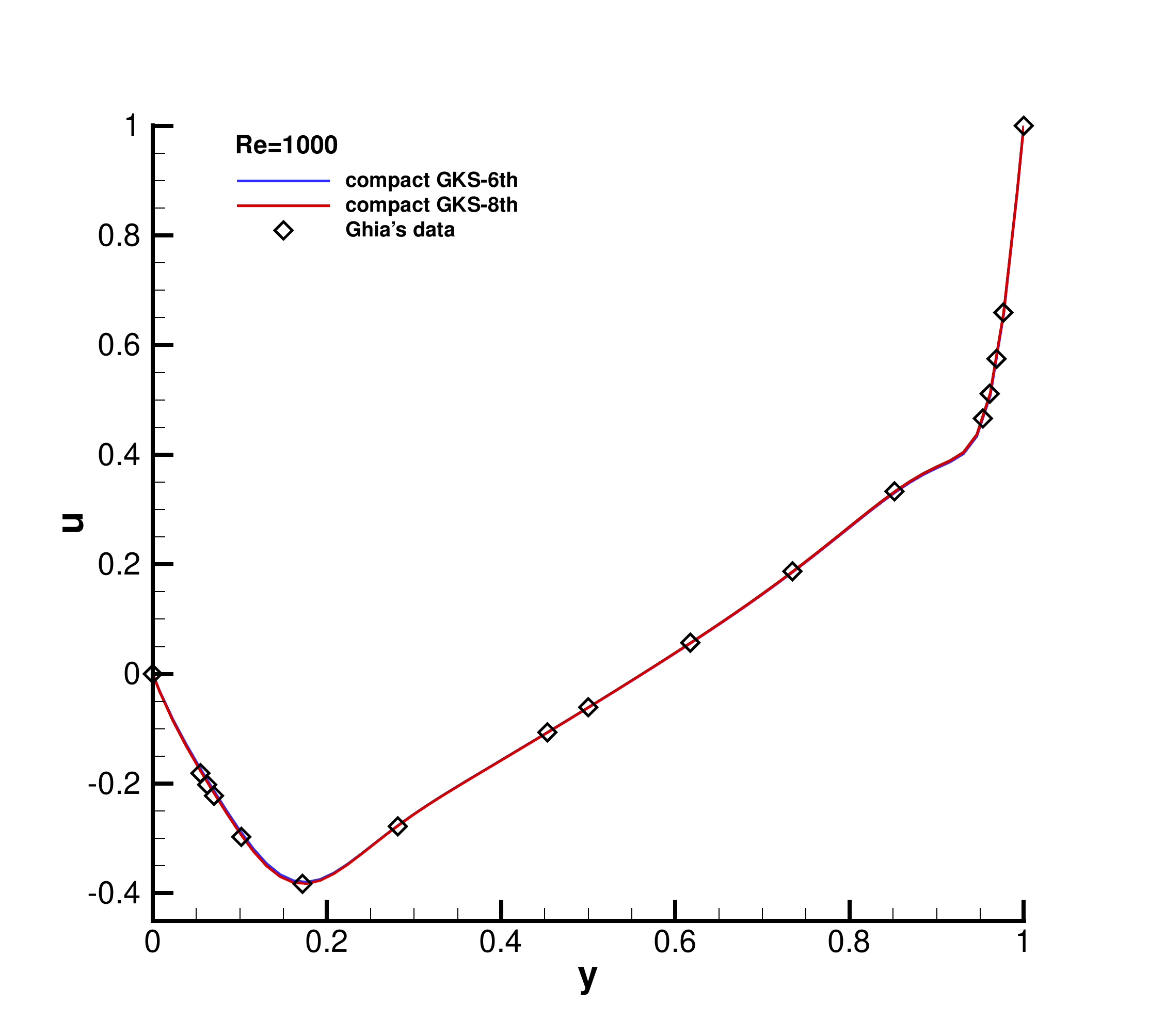}
\includegraphics[width=0.485\textwidth]{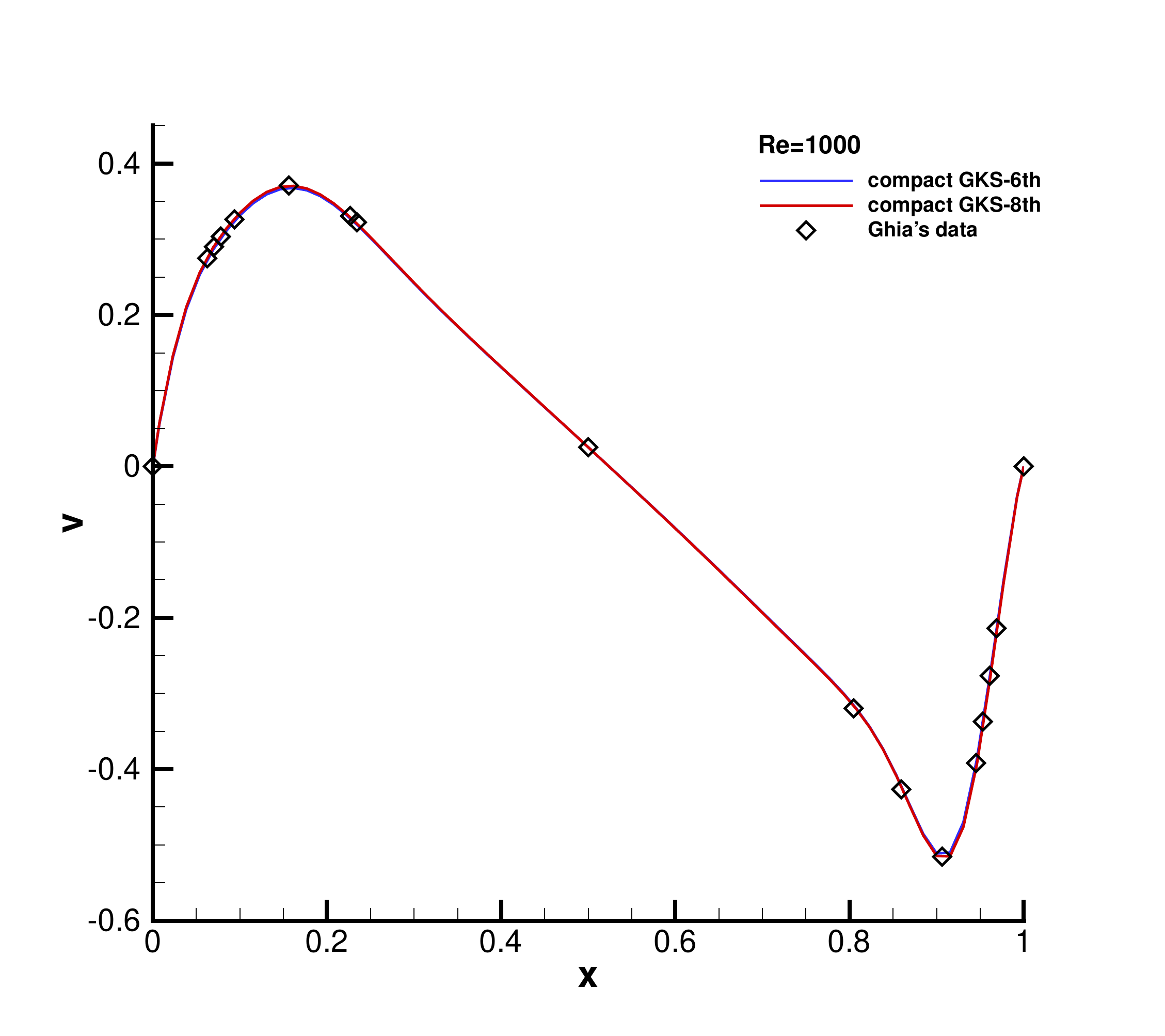}
\includegraphics[width=0.485\textwidth]{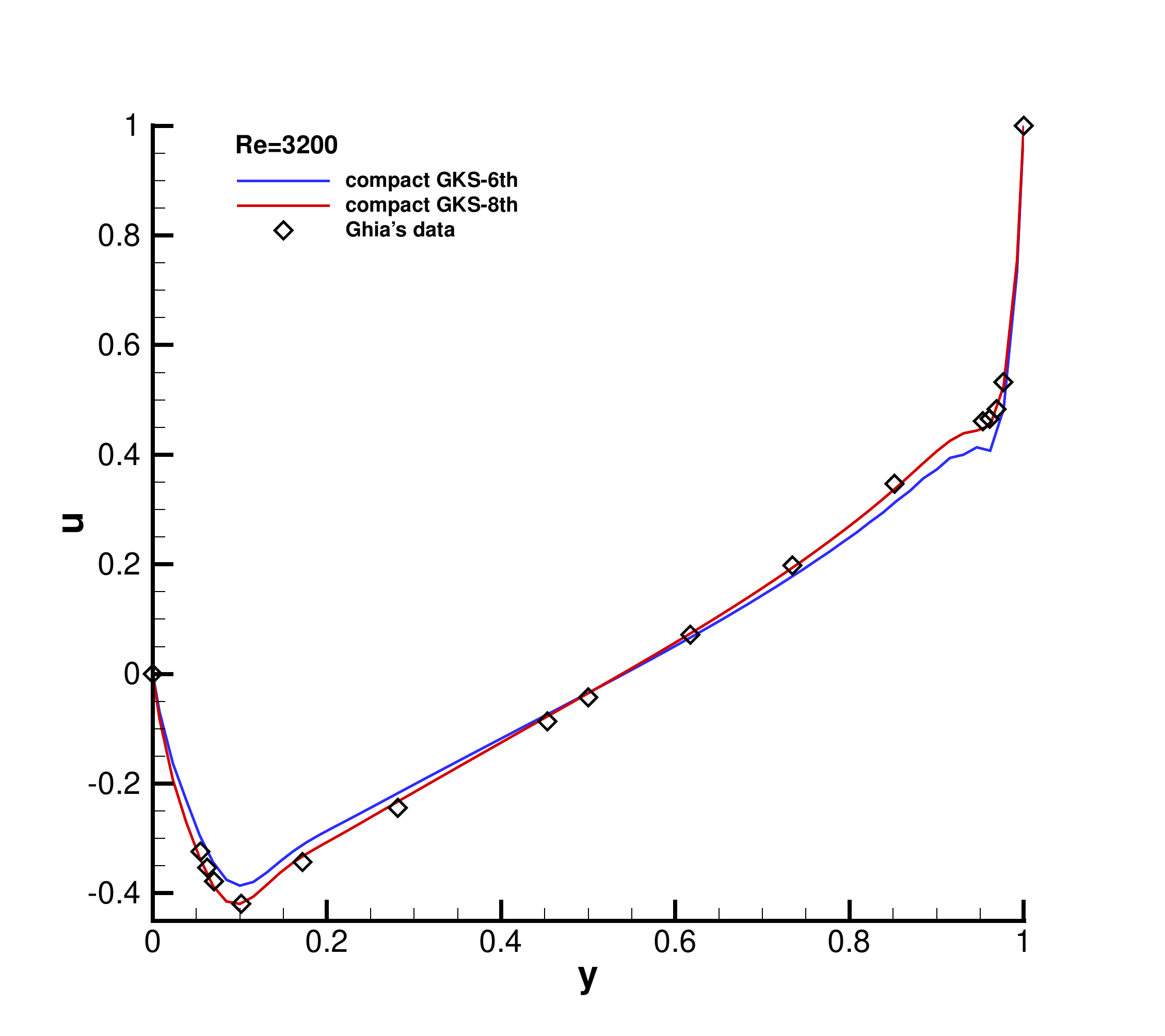}
\includegraphics[width=0.485\textwidth]{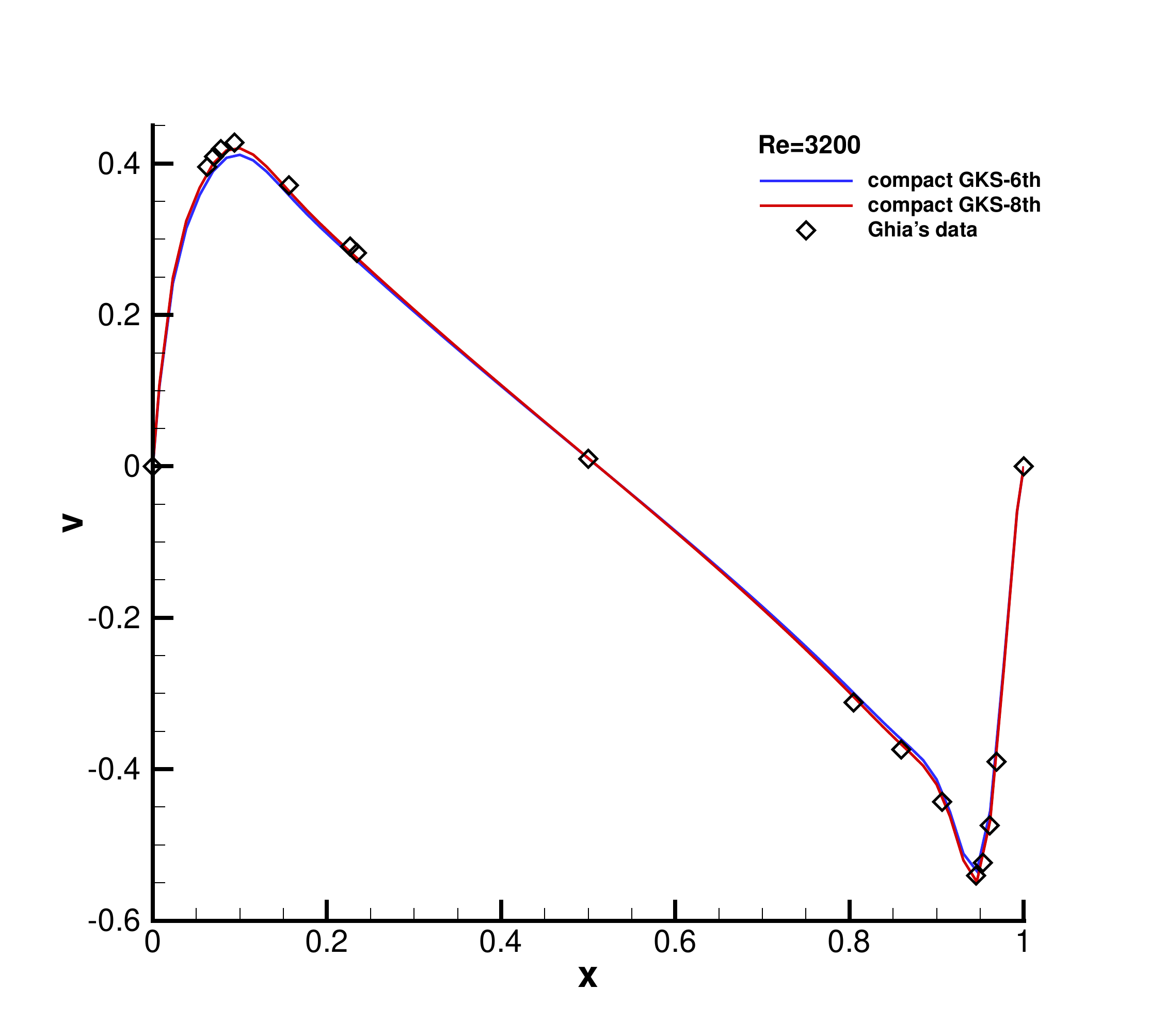}
\caption{\label{cavity-2} Lid-driven cavity flow: U-velocity
along vertical centerline and V-velocity along horizontal
center-line with $65\times65$ uniform mesh points at $Re=1000$ and $3200$. }
\end{figure}

\subsection{Lid-driven cavity flow}
Here we test the performance of the compact GKS for the capturing of viscous flow solutions.
The lid-driven cavity problem is one of the most important
benchmarks for validating incompressible Navier-Stokes flow
solvers. The fluid is bounded by a unit square and is driven by a
uniform translation of the top boundary. In this case, the flow is
simulated with Mach number $Ma=0.15$ and all boundaries are
isothermal and nonslip. The computational domain $[0, 1]\times[0,1]$
is covered with $65\times65$ mesh points. Numerical simulations are
conducted for two different Reynolds numbers, i.e., $Re=1000$ and
$3200$. The streamlines corresponding to $Re=1000$ are shown in
Fig. \ref{cavity-1}. The $U$-velocities along the center vertical line
and $V$-velocities along the center horizontal line are shown in
Fig. \ref{cavity-2}. The benchmark data \cite{Case-Ghia} for $Re=1000$
and $3200$ are also presented, and the simulation results match well
with these benchmark data. The results from 8th-order compact GKS are
almost identical to the reference solutions even for $Re=3200$ case.
The cavity case fully validates the higher-order accuracy of compact GKS for viscous flow.

\begin{figure}[!htb]
\centering
\includegraphics[width=0.485\textwidth]{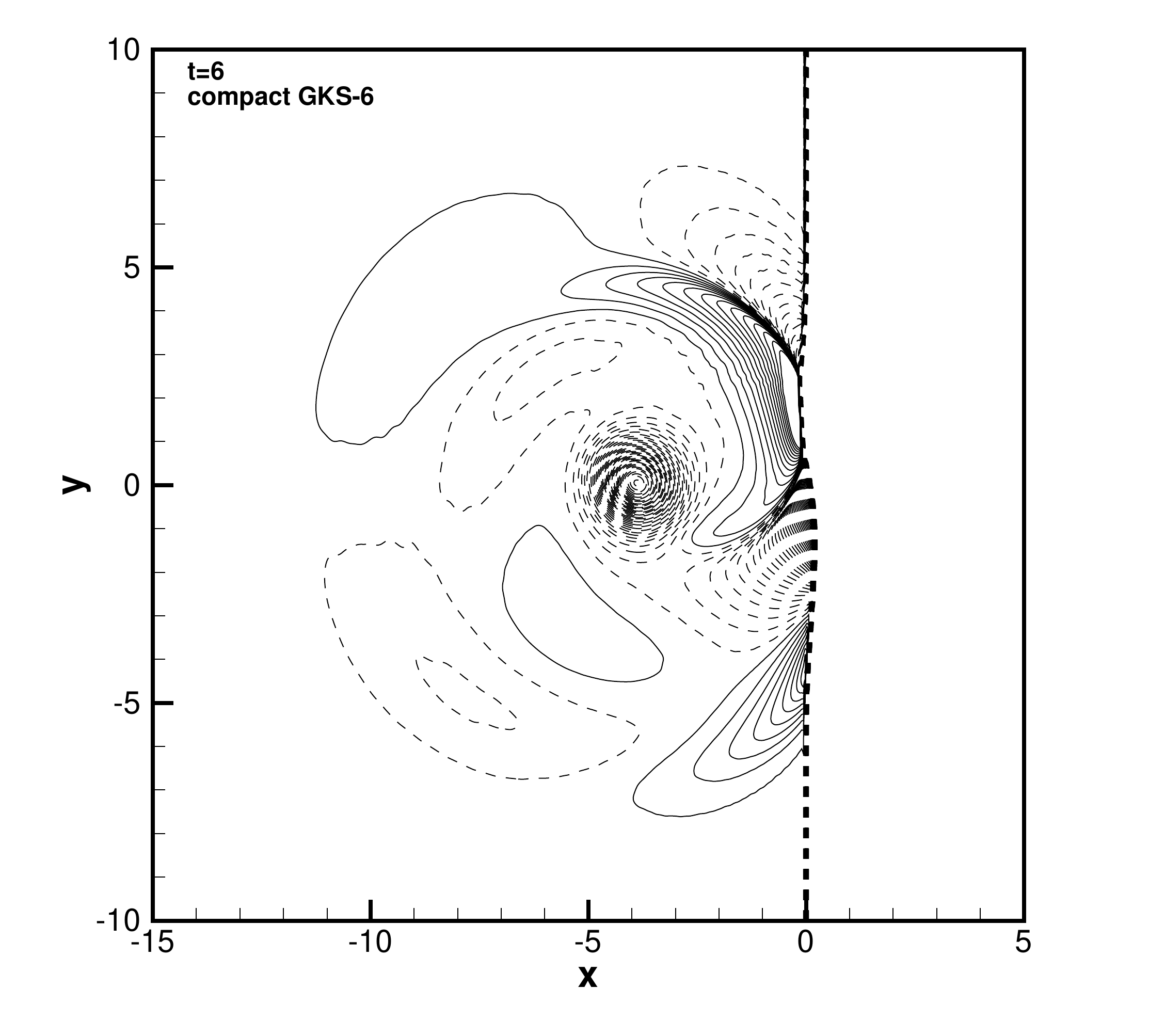}
\includegraphics[width=0.485\textwidth]{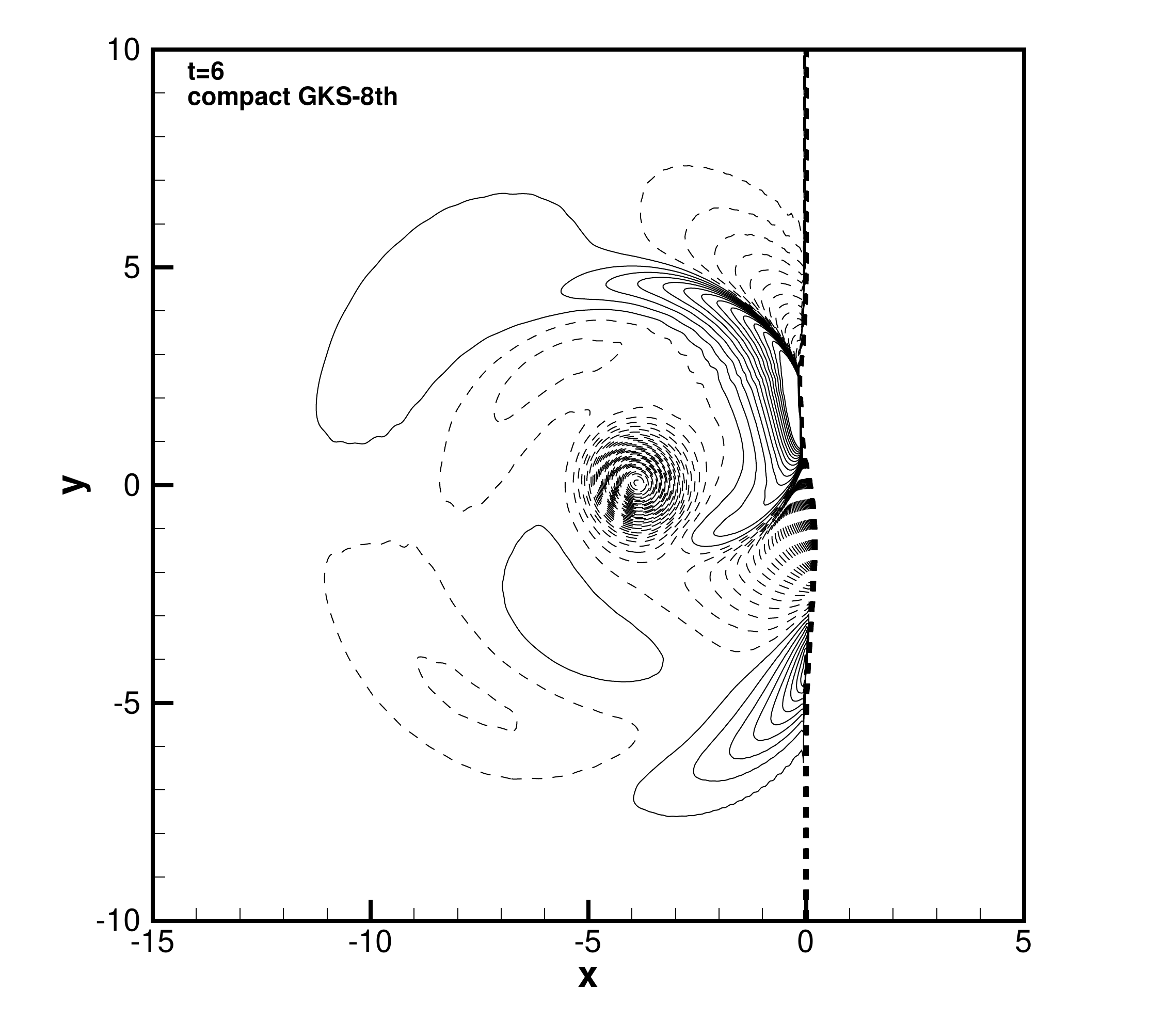}
\caption{\label{shock-vor-1} Two-dimensional acoustic problem: sound pressure contours obtained by 6th-order
and 8th-order compact GKS at $t=6$ with $700\times600$ mesh points. Shock and vortex Mach numbers
are $M_s=1.2$ and $M_v=1$. Here $150$ equal-spaced sound pressure contours from $-0.5$ to $0.05$
are plotted. The dash line represents rarefaction region, and the solid line represents the compression region.}
\end{figure}


\begin{figure}[!htb]
\centering
\includegraphics[width=0.485\textwidth]{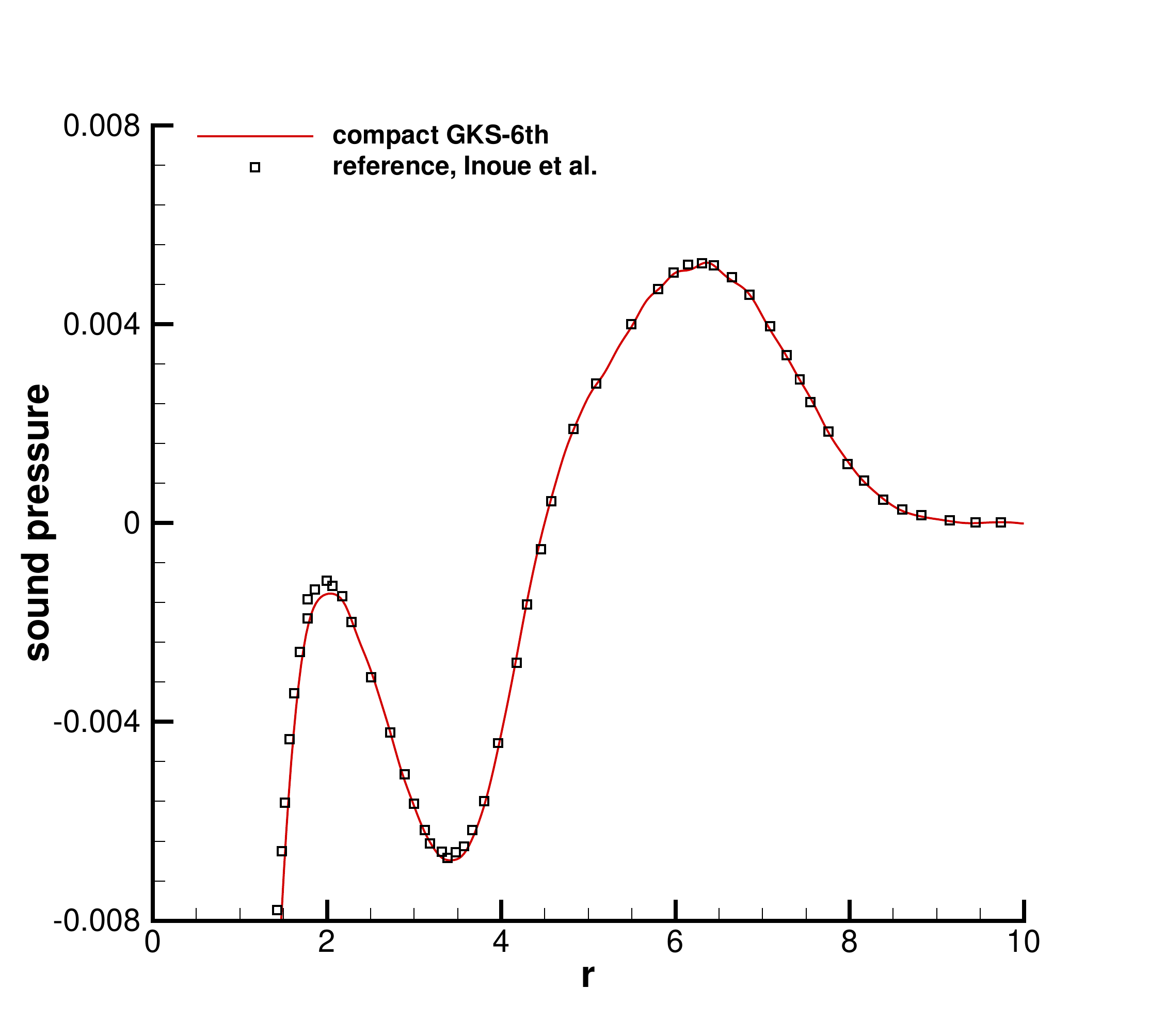}
\includegraphics[width=0.485\textwidth]{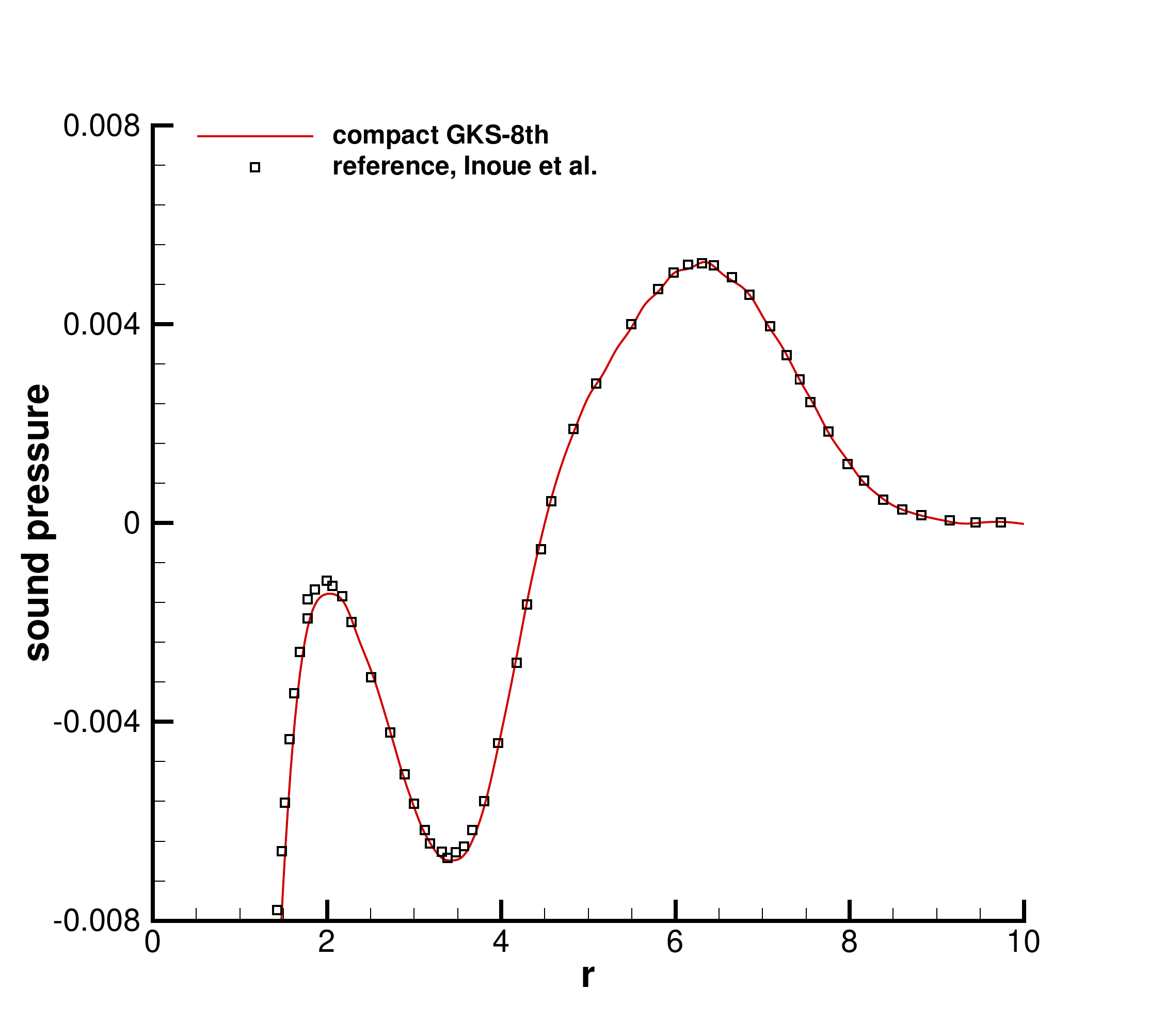}
\caption{\label{shock-vor-3} Two-dimensional acoustic problem: radial distribution of the sound pressure
$\triangle p$ obtained by 6th-order and 8th-order compact GKS at $t=6$ with $700\times600$ mesh points.
The vortex center at $t=6$ is approximately located at $(-3.9,0.0)$.}
\end{figure}

\subsection{Two-dimensional acoustic wave}
The case is an interaction of a shock wave with a single vortex as studied by Inoue and Hattori
\cite{shock-vor}, where the fluid is viscous. The computational domain is $[-20,8]\times[-12,12]$.
The initial counterclockwise vortex (the case C in \cite{shock-vor}) is set as follows
\begin{align*}
U_{\theta}(r)&=M_v r e^{(1-r^2)/2},~~U_r=0,
\end{align*}
and the distribution of pressure and density upstream of shock are
\begin{align*}
p(r)&=\frac{1}{\gamma}[1-\frac{\gamma-1}{2}M_v^2e^{(1-r^2)}]^{\gamma/(\gamma-1)},\\
\rho(r)&=[1-\frac{\gamma-1}{2}M_v^2e^{(1-r^2)}]^{1/(\gamma-1)},
\end{align*}
where $U_{\theta}$ and $U_r$ are the tangential and radial velocity respectively. Mach number $M_v$
of the vortex is $M_v=0.25$. The Mach number of shock wave is $M_s=1.2$. The Reynolds number is $Re=800$
defined by $Re=\rho_{\infty} a_{\infty}/\mu_{\infty}$, where the subscript $\infty$ denotes the
quantity downstream of the shock wave. The initial location of vortex is $(x_v,y_v)=(1,0)$,
and the stationary shock is at $x=0$.
In the computation, the supersonic inflow boundary conditions at $x=8$ as well as the periodic boundary
conditions at $y=\pm 12$ are imposed. The non-reflecting boundary conditions are adopted at $x=-20$.

In the case, the compact 6th-order and 8th-order GKS are tested on the uniform mesh with $700\times600$
cells. The sound pressure at $t=6$ are shown in Fig. \ref{shock-vor-1},
where the sound pressure is defined as $\triangle p=(p-p_{\infty})/p_{\infty}$.
The pressure distribution in Fig. \ref{shock-vor-1} clearly shows that the
incident shock wave and two reflected shock waves are connected at the triple point, and
the reflected shock waves emanate from the compression region of the incident shock wave.
In addition, both the first and second sound waves
have the clear quadruple nature with opposite sign, which is similar with the result in \cite{shock-vor}.
Both the 6th-order and 8th-order GKS perform well to resolve the shock wave and density wave, and the
solution is sufficiently smooth to resolve small amplitude waves.
The radial distribution of the sound pressure $\triangle p$ obtained by 6th-order and 8th-order compact GKS
at $t=6$ with $700\times600$ mesh points  are shown in Fig. \ref{shock-vor-3}. The radial distribution of
the sound pressure is subtracted from the vortex center with the angle $\theta= 135$ degree with the positive $x$ direction.
The vortex center at $t=6$ is approximately located at $(-3.9,0.0)$.
The results are concordant well with the results in \cite{shock-vor}, where the reference solution in \cite{shock-vor}
is obtained with $1044 \times 1170$ mesh points and local mesh refinement with approximate $1/10$ cell size reduction at the shock region.

\section{Conclusion}
In this paper, a class of compact high-order gas-kinetic scheme with WENO reconstruction has been proposed for the compressible
Euler and Navier-Stokes solutions. Based on the gas-kinetic theory, the current scheme depends solely on the time-accurate solution
of a gas distribution function at a cell interface for the flux and conservative flow variable evaluation.
With the update of pointwise values at cell interfaces, besides the cell-averaged conservative flow variables, their cell-averaged gradients can be
updated as well. Therefore, the compact linear and nonlinear reconstructions can be obtained from the
local cell averaged flow variables and their slopes.
With symmetrical stencils, the 6th-order and 8th-order compact linear reconstructions are given and used in the 1-D Fourier analysis.
The analysis elucidates spectral resolution of the current spatial discretization, even for very large wave number.
The GKS flux makes the scheme stable even for the initial reconstruction from symmetrical stencils.
In order to capture shock and other discontinuities, the nonlinear compact WENO reconstruction has been constructed as well.
Compared with other compact schemes, the current compact GKS is fully local and has explicit high-order discretization, without using any formulation for the globally connected flow variable and their slopes.
Equipped with both linear and nonlinear compact reconstruction, the GKS gas evolution model can make a smooth transition in dynamics
from the initial nonlinear reconstruction for non-equilibrium state to the final linear reconstruction for equilibrium state.
The transition rate depends on the local flow conditions. As a result, the current compact high-order scheme can naturally
capture both shock discontinuities and small amplitude acoustic waves in a single computation without using any trouble cell detection and additional
limiters.
Due to the existence of the time-derivative of the flux function,
the multi-stage and multi-derivative time stepping technique has been applied here for the achieving high-order temporal accuracy of the scheme.
More specifically, the two stage fourth-order time accuracy  has been achieved.
Both GKS and DG methods have the same compact stencils. But, the GKS uses a much large CFL number ($\geq 0.3$) for the 6th- and 8th-order schemes,
and it is much more robust in capturing shock discontinuity and is much more efficient in simulating NS equations than the DG method.
The extensive tests in this paper clearly demonstrate that the current 6th-order and 8th-order compact schemes advance the development of
high-order CFD methods to a new level of maturity.
The success of the current compact schemes depends solely on the high gas evolution model at a cell interface, which keeps and traces flow
dynamics as close as possible to a real physical time evolution process.
Any scheme based on the first order Riemann solver, which gives a time-independent gas evolution at a cell interface
from two constant states whatever the high-order initial reconstruction is, may have difficulty to possess
all these good properties of a high-order scheme, such as the compactness, large CFL number, robustness, high efficiency, and the dynamic unification
of linear and nonlinear reconstructions.

\section*{Acknowledgements}
The current research is supported by Hong Kong research grant council (16206617)
and National Science Foundation of China (11772281, 91852114).

\section*{Appendix}
For the WENO reconstruction, with the definition of smooth indicator in Eq.\eqref{smooth-indicator},
the reconstructed polynomials from the sub-stencils for the determination of the left state of the cell interface $x_{i+1/2}$ are presented.
Assume the candidate polynomial on a specific sub-stencil as
\begin{align}\label{candidate-polynomial}
w^{n}(x)\equiv\sum_{k=0}^{n} a^n_{k}x^{k}/k!,
\end{align}
where $n$ is the order of the polynomial and takes the values $n=2,3,4$ for the current 6th-order and 8th-order reconstruction schemes.
Substituting $w^{n}(x)$ into the smooth indicator $\beta$ in Eq.\eqref{smooth-indicator}, the result can be expressed
by the coefficient $a^n_{k}$ as
\begin{align*}
\beta =&(a^n_1\cdot \Delta x)^2+\frac{13}{12}(a^n_2 \cdot \Delta x^2)^2+\frac{1043}{960}(a^n_3 \cdot \Delta x^3)^2+ \\
&\frac{35045}{32256}(a^n_4 \cdot \Delta x^4)^2+\frac{1}{12}(a^n_1 \cdot \Delta x)\cdot(a^n_3 \cdot \Delta x^3)+\frac{7}{80}(a^n_2 \cdot \Delta x^2)\cdot(a^n_4 \cdot \Delta x^4).
\end{align*}
For $n=2$, $a^n_3=a^n_4=0$, and for $n=3$, $a^n_4=0$.
Therefore, once the candidate polynomial is given, the smoothing indicator can be fully obtained.

The polynomials of the sub-stencils for the 6th-order and 8th-order schemes are listed below.

\begin{scriptsize}
\begin{align*}
w_0(x)=&\frac{1}{12}(W_{i-1} + 11W_{i} + \Delta x W^{'}_{i-1}) + \frac{1}{\Delta x}(-2W_{i-1} + 2W_{i} - \Delta x W^{'}_{i-1})\cdot x - \frac{2}{\Delta x^2}(W_{i-1} - W_{i} + \Delta x W^{'}_{i-1})\cdot\frac{x^2}{2}, \\
\widetilde{w}_1(x)=&\frac{1}{24}(-W_{i-1} + 26W_{i} -W_{i+1} ) + \frac{1}{16\Delta x}(-23W_{i-1} + 20W_{i} + 3W_{i+1} - 10\Delta x W^{'}_{i-1})\cdot x + \\
&\frac{1}{\Delta x^2}(W_{i-1} - 2W_{i} + W_{i+1})\cdot\frac{x^2}{2} + \frac{3}{2\Delta x^3}(3W_{i-1} - 4W_{i} + W_{i+1} + 2\Delta x W^{'}_{i-1})\cdot\frac{x^3}{6}, \\
w_2(x)=&\frac{1}{48}(43W_{i} + 16W_{i+1} -11W_{i+2} + 6\Delta x W^{'}_{i+2}) + \frac{1}{16\Delta x}(-31W_{i} + 60W_{i+1} -29W_{i+2} + 14\Delta x W^{'}_{i+2})\cdot x + \\
&\frac{1}{2\Delta x^2}(5W_{i} - 16W_{i+1} + 11W_{i+2} - 6\Delta x W^{'}_{i+2})\cdot\frac{x^2}{2} +  \frac{3}{2\Delta x^3}(-W_{i} + 4W_{i+1} - 3W_{i+2} + 2\Delta x W^{'}_{i+2})\cdot\frac{x^3}{6}, \\
w_3(x)=&\frac{1}{24}(-W_{i-1} + 26W_{i} - W_{i+1}) + \frac{1}{2\Delta x}(-W_{i-1} + W_{i+1})\cdot x + \frac{1}{\Delta x^2}(W_{i-1} - 2W_{i} + W_{i+1})\cdot\frac{x^2}{2}, \\
w_4(x)=&\frac{1}{24}(23W_{i} + 2W_{i+1} - W_{i+2}) + \frac{1}{2\Delta x}(-3W_{i} + 4W_{i+1} - W_{i+2})\cdot x + \frac{1}{\Delta x^2}(W_{i} - 2W_{i+1} + W_{i+2})\cdot\frac{x^2}{2} , \\
w_5(x)=&\frac{1}{960}(-175W_{i-1} + 1148W_{i} -13W_{i+1} - 54\Delta x W^{'}_{i-1} - 108\Delta x W^{'}_{i}) + \frac{1}{8\Delta x}(W_{i-1} - W_{i+1} + 10\Delta x W^{'}_{i})\cdot x + \\
&\frac{1}{4\Delta x^2}(19W_{i-1} - 20W_{i} + W_{i+1} + 6\Delta x W^{'}_{i-1} + 12\Delta x W^{'}_{i})\cdot\frac{x^2}{2} - \frac{3}{\Delta x^3}(W_{i-1} - W_{i+1} + 2\Delta x W^{'}_{i})\cdot\frac{x^3}{6}- \\
&\frac{6}{\Delta x^4}(5W_{i-1} - 4W_{i} - W_{i+1} + 2\Delta x W^{'}_{i-1} + 4\Delta x W^{'}_{i})\cdot\frac{x^4}{24}, \\
w_6(x)=&\frac{1}{960}(707W_{i} - 292W_{i+1} + 545W_{i+2} - 612\Delta x W^{'}_{i+1} - 186\Delta x W^{'}_{i+2}) + \\
&\frac{1}{8\Delta x}(-21W_{i} + 8W_{i+1} + 13W_{i+2} - 22\Delta x W^{'}_{i+1} - 4\Delta x W^{'}_{i+2})\cdot x + \\
&\frac{1}{4\Delta x^2}(25W_{i} + 28W_{i+1} - 53W_{i+2} + 60\Delta x W^{'}_{i+1} + 18\Delta x W^{'}_{i+2})\cdot\frac{x^2}{2} - \\
&\frac{3}{\Delta x^3}(3W_{i} + 8W_{i+1} - 11W_{i+2} + 10\Delta x W^{'}_{i+1} + 4\Delta x W^{'}_{i+2})\cdot\frac{x^3}{6} \\
&\frac{6}{\Delta x^4}(W_{i} + 4W_{i+1} - 5W_{i+2} + 4\Delta x W^{'}_{i+1} + 2\Delta x W^{'}_{i+2})\cdot\frac{x^4}{24}, \\
\end{align*}
\end{scriptsize}


\section*{References}
\begin{thebibliography}{}

\bibitem{bai} Z. Bai, X. Zhong, New very high-order upwind multi-layer compact (MLC) schemes with spectral-like resolution for flow simulations,
   J. Comput. Phys. 378 (2019) 63-109.
\bibitem{balsara} D.S. Balsara, C. Altmann, C.-D. Munz, M. Dumbser, A sub-cell based indicator for troubled zones in RKDG schemes
   and a novel class of hybrid RKDG+HWENO schemes, J. Comput. Phys. 226 (2007) 586-620.

\bibitem{grp1} M. Ben-Artzi, J. Falcovitz, A second-order Godunov-type scheme for compressible fluid dynamics, J. Comput. Phys. 55 (1984) 1-32.
\bibitem{grp3} M. Ben-Artzi, J. Li, Hyperbolic conservation laws: Riemann invariants and the generalized Riemann problem,
   Numer. Math. 106 (2007) 369-425.

\bibitem{grp2} M. Ben-Artzi, J. Li, G. Warnecke, A direct Eulerian GRP scheme for compressible fluid flows, J. Comput. Phys. 218 (2006) 19-43.
\bibitem{BGK-1} P.L. Bhatnagar, E.P. Gross, M. Krook, A model for collision processes in gases I:
   small amplitude processes in charged and neutral one-component systems, Phys. Rev. 94 (1954), 511-525.

\bibitem{WENO-Z} R. Borges, M. Carmona, B. Costa, W. S. Don, An improved weighted essentially non-oscillatory scheme for hyperbolic conservation laws, J. Comput. Phys. 227 (2008) 3191-3211.
\bibitem{choi} J.J. Choi, Hybrid spectral difference/embedded finite volume method for conservation laws, J. Comput. Phys. 295 (2015) 285-306.


\bibitem{christlieb} A.J. Christlieb, S. Gottlieb, Z. Grant, D.C. Seal, Explicit strong stability preserving multistage
two-derivative time-stepping scheme, J. Sci. Comput. 68 (2016) 914-942.

\bibitem{cockburn1} B. Cockburn, C. W. Shu, TVB Runge-Kutta local projection discontinuous Galerkin finite element method for conservation laws II:
  general framework, Mathematics of Computation, 52 (1989) 411-435.
\bibitem{cockburn2} B. Cockburn, C. W. Shu, The Runge-Kutta discontinuous Galerkin method for conservation laws V:
  multidimensional systems, J. Comput. Phys. 141 (1998) 199-224.
\bibitem{WCNS} X. G. Deng, H. X. Zhang, Developing high-order weighted compact nonlinear schemes, J. Comput. Phys. 165 (2000) 22-44.

\bibitem{dumbser} M. Dumbser, D.S. Balsara, E.F. Toro, C.D. Munz, A unified framework for the construction of one-step
finite volume and discontinuous Galerkin schemes on unstructured meshes, J. Comput. Phys. 227 (2008) 8209-8253.
\bibitem{Case-Ghia} U. Ghia, K. N. Ghia, C.T Shin, High-Re solutions for incompressible flow using the Navier-Stokes equations and a multigrid method, J. Comput. Phys. 48 (1982) 387-411.


\bibitem{hairer} E. Hairer, G. Wanner, Multistep multistage multiderivative methods for ordinary differential equations, Computing 11 (1973) 287-303.

\bibitem{harten1} A. Harten, High resolution schemes for hyperbolic conservation laws, J. Comput. Phys. 49 (1983) 357-393.
\bibitem{harten2} A. Harten, B. Engquist, S. Osher, S.R. Chakravarthy, Uniformly high order accurate essentially non-oscillatory schemes,
III, J. Comput. Phys. 71 (1987) 231-303.
\bibitem{WENO-M} A. K. Henrick, T. D. Aslam, J. M. Powers, Mapped weighted essentially non-oscillatory schemes:
  achieving optimal order near critical points, J. Comput. Phys. 207 (2005) 542-567.

\bibitem{hill} D.J. Hill, D.I. Pullin, Hybrid tuned center-difference-WENO method for large eddy simulations in the presence of strong shocks,
J. Comput. Phys. 194 (2004) 435-450.
\bibitem{shock-vor} O. Inoue and Y. Hattori, Sound generation by shock-vortex interactions, J. Fluid Mech. 380 (1999) 81-116.
\bibitem{ji2} X. Ji, L. Pan, W. Shyy, K. Xu, A compact fourth-order gas-kinetic scheme for the Euler and Navier-Stokes equations
  J. Comput. Phys. 372 (2018) 446-472.
\bibitem{ji1} X. Ji, F.X. Zhao, W. Shyy, K. Xu, A family of high-order gas-kinetic schemes and its comparison with Riemann solver
based high-order methods, J. Comput. Phys. 356 (2018) 150-173.


\bibitem{jiang} G.-S. Jiang, C.-W. Shu, Efficient implementation of weighted ENO schemes, J. Comput. Phys. 126 (1996) 202-228.
\bibitem{Case-Riemann-1} P. D. Lax,  X.D. Liu,  Solution of two-dimensional riemann problems of gas dynamics by positive schemes,
SIAM J. Sci. Comput. 19 (1998) 319-340.
\bibitem{lax} P.D. Lax, B. Wendroff, Systems of conservation laws, Commun. Pure Appl. Math. 13 (1960) 217-237.

\bibitem{lele} S.K. Lele, Compact finite difference schemes with spectral-like resolution, J. Comput. Phys. 103 (1992) 16-42.
\bibitem{li} J. Li, Z. Du, A two-stage fourth order time-accurate discretization for Lax-Wendroff type flow solvers I.
hyperbolic conservation laws, SIAM J. Sci. Comput. 38 (2016) 3046-3069.
\bibitem{li2} J. Li, Q.B. Li, K. Xu, Comparison of the generalized Riemann solver and the gas-kinetic scheme
for inviscid compressible flow simulations, J. Comput. Phys. 230 (2011) 5080-5099.

\bibitem{liu} X.-D. Liu, S. Osher, T. Chan, Weighted essentially non-oscillatory schemes, J. Comput. Phys. 115 (1994) 200-212.

\bibitem{luo1} H. Luo, J.D. Baum, R. Lohner, A Hermite WENO-based limiter for discontinuous Galerkin method on unstructured grids, J. Comput. Phys.
225 (2007) 686-713.
\bibitem{luo2} H. Luo, Y. Xia, S. Li, R. Nourgaliev, C. Cai, A Hermite WENO reconstruction-based discontinuous Galerkin method for the Euler
equations on tetrahedral grids, J. Comput. Phys. 231 (2012) 5489-5503.
\bibitem{luo3} H. Luo, Y. Xia, S. Spiegel, R. Nourgaliev, Z. Jiang, A reconstructed discontinuous Galerkin method based on a Hierarchical WENO
reconstruction for compressible flows on tetrahedral grids, J. Comput. Phys. 236 (2013) 477-492.

\bibitem{GKS-Luo} J. Luo, K. Xu, A high-order multidimensional gas-kinetic scheme for hydrodynamic equations,
SCIENCE CHINA Technological Sciences. 56 (2013) 2370-2384.

\bibitem{mahesh} K. Mahesh, A family of high order finite difference schemes with good spectral resolution, J. Comput. Phys. 145 (1998) 332-358.


\bibitem{pan2} L. Pan, J.X. Cheng, S.H. Wang, K. Xu, A two-stage fourth-order gas-kinetic scheme for compressible multicomponent flows,
Commun. Comput. Phys. 22 (2017) 1123-1149.
\bibitem{pan0} L. Pan, J.Q. Li, K. Xu, A few benchmark test cases for higher-order Euler solvers,
   Numer. Math. Theor. Meth. Appl. 10 (2017) 711-736.
\bibitem{pan3} L. Pan, K. Xu, A compact third-order gas-kinetic scheme for compressible Euler and Navier-Stokes equations,
Commun. Comput. Phys. 18 (2015) 985-1011.
\bibitem{pan4} L. Pan, K. Xu, A third-order compact gas-kinetic scheme on unstructured meshes for compressible Navier-Stokes solutions,
J. Comput. Phys. 318 (2016) 327-348.
\bibitem{pan1} L. Pan, K. Xu, Q. Li, J. Li, An efficient and accurate two-stage fourth-order gas-kinetic scheme for the Navier-Stokes equations,
J. Comput. Phys. 326 (2016) 197-221.

\bibitem{qiu1} J. Qiu, C.-W. Shu, Hermite WENO schemes and their application as limiters for Runge-Kutta discontinuous Galerkin method:
onedimensional case, J. Comput. Phys. 193 (2004) 115-135.
\bibitem{qiu2} J. Qiu, C.-W. Shu, Hermite WENO schemes and their application as limiters for Runge-Kutta discontinuous Galerkin method II:
Two dimensional case, Comput. Fluids. 34 (2005) 642-663.

\bibitem{seal} D.C. Seal, Y. Guclu, A.J. Christlieb, High-order multiderivative time integrators for hyperbolic conservation laws,
J. Sci. Comput. 60 (2014) 101-140.


\bibitem{reed} W.H. Reed, T.R. Hill, Triangular mesh methods for the neutron transport equation, Technical Report LA-UR-73-479 (1973),
Los Alamos Scientific Laboratory, Los Alamos.

\bibitem{ren} Y.-X. Ren, M. Liu, H. Zhang, A characteristic-wise hybrid compact-WENO scheme for solving hyperbolic conservation laws, J. Comput.
Phys. 192 (2003) 365-386.

\bibitem{shu1} C.-W. Shu, High order weighted essentially nonoscillatory schemes for convection dominated problems, SIAM Rev. 51 (2009) 82-126.

\bibitem{shu} C.-W. Shu, High order WENO and DG methods for time-dependent convection-dominated PDEs: A brief survey of several recent
developments, J. Comput. Phys. 316 (2016) 598-613.

\bibitem{Case-Shu-Osher} C. W. Shu, S. Osher, Efficient implementation of essentially nonoscillatory shock-capturing schemes II,
J. Comput. Phys. 83 (1989) 32-78.

\bibitem{taylor} E.M. Taylor, M. Wu, M.P. Mart¨ªn, Optimization of nonlinear error for weighted essentially non-oscillatory methods
in direct numerical simulations of compressible turbulence, J. Comput. Phys. 223 (2007) 384-397.
\bibitem{Titarev-Toro} V. A. Titarev and E. F. Toro, Finite volume WENO schemes for three-dimensional conservation laws,
J. Comput. Phys. 201 (2014) 238-260.

\bibitem{toro} E. Toro, Riemann Solvers and Numerical Methods for Fluid Dynamics, third edition, Springer (2009).

\bibitem{Case-Woodward} P. Woodward and P. Colella, Numerical simulations of two-dimensional fluid flow with strong shocks,
J. Comput. Phys. 54 (1984) 115-173.
\bibitem{xu} K. Xu, A slope-update scheme for compressible flow simulation, J. Comput. Phys. 178 (2002) 252-259.
\bibitem{xu2} K. Xu, A gas-kinetic BGK scheme for the Navier-Stokes equations and its connection with artificial dissipation and Godunov method,
J. Comput. Phys. 171 (2001) 289-335.

\bibitem{xu1} K. Xu, Direct modeling for computational fluid dynamics: construction and application of unified gas kinetic schemes,
World Scientific (2015).
\bibitem{xu3} K. Xu, J.C. Huang, A unified gas-kinetic scheme for continuum and rarefied flows, J. Comput. Phys. 229 (2010) 7747-7764.


\bibitem{zhang} L. Zhang, W. Liu, L. He, X. Deng, H. Zhang, A class of hybrid DG/FV methods for conservation laws I:
Basic formulation and onedimensional systems, J. Comput. Phys. 231 (2012) 1081-1103.
\bibitem{Case-Riemann-2} T. Zhang, Y. Zheng, Conjecture on the structure of solutions of the Riemann problem for
two-dimensional gas dynamics systems, SIAM J. Math. Anal. 21 (1990) 593-630.

\bibitem{zhou} G. Zhou, K. Xu, and F. Liu, Simplification of the flux function for a high-order gas-kinetic evolution model,
J. Comput. Phys. 339 (2017) 146-162.

\bibitem{qiu3} J. Zhu, J. Qiu, Hermite WENO schemes and their application as limiters for Runge-Kutta discontinuous Galerkin method, III:
unstructured meshes, J. Sci. Comput. 39 (2009) 293-321.


\end{thebibliography}
\end{document}